\documentclass[epj,nopacs,referee]{svjour}
\usepackage{latexsym}
\usepackage{graphics}
\usepackage{longtable}

\usepackage{bm}
\usepackage{multirow}
\usepackage{amssymb}

\begin{document}
\title{Search for a Standard Model Higgs Boson in CMS 
        via Vector Boson Fusion in the H$\rightarrow$WW$\rightarrow l\nu l\nu$ Channel}
\author{E. Yazgan\inst{1,2}, J. Damgov\inst{3,2}, N. Akchurin\inst{4}, V. Genchev\inst{3}, D. Green\inst{2}, S. Kunori\inst{5}, M. Schmitt\inst{6}, W. Wu\inst{2} \and M. T. Zeyrek\inst{1}
\thanks{\emph{efe@fnal.gov}}%
}                     
\offprints{}          

\institute{Middle East Technical University, Ankara, Turkey \and Fermi National Accelerator Laboratory, Batavia, Illinois, USA \and Institute for Nuclear Research and Nuclear Energy, Bulgaria Academy of Science, Sofia, Bulgaria \and Texas Tech University, Lubbock, Texas, USA \and University of Maryland, College Park, Maryland, USA \and Northwestern University, Evanston, Illinois, USA}
\date{Received: date / Revised version: date}
%

\abstract{
We present the potential for discovering the Standard Model Higgs 
boson produced by the vector-boson fusion mechanism.  
We considered the decay of Higgs bosons into the $W^+W^-$ final 
state, with both $W$-bosons subsequently decaying leptonically.  
The main background is $t\bar{t}$ with one or more jets produced. 
This study is based on a full simulation of the CMS detector, and up-to-date reconstruction codes.  
The result is that a signal of $5\sigma$ significance can be obtained with an integrated luminosity of $12-72~fb^{-1}$ for Higgs boson masses between $130 < m_H < 200~GeV$.
In addition, the major background can be measured directly to 7\% from the data with an integrated luminosity of $30~fb^{-1}$.
In this study, we suggested a method to obtain information in Higgs mass using the transverse mass distributions. 
\PACS{
      {14.80.Bn}{Standard-model Higgs bosons}   \and
      {PACS-key}{discribing text of that key}
     } 
} 

\authorrunning{E. Yazgan, J. Damgov, et al.}
\titlerunning{Search for a SM Higgs Boson in CMS 
        via VBF in the H$\rightarrow$WW$\rightarrow l\nu l\nu$ Channel}

\maketitle
\section{Introduction}
\label{intro}
One of the primary goals of CMS is to prove or 
disprove the existence of the Higgs boson. The LEP experiments 
set a lower limit on the Standard Model (SM) Higgs boson at
114.4 GeV for a 95\%~C.L.~\cite{LEP}, and unitarity puts an 
upper limit of about 1~TeV.   Even more constraining are the results
of fits to precision electroweak measurements,
which limit the mass of a Standard Model-like Higgs boson 
to be less than 194~GeV~\cite{PDG} at 95\%~C.L.
In extended Higgs sectors, there is often one scalar
boson that resembles the Higgs boson of the Standard Model,
and is responsible for electroweak symmetry-breaking.
The mass of such a Higgs must also satisfy these constraints approximately.
In the Minimal Supersymmetric extension of the Standard Model (MSSM),
there is a more stringent bound coming from
the internal constraints of the theory; the lightest
Higgs boson must have a mass less than about~$135$~GeV.
For these reasons, we focus on the
mass region $120 < m_H < 200$~GeV.
\begin{figure}
     \resizebox{8cm}{!}{\includegraphics{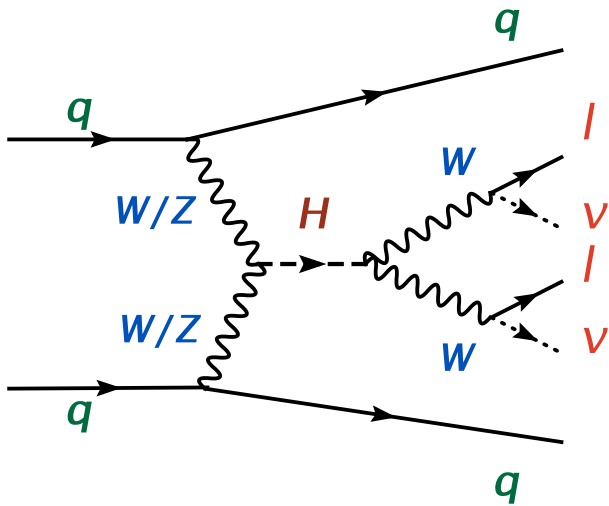}}
     \caption{\label{fig:vbf}\em
     Feynman diagram for Higgs boson production through Vector Boson Fusion. 
The Higgs boson decays into W's which further decay into electron/muon-neutrino pairs.}
\end{figure}
\par
The two main decay modes of the Standard Model Higgs boson
in this mass range
are $h \rightarrow b\bar{b}$ and $h \rightarrow W^+W^-$.
In the latter case, one of the $W$ bosons may be off the
mass shell.  If the Higgs boson is heavier than about
$135$~GeV, the $WW^{*}$ branching fraction will dominate,
but it can be important for masses as low as $120$~GeV.
In this study, we consider the decay $h \rightarrow WW^{*}$
with the subsequent decay of the $W$-bosons to two charged leptons.
\par
Higgs bosons may be produced in $pp$ collisions when radiated
off the virtual $W$-boson that is exchanged in the $t$-channel -
this is called ``Vector Boson Fusion'' (VBF). The Feynman diagram for 
this process is shown in Fig.~\ref{fig:vbf}. This channel has good
prospects for the discovery of a Standard Model Higgs
boson, especially if it is not too heavy because of the distinctive
VBF topology which contains two jets with small angles with respect to the beam axis.  
Furthermore, when
the Higgs decays to two $W$-bosons, the presence of the $hWW$
vertex both in production and decay of the Higgs boson gives
a relatively clean determination to the $hWW$ coupling.
Given the Higgs mass the Standard Model(SM) is completely determined,
so that a measure of $hWW$ coupling over-constrains the SM.
This will be crucial to establishing the origin of
electroweak symmetry breaking.
\par
The VBF mechanism was proposed as a potential discovery channel 
several years ago~\cite{krplrwzep01}. Our initial study of this 
channel for the CMS detector was carried out in~2002~\cite{zey02}, 
with a number of simplifications. The conclusion of this previous 
CMS study was that a convincing signal for a Higgs boson with a mass 
of~$120$~GeV would be observed with about $70~{\mathrm{fb}}^{-1}$.
In the present study, we repeat the entire analysis in 
the mass range $120$--$200$~GeV, using the latest  
simulation and reconstruction software for CMS in order to verify 
and improve the 2002~study. A similar study of this channel for the ATLAS 
detector was performed in 2004 using different generators and 
slightly different cuts ~\cite{atlas}.
\par
The VBF process is characterized by two forward jets with modest
transverse momentum, $E_{T}\approx m_{W}/2$, separated by a large 
rapidity difference. The Higgs boson signature is at low rapidity, with a 
pair of clean, isolated leptons and missing energy.  
The main backgrounds for this channel are the irreducible continuum $W^+W^-$ 
production, and $t\bar{t}$ in which both top quarks
decay semi-leptonically.  These backgrounds are particularly
troublesome when there are extra jets, j, in the event, so we have taken
particular care with the generation of $W^+W^-jj$ and $t\bar{t}$
events.

\section{Event Generation}
\label{sec:1}
The signal process and the $W^+W^-jj$ background have been simulated
on the basis of a matrix-element calculation using MadGraph~\cite{MadGraph}. 
For the $t\bar{t}j$ background, we used the AlpGen~\cite{AlpGen} package  
which correctly simulates spin correlations.  
We simulated the parton showers using Pythia~\cite{Pythia}. 
MadGraph and AlpGen calculations are made leading order (LO). 
The parton distribution functions used by MadGraph and AlpGen are 
CTEQ6L1 and CTEQ5L1 respectively. The minimum transverse momentum cut 
on jets is~$15$~GeV, and the jet pseudo-rapidity is limited to $|\eta| < 5$.
We required a separation of any jet pair, namely, $\Delta R > 0.5$,
where $\Delta R = \sqrt{(\Delta\eta)^2+(\Delta\phi)^2}$.
\par
Next-to-leading order (NLO) cross-sections differ from 
LO cross-sections by $\sim30\%$ for a 120 GeV Higgs boson and 
$\sim10\%$ for a 200 GeV Higgs boson~\cite{nikitenko}. However, 
since there are no NLO cross-section calculations for the backgrounds, 
the LO cross-sections are used consistently for both signal and background 
processes in this study. 
The cross sections are listed in Table~\ref{table:cross}. 
The `electroweak' (EW) part of the $W^+W^-jj$ process is defined as the subsample with no 
$\alpha_s$-dependent vertex in the diagrams, and the `QCD' part 
is the rest of this process. Note that the EW part is topologically very 
similar to the signal and hence is almost irreducible.

\begin{table}[htb]
\caption{\label{table:cross}
Production cross-section for the signal and main 
backgrounds}
\begin{minipage}{0.75\linewidth}
\footnotesize
\begin{center}
\begin{tabular}{llll} \hline\noalign{\smallskip}
Channel &cross-section [pb] & WW BR & $\sigma\times$BR 
[pb]\\ 
\noalign{\smallskip}\hline\noalign{\smallskip}
         qqH m=120 &  4.549 & 0.133 &  0.605 \\ 
         qqH m=130 &  4.060 & 0.289 &  1.173  \\ 
         qqH m=140 &  3.648 & 0.486 &  1.773 \\ 
         qqH m=160 &  3.011 & 0.902 &  2.715 \\ 
         qqH m=180 &  2.542 & 0.935 &  2.376 \\ 
         qqH m=200 &  2.177 & 0.735 &  1.600 \\ \hline
         ttj       & 736.5  & 1. & 736.5 \\ 
         WWjj QCD  & 43.6 & 1. & 43.6 \\ 
         WWjj EW   & 0.933 & 1. & 0.933 \\ 
\noalign{\smallskip}\hline
\end{tabular}
\end{center}
\end{minipage}
\end{table}

\section{Detector Simulation and Event Reconstruction}
\par
We processed the generated events through the CMS detector simulation software 
which is based on the Geant-4 simulation of the CMS detector. 
We simulated pile-up from out-of-time interactions representing the low luminosity LHC running condition 
($\sim2\times10^{33}~{\mathrm{cm}}^{-2} {\mathrm{s}}^{-1}$).
Subsequently, we processed digitized information (digis) using the CMS event reconstruction
software. 

\subsection{Trigger}
\par
We refer to Ref.~\cite{triggers} for the presently planned trigger table.
The inclusive single electron trigger has an $E_T$-threshold of $26$~GeV,
which is too high for our purposes.  Therefore we will augment
this trigger with the di-electron trigger, which has a threshold 
of~$12$~GeV for both electrons.
The $p_T$-threshold for the inclusive single muon trigger is~$19$~GeV,
which is well suited to this analysis.  Concerning the $e$-$\mu$
channel, we plan to use the $e$+$\mu$ di-lepton trigger, which
will have a threshold of~$10$~GeV for each lepton.
The efficiency for the L1+HLT trigger with respect to our
offline cuts varies from about $95\%$ to $99\%$ based on Ref.~\cite{DADIGI}.
This presents no significant effect at the current state of our
analysis.
\par
There will be lepton+jet triggers that should be very useful
for this analysis if lower lepton thresholds are needed.  
However, since the details for these triggers
are not available at this time, we have based our study solely 
on the leptonic triggers.
\subsection{Lepton Reconstruction and Identification}
\par
We have used standard packages and selection criteria for muon and electron identification.  
Below, we describe our assessment of the identification efficiency.

\subsubsection{Muons}
\par
We use the ``global'' muon reconstruction, which
takes muons found in the muon chambers (drift tubes, 
cathode strip chambers, and RPC's), and extrapolates them
into the silicon tracker to pick up additional hits and
better define the kinematics.  This extrapolation takes
into account the energy lost by the muon as well as
multiple scattering.  
\par
Muons are found within $|\eta| < 2.4$. 
The overall muon reconstruction efficiency in this angular
range is $\approx 95\%$ for 
$10 < p_{T} < 30~GeV$ and $97\%$ for 
$p_{T} > 30~GeV$.

\subsubsection{Electrons}
\par
Electrons are reconstructed by combining super-clusters  ~\cite{ref_sc1,ref_sc2}
and Kalman tracks~\cite{cmskalman}. The track -- super-cluster (SC) matching 
condition is $\Delta R<0.15$.  Such tracks should have at least
four hits, and transverse momentum $p_{T}>5$~GeV. If several 
tracks satisfy these conditions, then the one having the least 
difference $|p_T - E_T|$ is taken.
We reject the electron candidates if $E_{T}^{\mathrm{SC}}<10$~GeV or 
$|\eta^{\mathrm{SC}}|>2.0$. The probability for a generator level electron 
with $p_{T}>10$~GeV and $|\eta|<2.0$ to be reconstructed within 
$\Delta R < 0.2$ is $\sim 92$--$98\%$ for $10<p_{T}(gen)<20$ GeV and 
$\sim98$--$99\%$ for $p_{T}>20$~GeV. These reconstructed electrons 
are said to be identified if they satisfy 
$E_{\mathrm{HCAL}}/E_{\mathrm{ECAL}}<0.05$, 
$|\Delta\eta({\mathrm{trk,SC}})|<0.005$, 
$E^{\mathrm{SC}}/p^{\mathrm{trk}}>0.8$ and
$|1/E^{\mathrm{SC}}-1/p^{\mathrm{trk}}|<0.06$. 
\par
An isolation variable is defined by taking the sum of the $p_{T}$ 
of all the tracks (except the electron candidate) within 
a cone of $\Delta R^{\mathrm{SC}} < 0.2$, and dividing by 
the $E_{T}^{\mathrm{SC}}$.  The tracks
entering this sum must have at least four hits, $p_T > 0.9$~GeV,
and $|z^{\mathrm{trk}}-z^{e}|<0.4$ cm, where z is the position of the track along the beam line.  
We place the requirement
that this isolation ratio be smaller than~0.2.
The overall single electron efficiency for electron isolation and 
identification is $\approx 80\%$ for $10<p_{T}<30~GeV$ 
and $\approx 90\%$ for $p_{T}>30$~GeV.
The electron fake rate per jet is $\approx3\%$ for 10$<p_T^j<$30 GeV and 
less than $\approx0.1\%$ for $p_T^j>$ 120 GeV calculated using the jets from
 W decay in the associated production and using the forward jets in the $qqH$ sample.

\subsection{Jet and Missing E$_T$ Reconstruction and Correction}
\par
The cell-level thresholds are set at least 2$\sigma$ above the noise level 
to remove the effects of calorimeter noise fluctuations in jet reconstruction.
This is important since we are mainly dealing with quite low-$p_T$ jets in the 
current study.
\par
We reconstructed the jets using the ``Iterative Cone'' algorithm, with a cone 
size of $\Delta R = 0.5$ and a cone seed $E_T$ cut of $1$~GeV. We removed the jets 
from an event if they match a reconstructed electron within a cone
of $\Delta R<0.45$.
\par
We calibrated the reconstructed jets using the $qqH$ signal sample. 
Reconstructed jets are first matched to generator level jets
within a cone of $\Delta R<0.12$.
We fit the jet response  to second-order polynomials as 
a function of generator-level jet $E_{T}$ for 20 different $\eta$ regions 
covering $\eta=0$ to $\eta=4$ in bins of $\Delta\eta = 0.2$. 
The difference between the corrected and uncorrected 
responses varies by 10\% to 30\% depending on the jet $E_T$ and $\eta$ 
values.    
When applying the correction to jets with $|\eta|>4$, we used the correction parameters 
for the last interval $|\eta| = 3.8$ -- $4.0$. 
The polynomial
extrapolation is unreliable beyond $p_T = 200$~GeV, so we fixed the corrections above 200 GeV
to those obtained at $200$~GeV.
The response to jets in the QCD di-jet sample is lower than the response to jets in the
qqH sample. This produces different correction functions. However, in the 
current study, VBF tag jets are at high $\eta$ and have at least $p_T>30$ GeV and
for this part of phase space the differences between responses(or equivalently, 
the jet correction functions) are very small.

In the analysis, we used missing $E_T$($\not\!\!\!E_T$) calculated from calorimeter hits. 
We corrected the $\not\!\!\!E_T$ using the sum of the $E_T$ difference between the 
corrected and uncorrected jets for which the corrected jets have 
$E_T>30$~GeV. 
\section{Event Selection}
\par
The strategy of the analysis is not complicated. We select events
with two forward jets separated by a large rapidity difference, veto
any event with additional central jets, and demand two energetic,
isolated leptons in the central region.  Finally, we apply additional cuts 
on the kinematics and the event topology.

\subsection{Forward Jet Tagging}
\par
\begin{figure*}
    \resizebox{9cm}{!}{\includegraphics{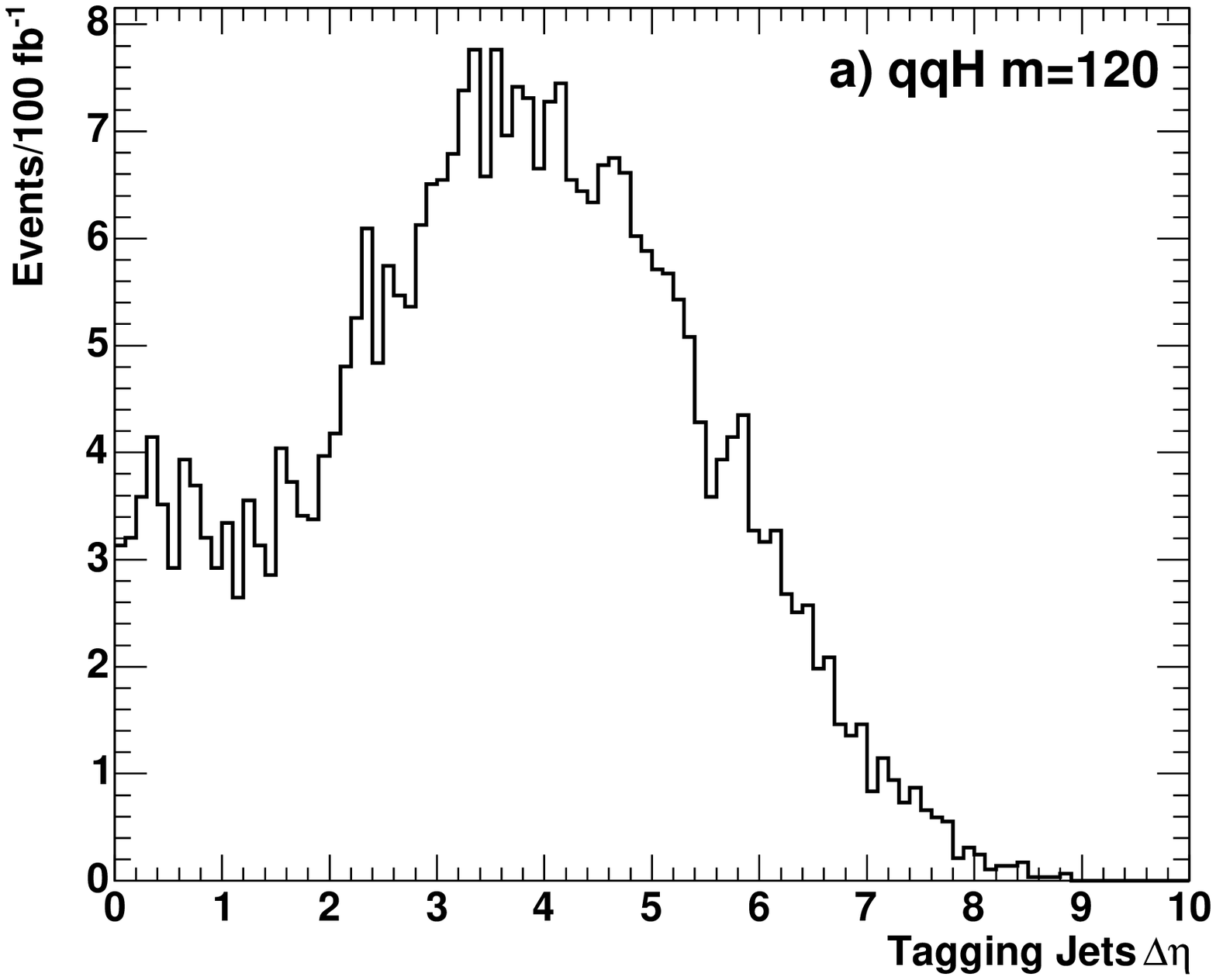}}    
    \resizebox{9cm}{!}{\includegraphics{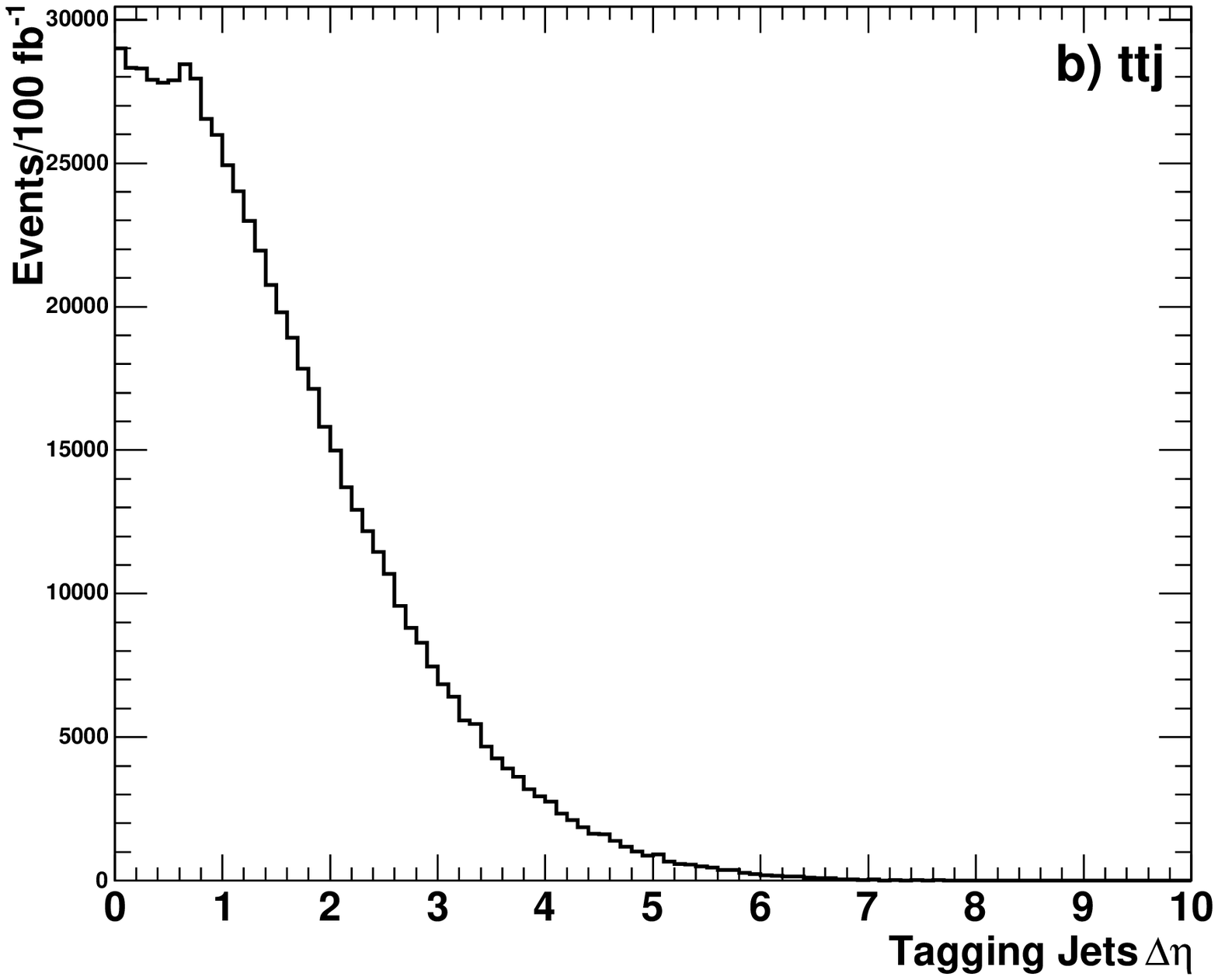}}    
    \resizebox{9cm}{!}{\includegraphics{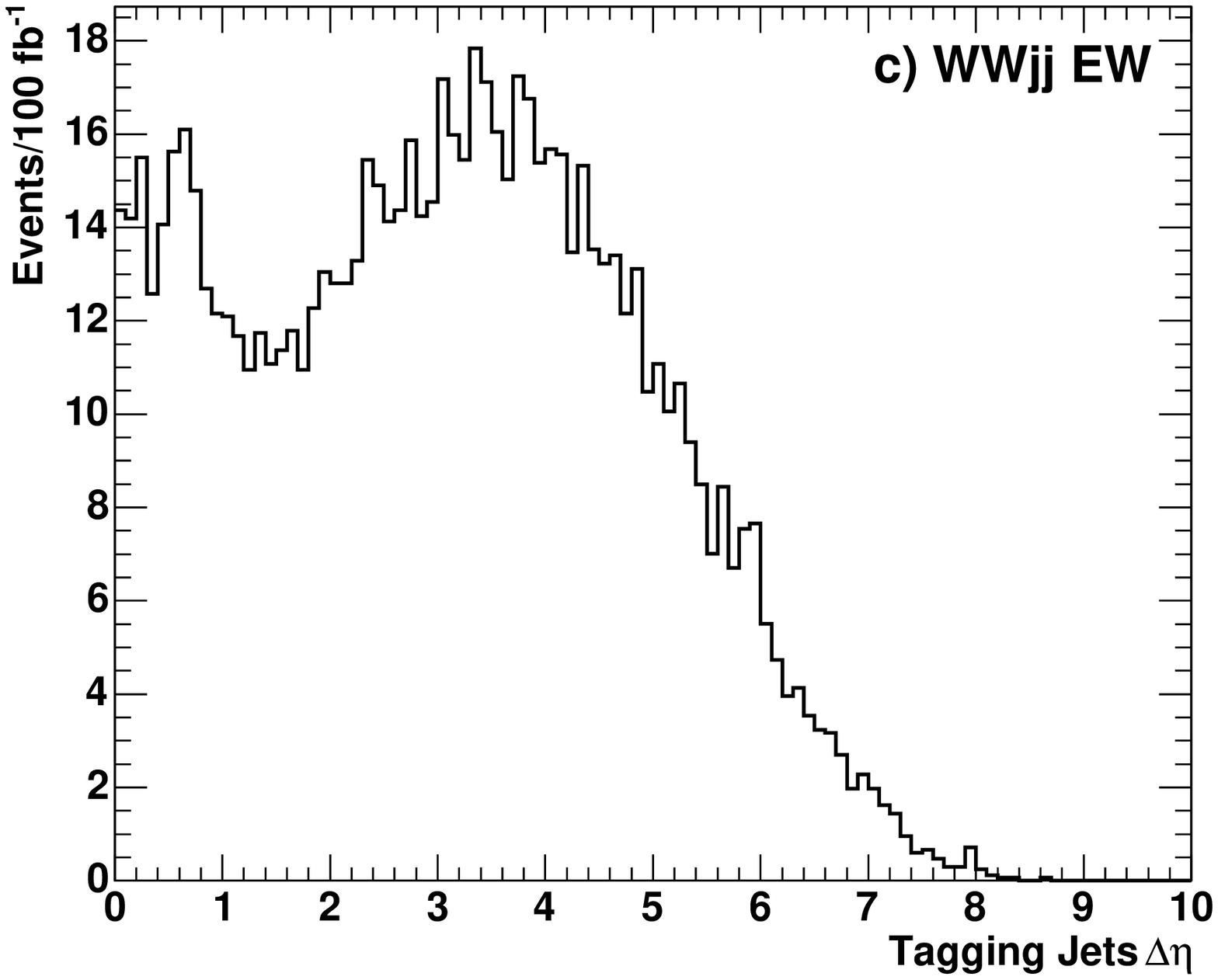}}    
    \resizebox{9cm}{!}{\includegraphics{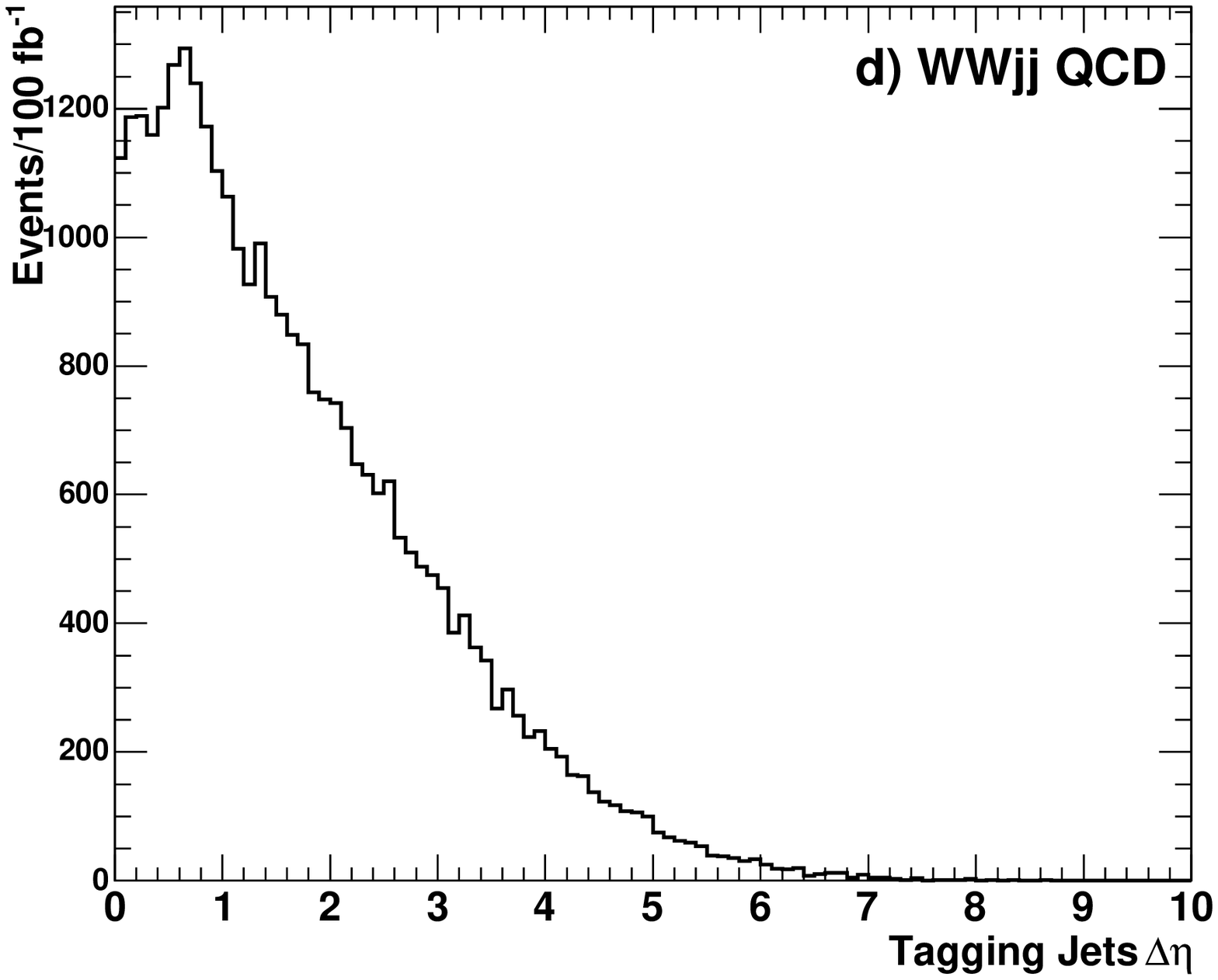}}    
     \caption{\label{fig:deltaeta}\em
     $\Delta\eta=|\eta_1-\eta_2|$ distribution for the forward tagging jets which 
     have $E_{T1} > 50~GeV$ and $E_{T2} > 30~GeV$ for a) qqH, $m_{H}=120$ GeV and backgrounds
b)$t\bar{t}j$, c) EW WWjj and d) QCD WWjj. Note that the EW WWjj background is basically irreducible.}
\end{figure*}

The jets are ordered in $E_T$ after the corrections have been applied.
The first two tag jets should be energetic, so we require
$E_{T1} > 50$~GeV and $E_{T2} > 30$~GeV.  Fig.~\ref{fig:deltaeta}
shows the rapidity separation $|\Delta\eta|$ between these
two most energetic jets, for the signal(a) and
the backgrounds(b-d).  It is clear that the jets for
signal events are well separated in rapidity, and we apply the
cut $|\Delta\eta| > 4.2$.  We also make sure that they
fall in opposite laboratory hemispheres by requiring $\eta_1 \cdot \eta_2 < 0$.

\subsection{Central Jet Veto}
\par
\begin{figure*}
    \resizebox{9cm}{!}{\includegraphics{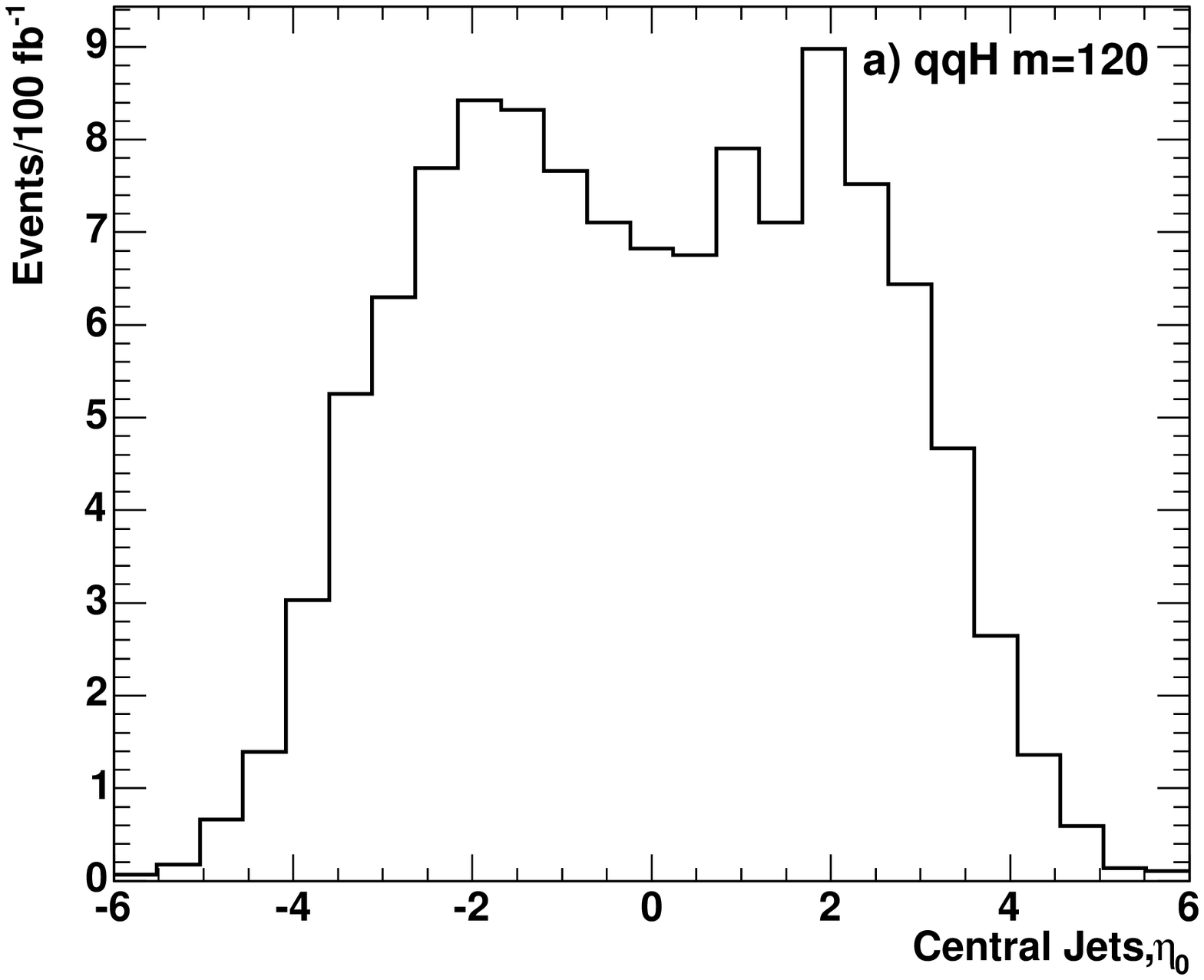}}
    \resizebox{9cm}{!}{\includegraphics{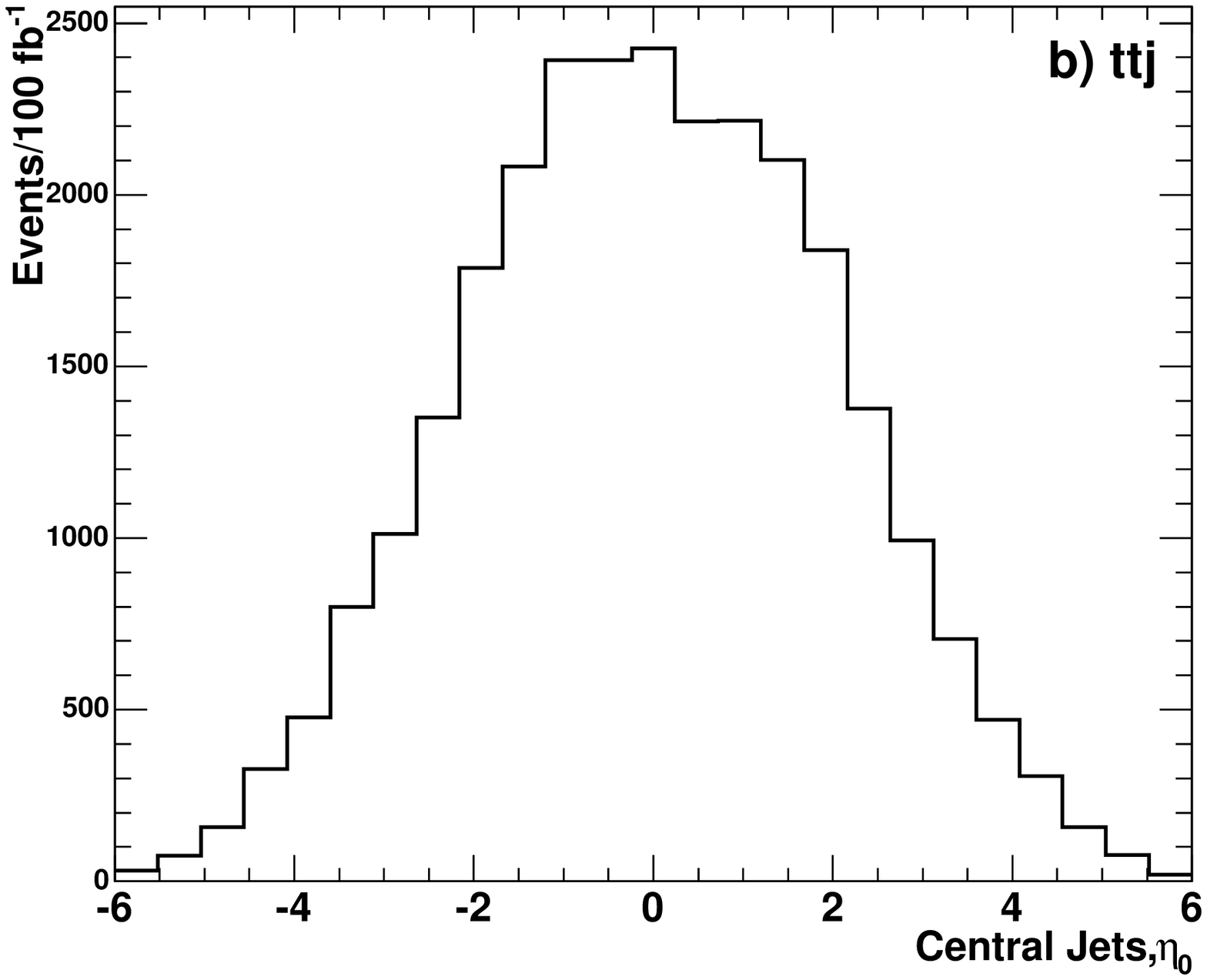}}
    \resizebox{9cm}{!}{\includegraphics{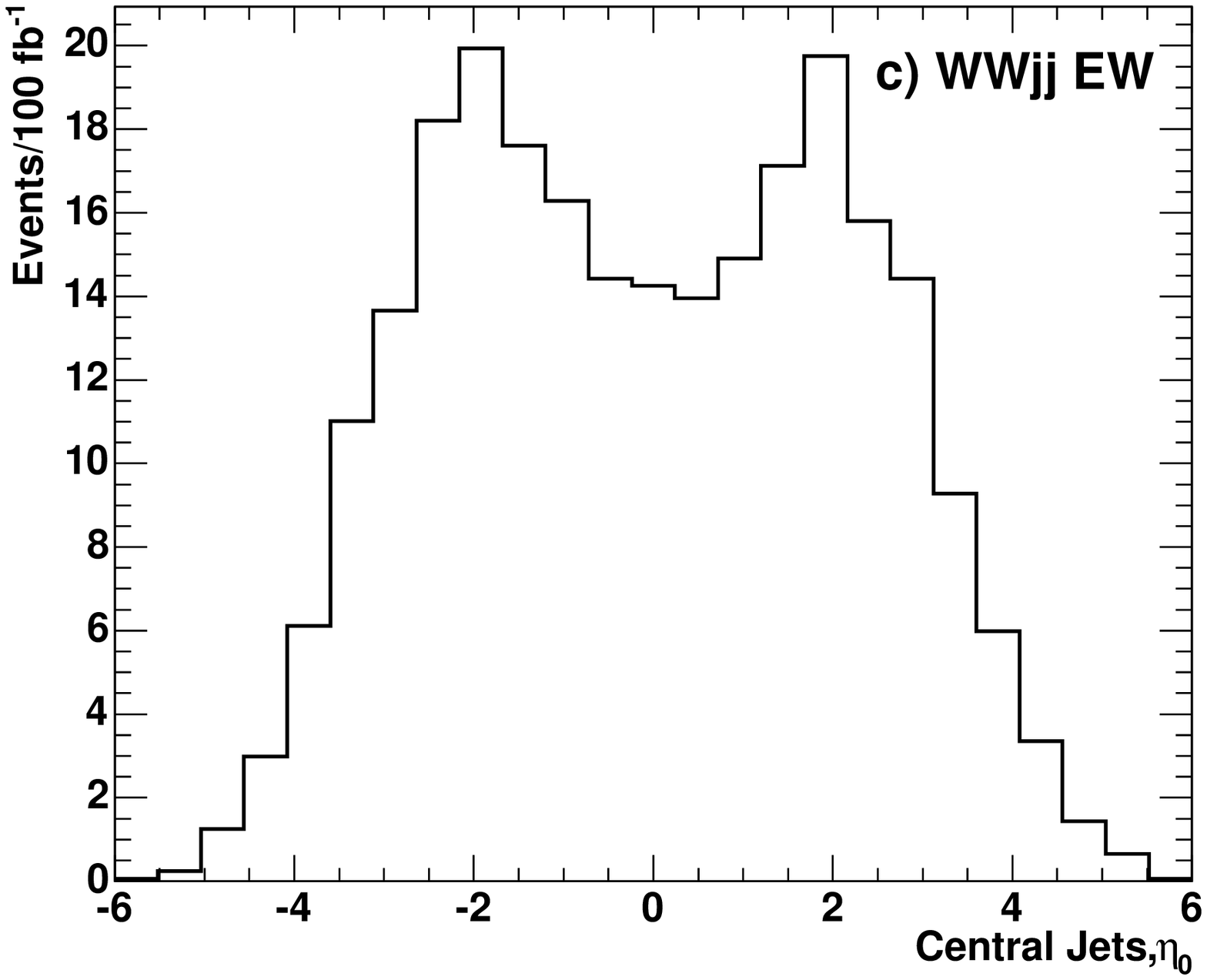}}
    \resizebox{9cm}{!}{\includegraphics{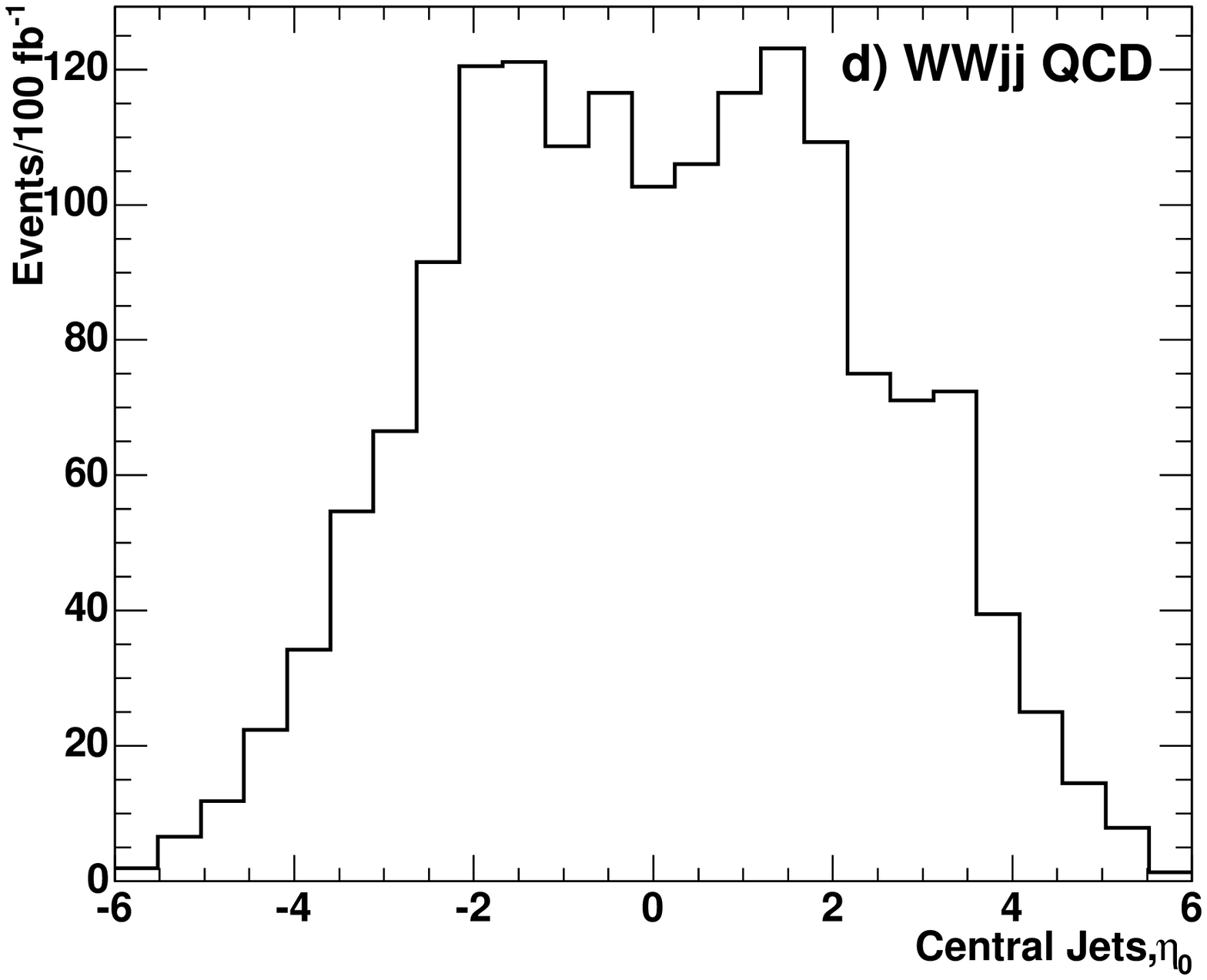}}
     \caption{\label{fig:jeta0}
      $\eta_0 = \eta_3 - (\eta_1 + \eta_2)/2$ for the third jet. $\eta$ of the third jet with respect
to the average of the two forward jets. For signal a) qqH, $m_H=120$ GeV and backgrounds b) $t\bar{t}j$, c) EW WWjj
and d) QCD WWjj.}
\end{figure*}
\begin{figure}
\resizebox{9cm}{!}{\includegraphics{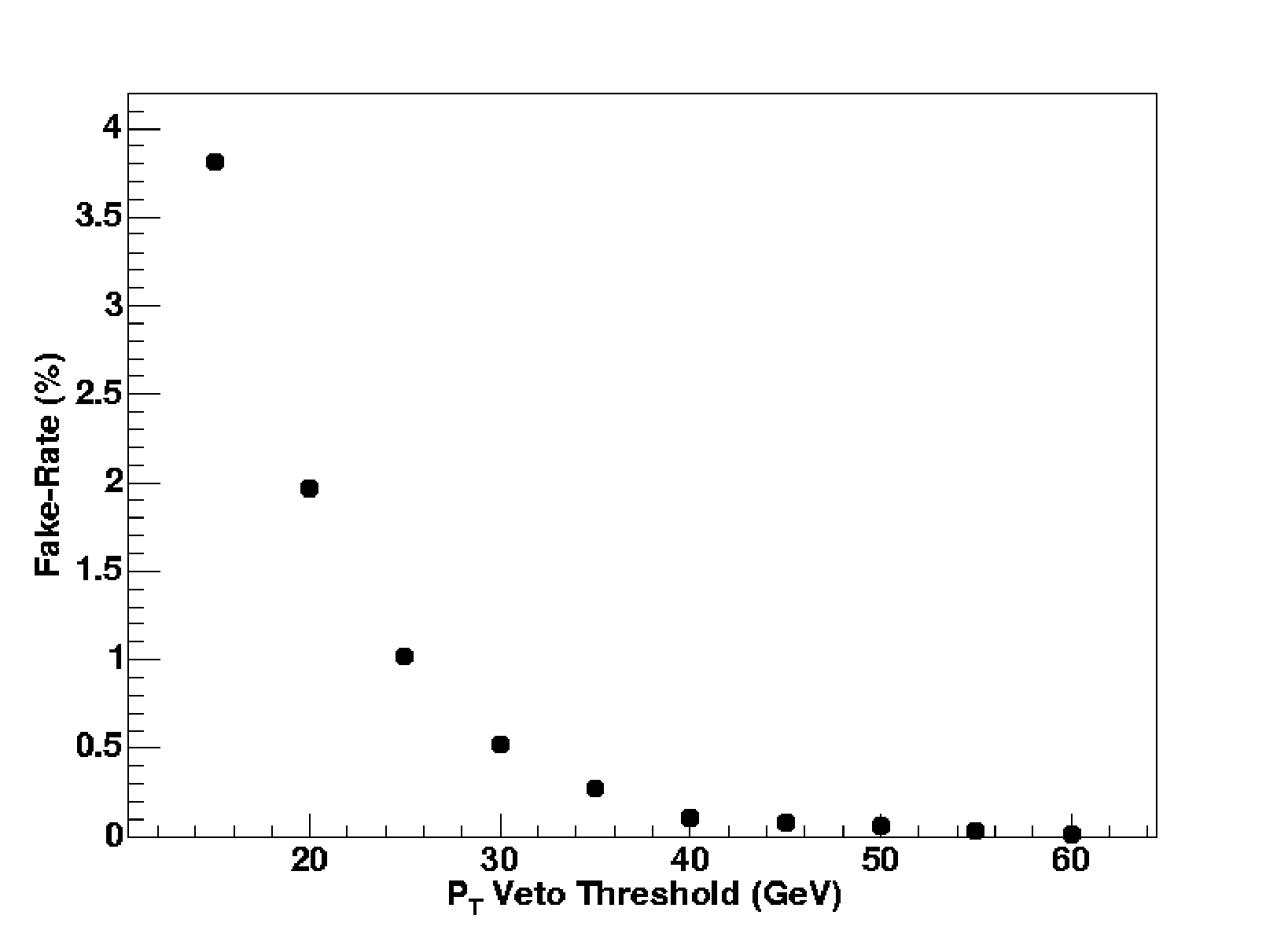}}
\caption[.]{\label{fig:fakecentral}
	Fake central jets fraction per event as a function of 
$E_T$ veto threshold. A fake is defined 
as the probability to find at least one jet(due to pile-up) 
satisfying the central jet 
veto conditions, with no "real" jets satisfying the central jet veto 
condition in that event.}
\end{figure}

In the signal process, there is no color exchange between the protons,
and consequently any additional jets will tend to be radiated in the forward direction.  
In contrast, the backgrounds will tend to have additional jets
in the central region, especially the $t\bar{t}j$ process.
We take advantage of this distinction by vetoing events with
additional jets in the central region.  In particular, we
consider any jet with $E_{T3} > 20$~GeV and
compute the rapidity with respect to the average of the
two forward jets: $\eta_0 = \eta_3 - (\eta_1 + \eta_2)/2$.
We veto the event if $|\eta_0| < 2$.
See Fig.~\ref{fig:jeta0} for distributions of both signal and background. 
The probability to find
a fake jet from pile-up events for low luminosity LHC running is 
shown in Fig.~\ref{fig:fakecentral} as a function of the $E_T$ threshold 
for the central jet veto. The fake rate is defined as the rate for pile-up jets 
satisfying the central jet veto condition in an event where there are no
real jets satisfying those conditions. Therefore, the fake rate is just the 
rate of events mistakenly rejected due to pile-up.  
The loss of events for a $E_T$ threshold of $20$~GeV is only about~$2\%$. 

\begin{figure*}
\begin{center}
\resizebox{16cm}{!}{\includegraphics{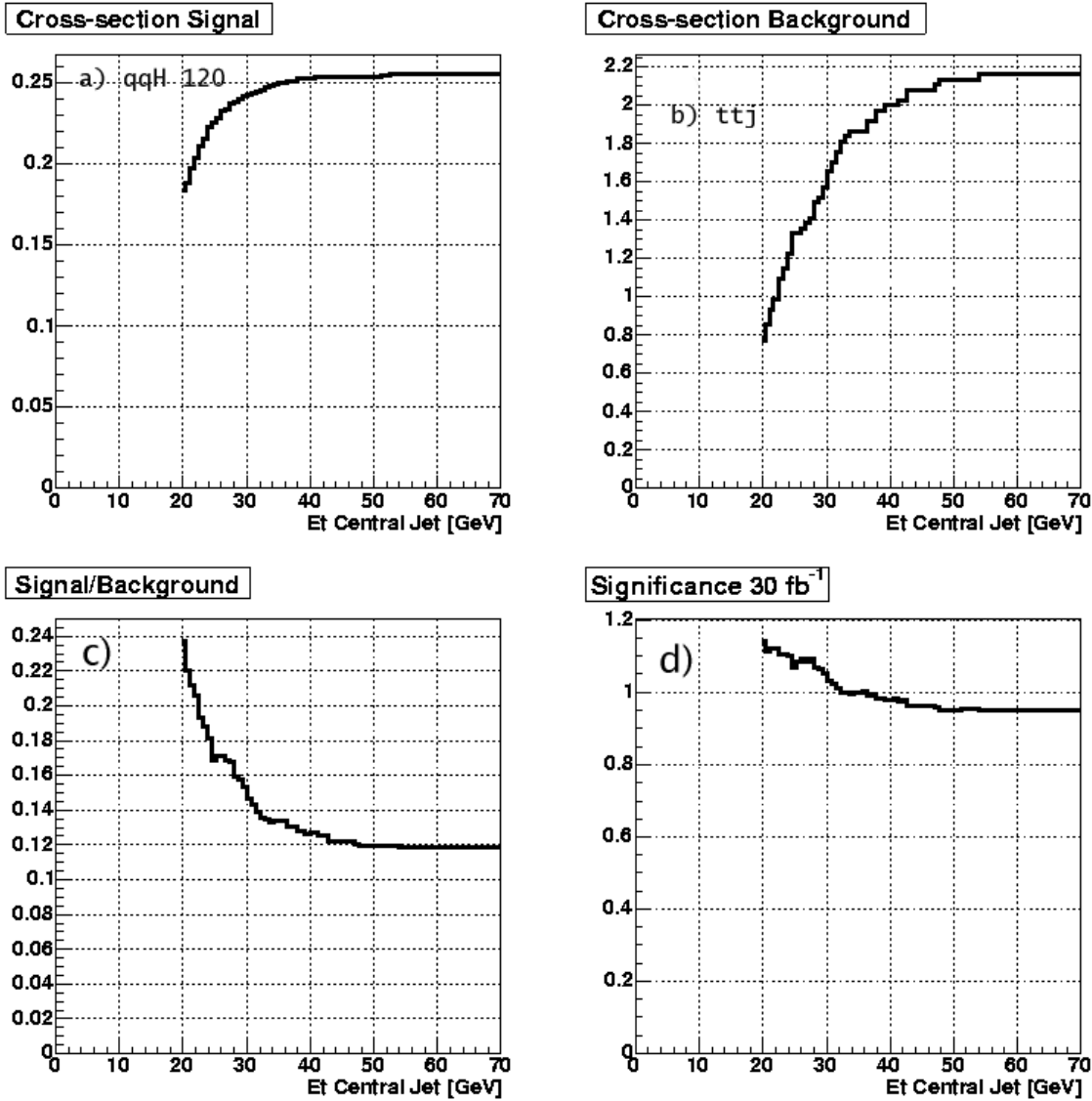}}
\caption[.]{\label{fig:centraltune}
The effect of the $E_T$ threshold for the central jet veto.
For a) signal evernts, qqH with $m_H=120$ GeV and background events b) $t\bar{t}j$.
c) the S/B ratio and d) the significance for a 30 $fb^{-1}$ integrated luminosity.}
\end{center}
\end{figure*}

The effect of the $E_T$ threshold for the central jets on the final cross 
sections and significances for the $120$~GeV signal and for the background 
are displayed in Fig.~\ref{fig:centraltune}. Here, the significance is 
defined as $S/\sqrt{B}$, where~$S$ and~$B$ represent the numbers of signal 
and background events.

\subsection{Lepton Kinematics}
\par
\begin{figure}
    \resizebox{9cm}{!}{\includegraphics{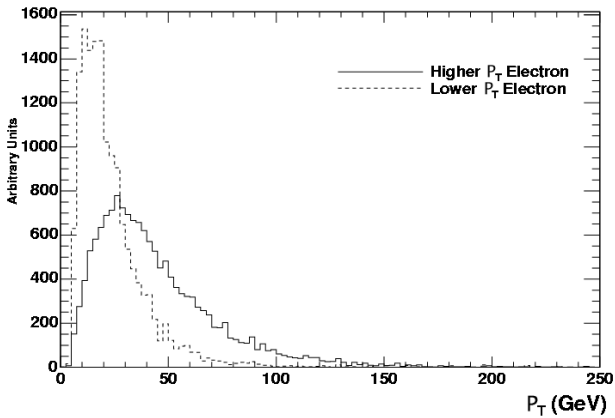}}
     \caption{\label{fig:etspectra}\em
     Electron $p_T$ spectra for the signal process,qqH, when $m_H = 120$~GeV}
\end{figure}

We require two opposite-sign leptons in an event.  
The most energetic lepton
must have $p_{T1} > 20$~GeV, and the other, $p_{T2} > 10$~GeV.
The  $p_T$-threshold for the second lepton must be low
since one of the two $W$'s in the Higgs decay is off the mass shell for low Higgs masses.
Fig.~\ref{fig:etspectra} shows the $p_T$ spectra for electrons
in the signal process ($M_H = 120$~GeV).
We reject events with more than two leptons.  
The two leptons must be well separated from all jets with $\Delta R_{\ell j} > 0.7$.
\par

In light of the thresholds for the electron triggers, we
modified our $p_T$ requirements slightly in the di-electron channel.  
An event is selected if it has two electrons which satisfy:
\begin{center}
\begin{tabular}{c}
 $(p_{T1} > 26~GeV {\mathrm{~AND~}} p_{T2} > 10~GeV)$ \\
    OR \\
 $(p_{T1} > 20~GeV {\mathrm{~AND~}} p_{T2} > 12~GeV)$ .
\end{tabular}
\end{center}
\par
\begin{figure*}
    \resizebox{9cm}{!}{\includegraphics{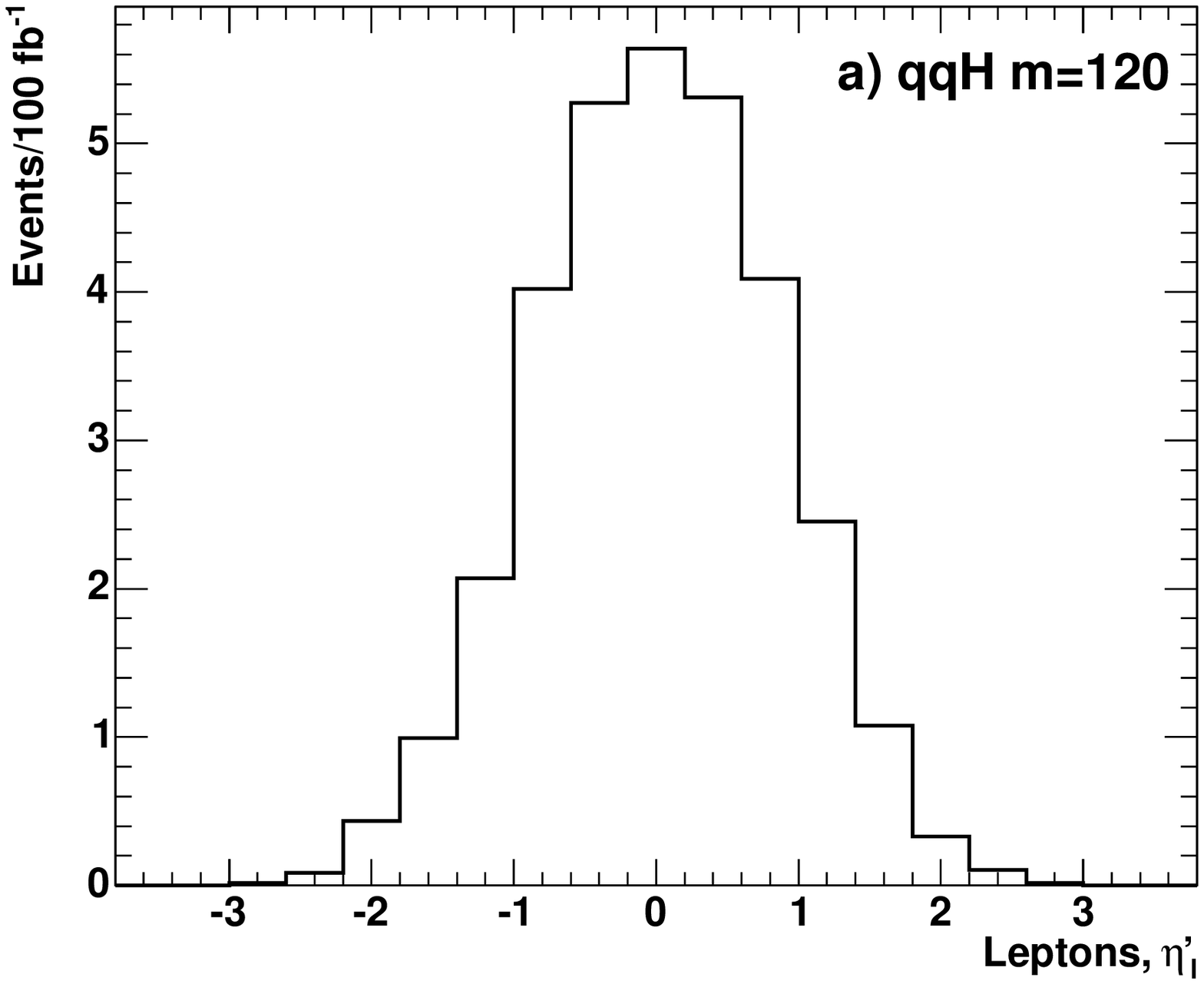}}
    \resizebox{9cm}{!}{\includegraphics{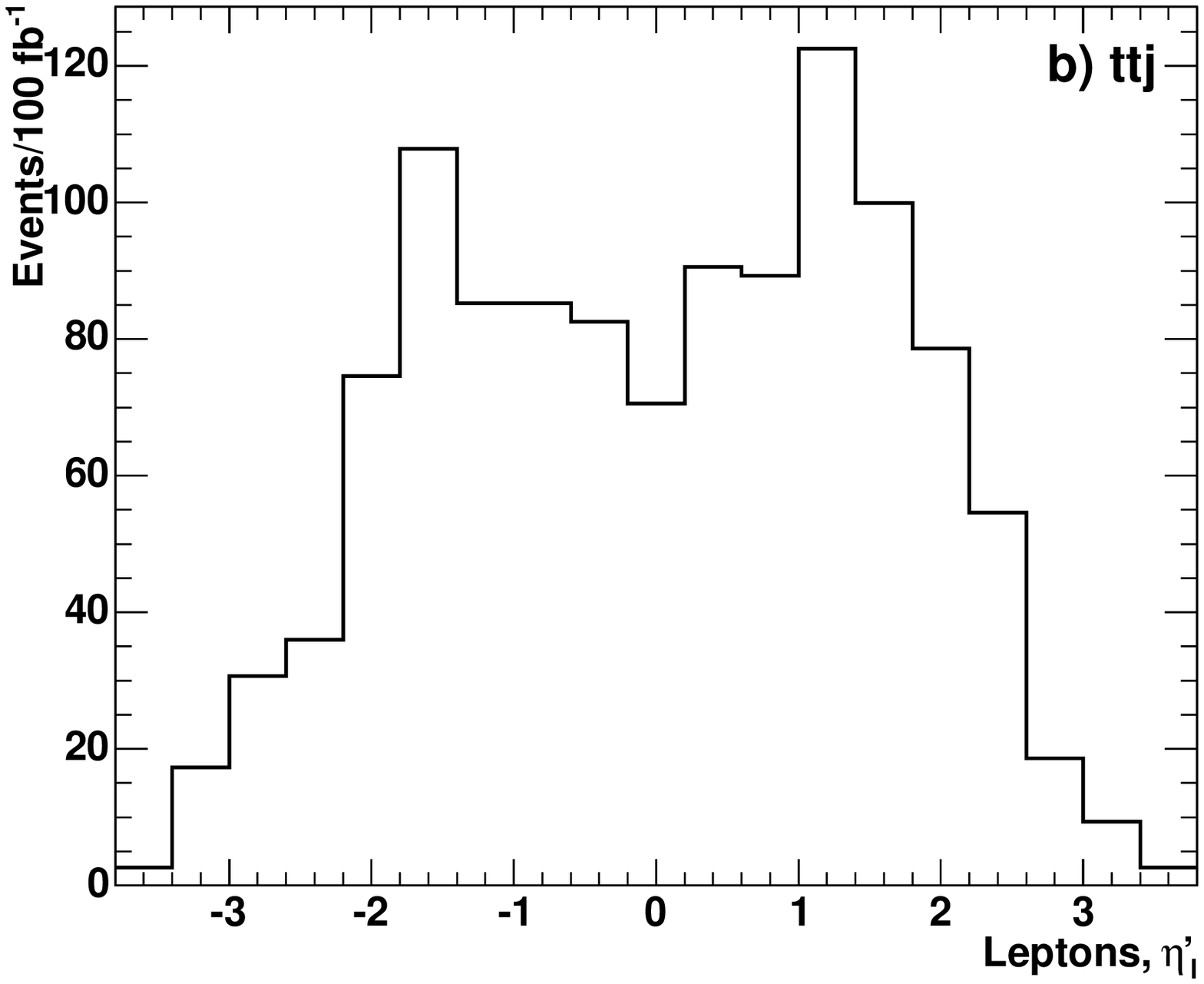}}
    \resizebox{9cm}{!}{\includegraphics{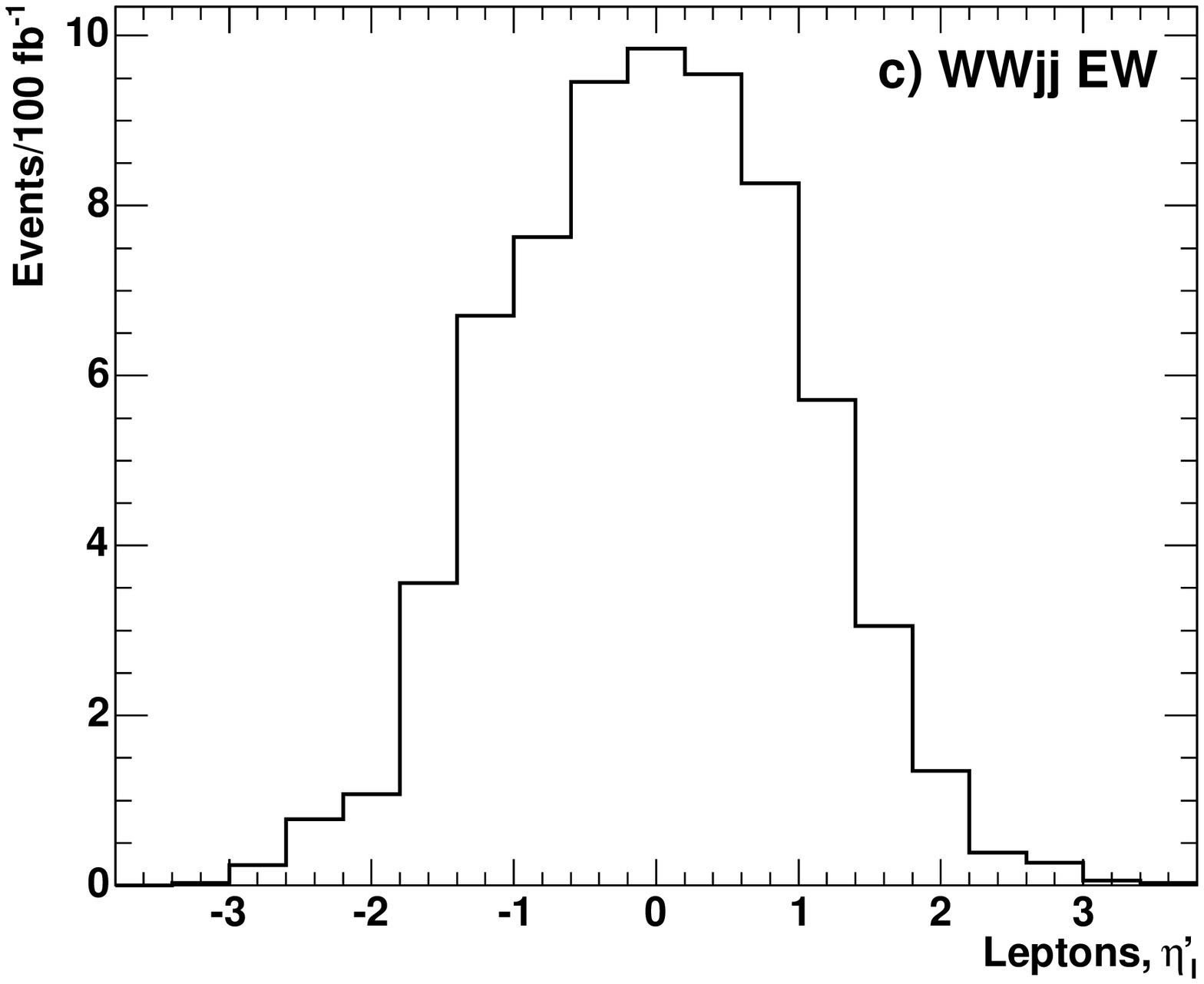}}
    \resizebox{9cm}{!}{\includegraphics{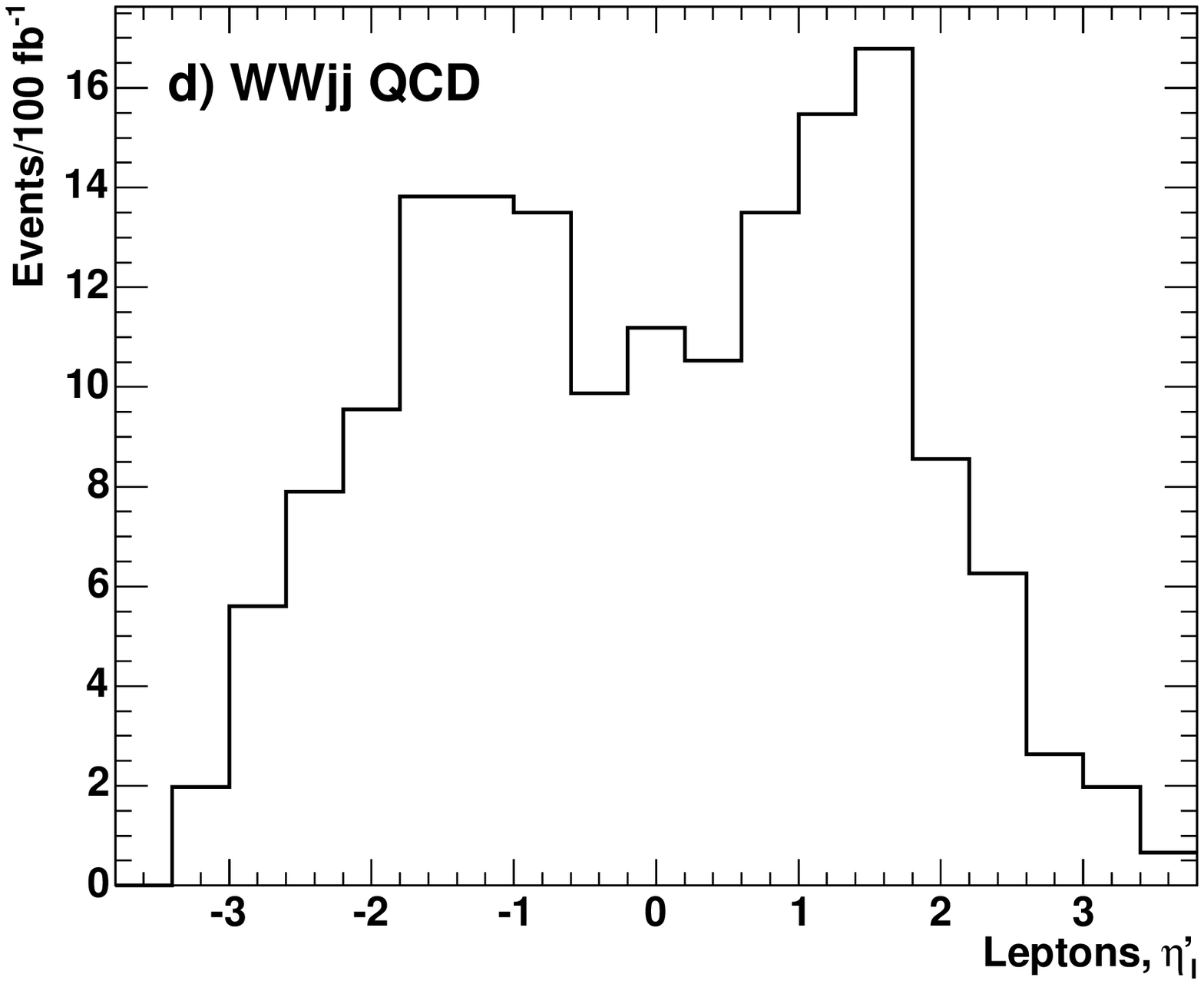}}
     \caption{\label{fig:leta0_prim}\em
     Centrality of the leptons, using the quantity $\eta_\ell^\prime$ defined in the text
for a) qqH, $m_{H}=120$ GeV and backgrounds b) $t\bar{t}j$, c) EW WWjj and d) QCD WWjj.}
\end{figure*}

Since the leptons come from the $W$'s that come from
the centrally-produced Higgs boson, we require them to
be central.  If $\eta_{\mathrm{hi}}$ is the forward-tag jet
having higher-rapidity, and $\eta_{\mathrm{lo}}$
is that of the lower-rapidity forward-tag jet, then
our requirement can be written
$\eta_{\mathrm{lo}}+0.6 < \eta_{\ell} < \eta_{\mathrm{hi}}-0.6$.
This condition must be satisfied by both leptons.
Fig.~\ref{fig:leta0_prim} shows the distributions of the
related quantity,  $\eta_\ell^\prime = (\eta_\ell - (\eta_{1}+\eta_{2})/2) \times
4.2 / \Delta\eta$. This quantity is sensitive to the $\eta$ distribution
of leptons with respect to the forward tag jets.  

\subsection{Further Kinematic Requirements}
\par
After the forward-jet tag, the central jet veto, and
the lepton kinematics cuts, we are left with a sample which
still has a large contamination from background processes.  
We can further reduce this contamination with some additional kinematic cuts.  
\par
\begin{figure*}
    \resizebox{9cm}{!}{\includegraphics{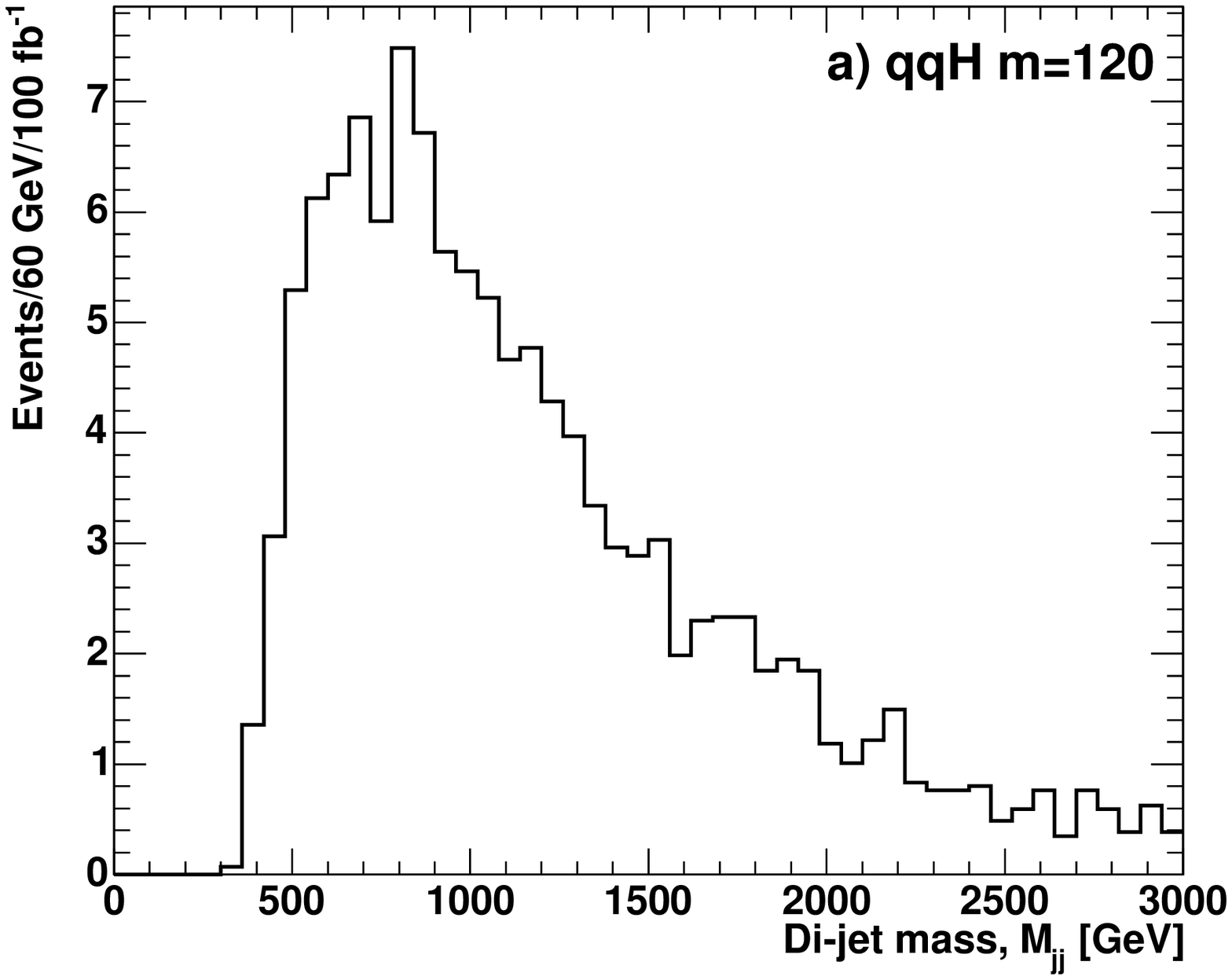}}
    \resizebox{9cm}{!}{\includegraphics{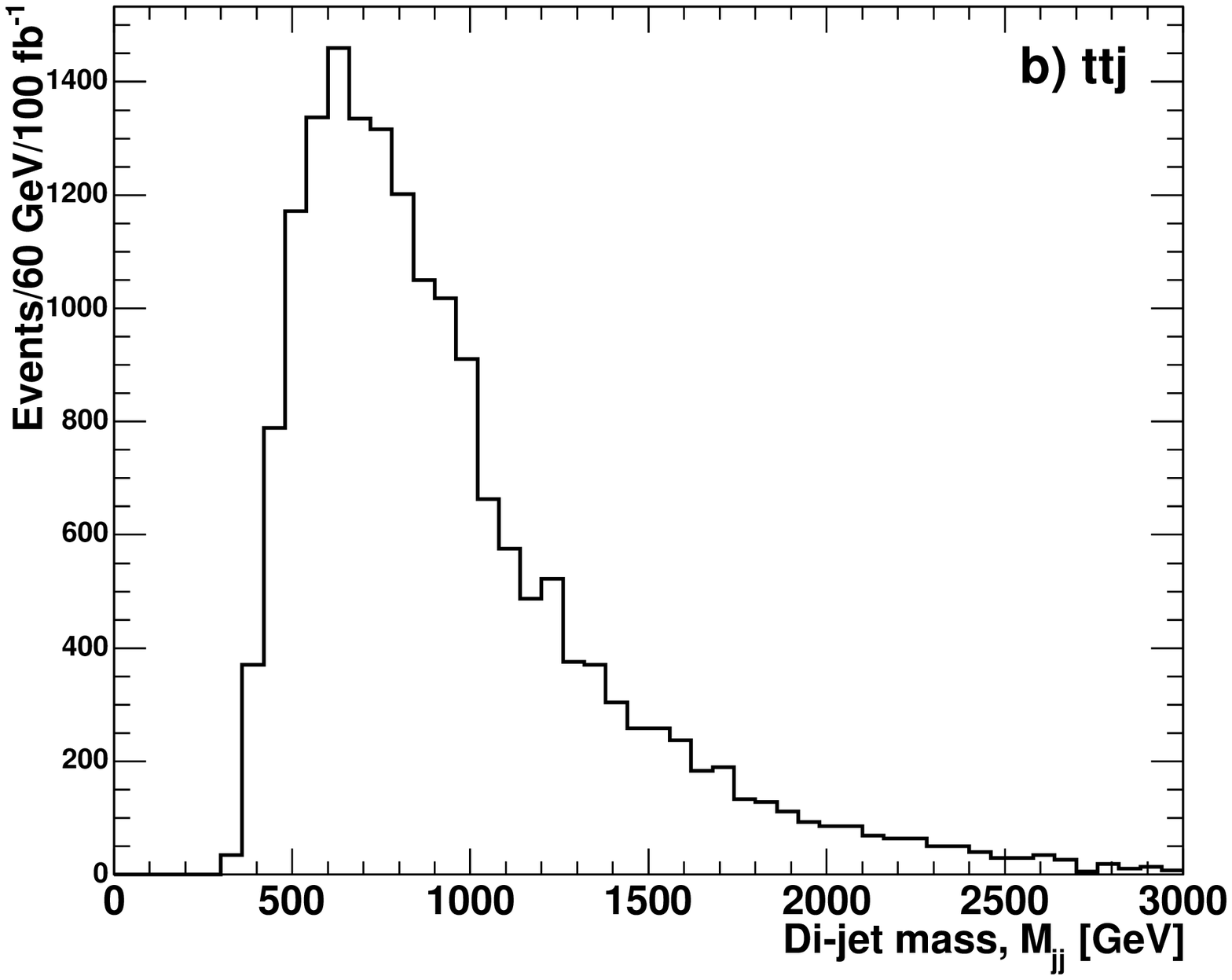}}
    \resizebox{9cm}{!}{\includegraphics{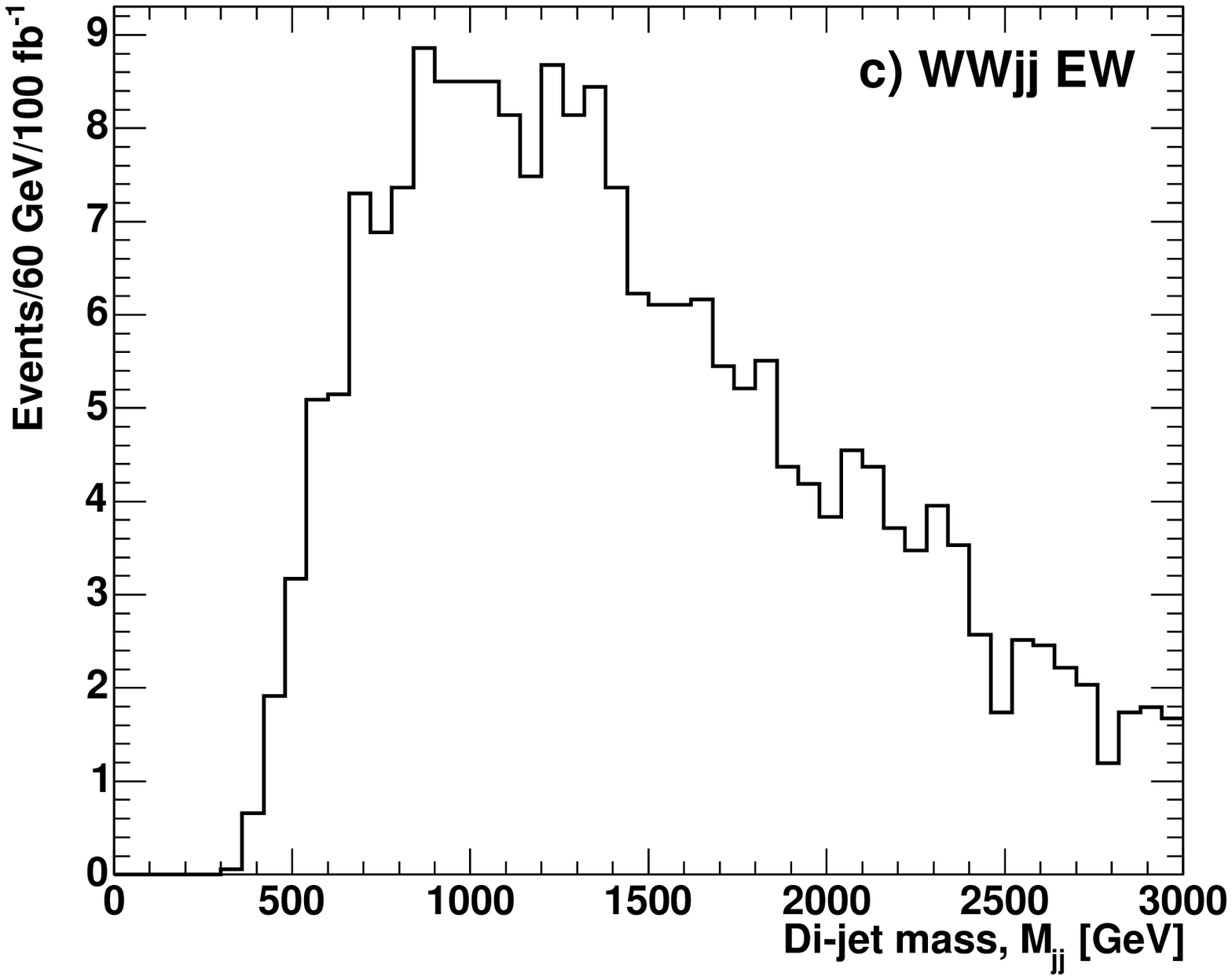}}
    \resizebox{9cm}{!}{\includegraphics{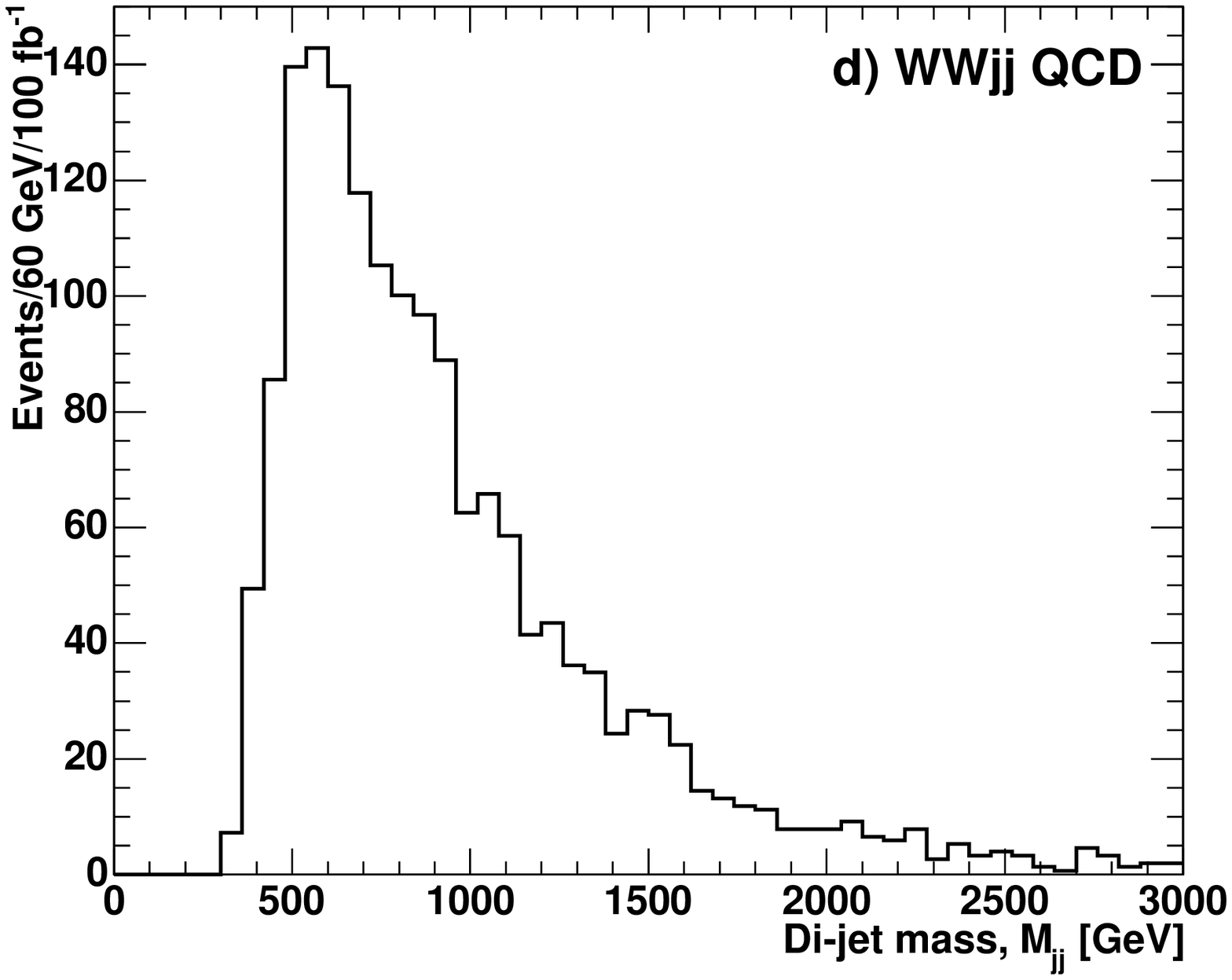}}
     \caption{\label{fig:Mjj}\em
     Invariant mass distributions for the two forward tag jets, for a) qqH, $m_{H}=120$ GeV and backgrounds
b) $t\bar{t}j$, c) EW WWjj and d) QCD WWjj.}
\end{figure*}
\begin{figure*}
    \resizebox{9cm}{!}{\includegraphics{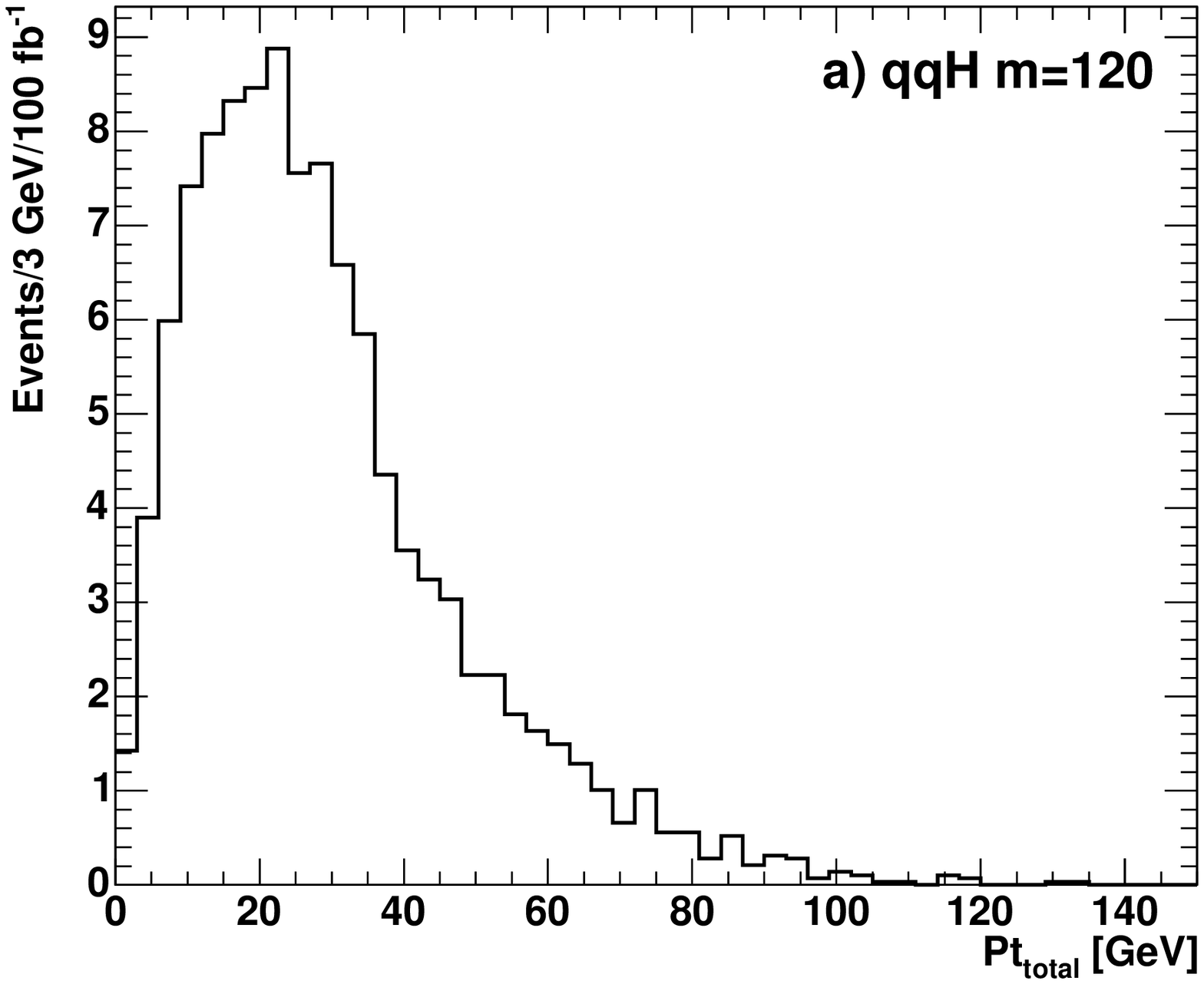}}
    \resizebox{9cm}{!}{\includegraphics{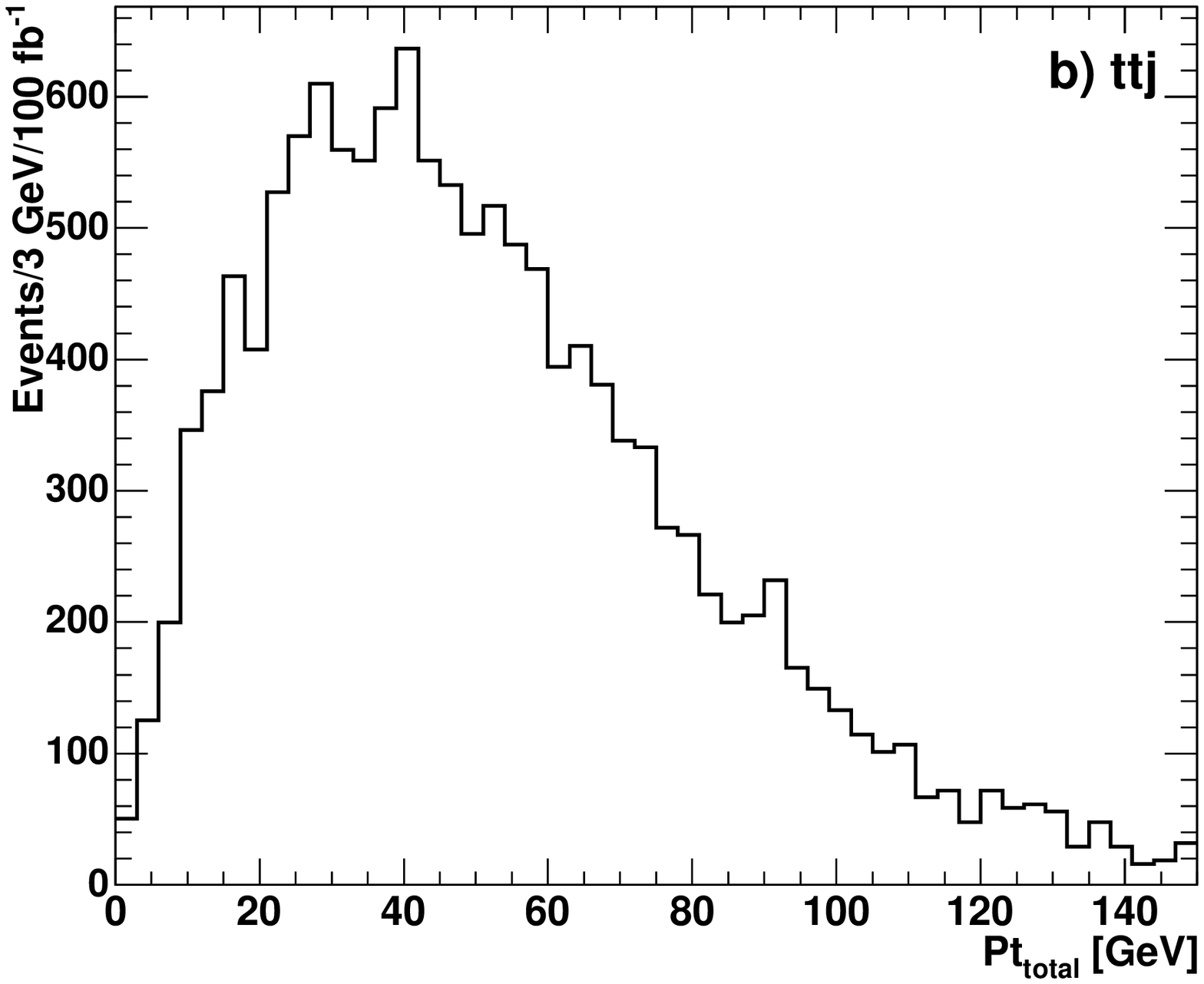}}
    \resizebox{9cm}{!}{\includegraphics{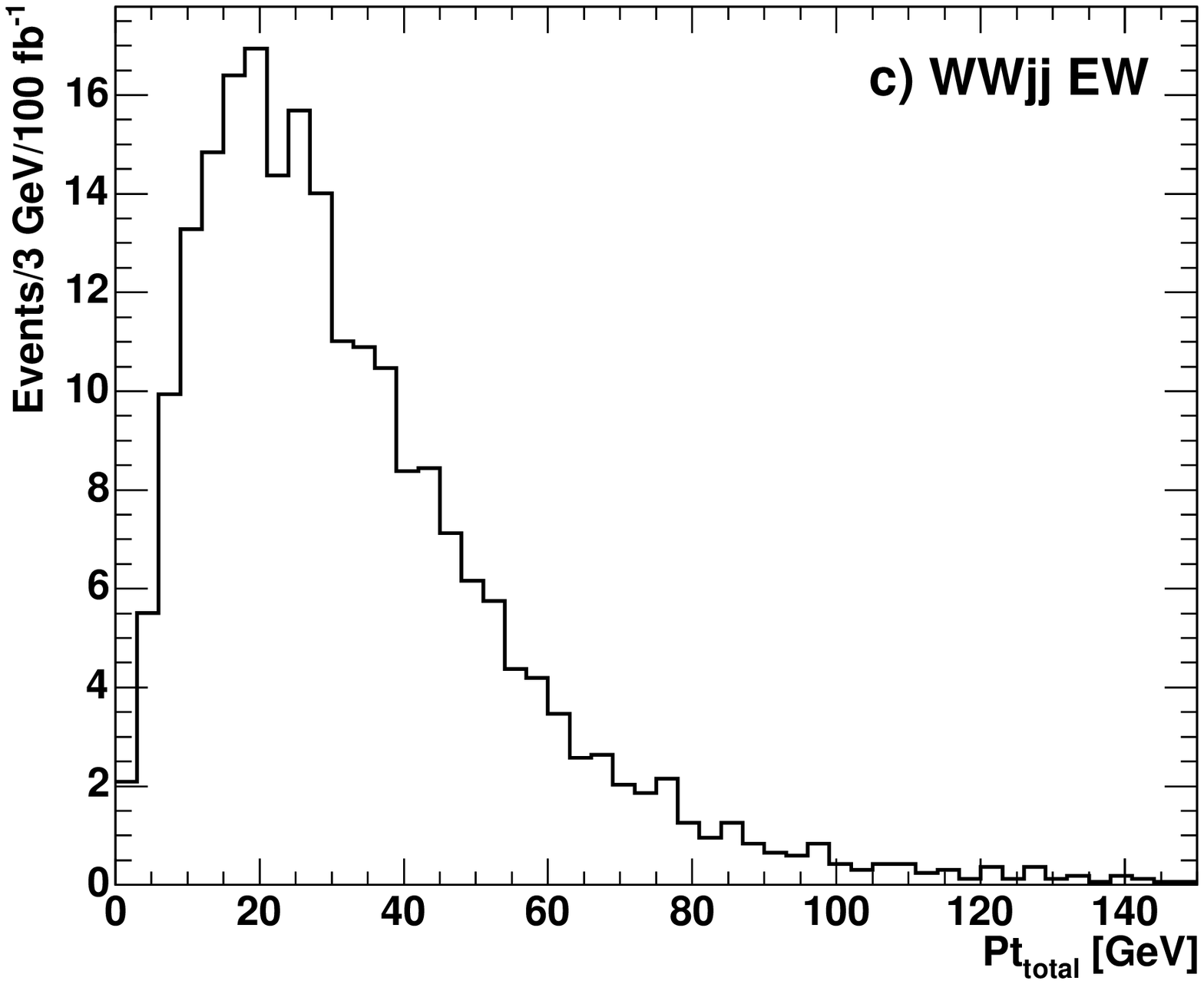}}
    \resizebox{9cm}{!}{\includegraphics{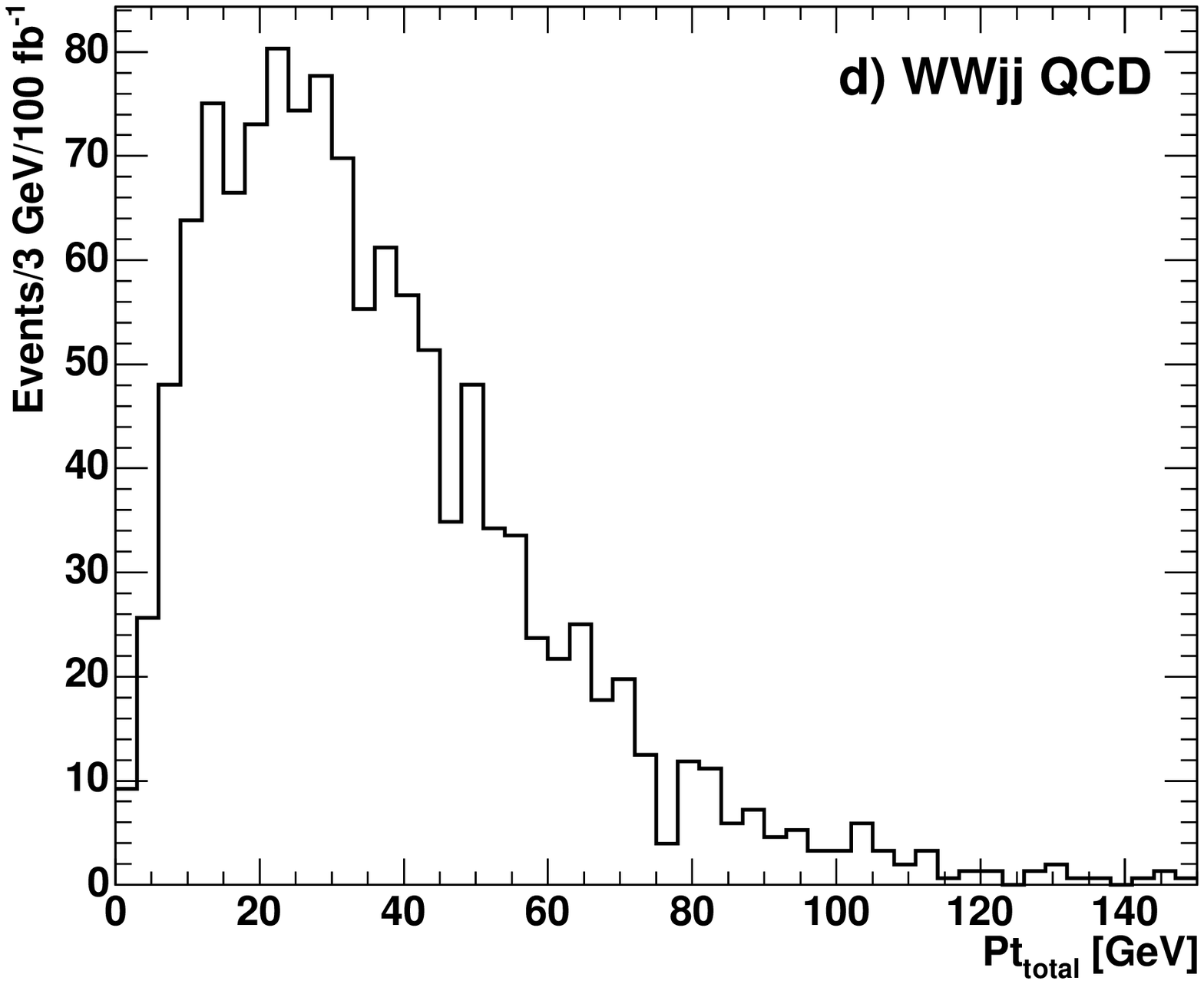}}
     \caption{\label{fig:ptbalance}\em
     The overall $p_T$-balance in the event.  See the text for an explanation.
for a) qqH, $m_{H}=120$ GeV and backgrounds b) $t\bar{t}j$, c) EW WWjj and d) QCD WWjj.}
\end{figure*}

First, we require the di-jet mass to be greater than~$600$~GeV (see Fig.~\ref{fig:Mjj}).  
Next, we look at the overall $p_T$-balance in the event, by
computing the vector sum of the transverse momenta of the
two leading jets, the leptons, and the missing energy.
The magnitude of that sum should be less than $40$~GeV (see Fig.~\ref{fig:ptbalance}).
\par
\begin{figure*}
    \resizebox{9cm}{!}{\includegraphics{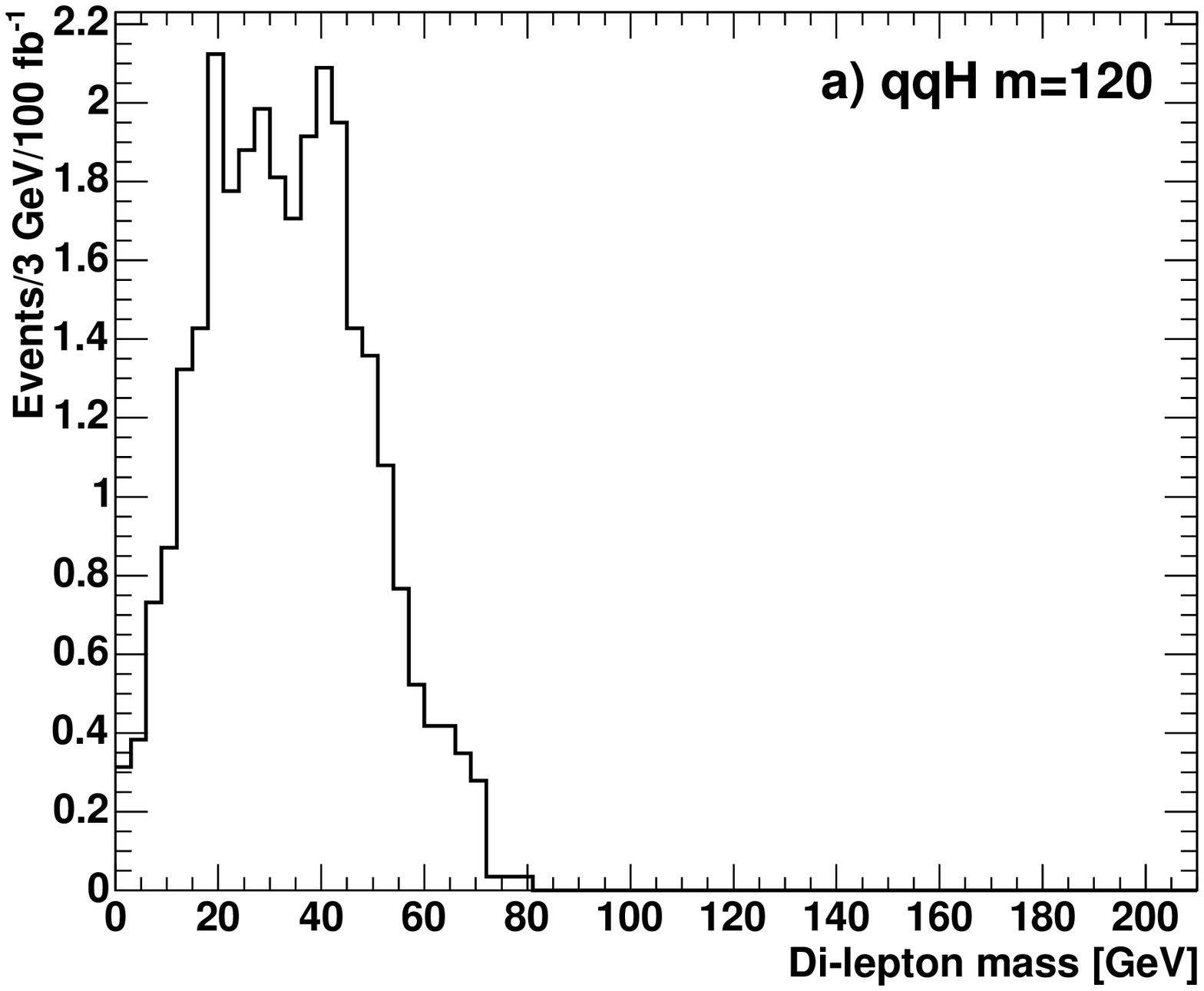}}     
    \resizebox{9cm}{!}{\includegraphics{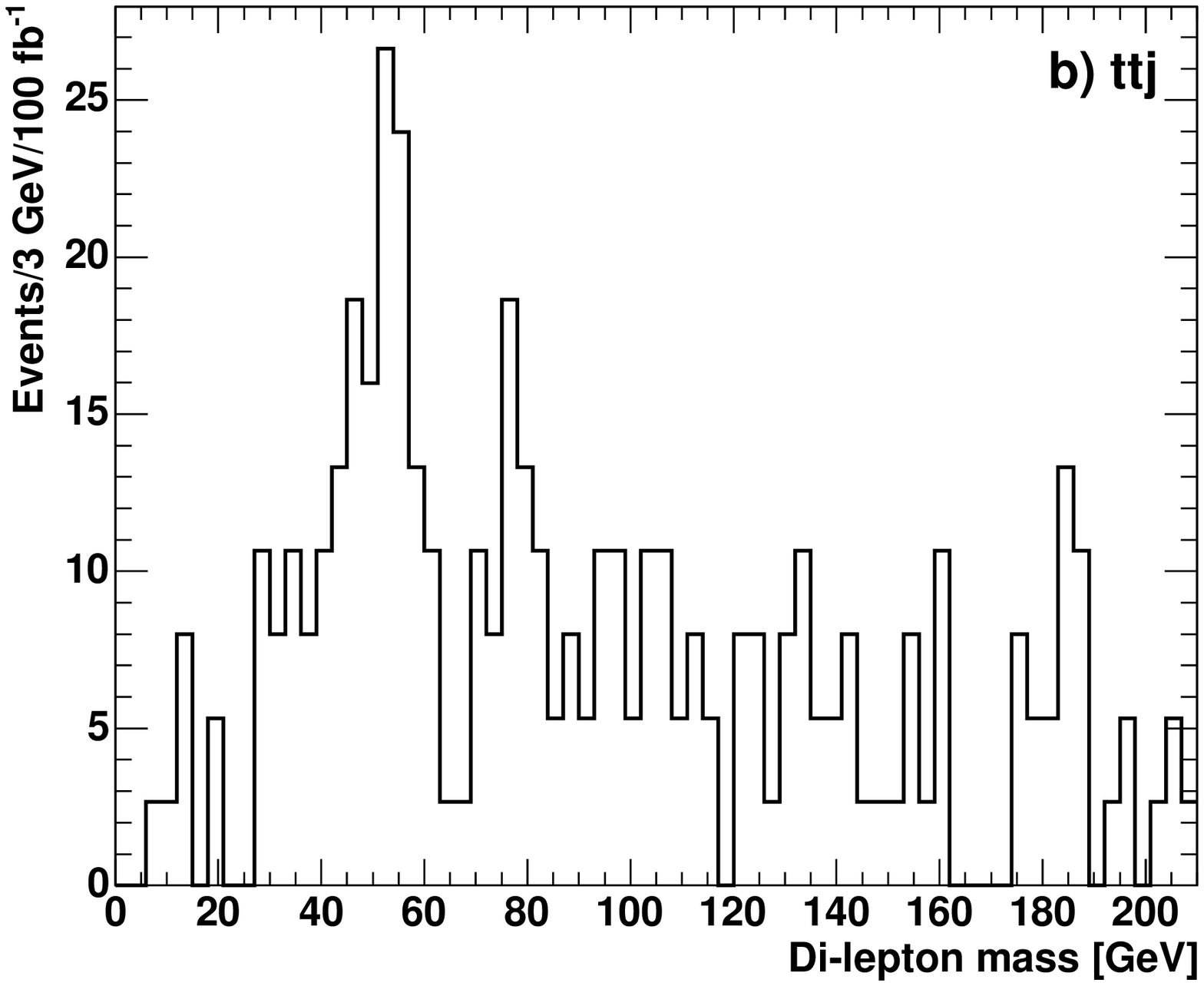}}
    \resizebox{9cm}{!}{\includegraphics{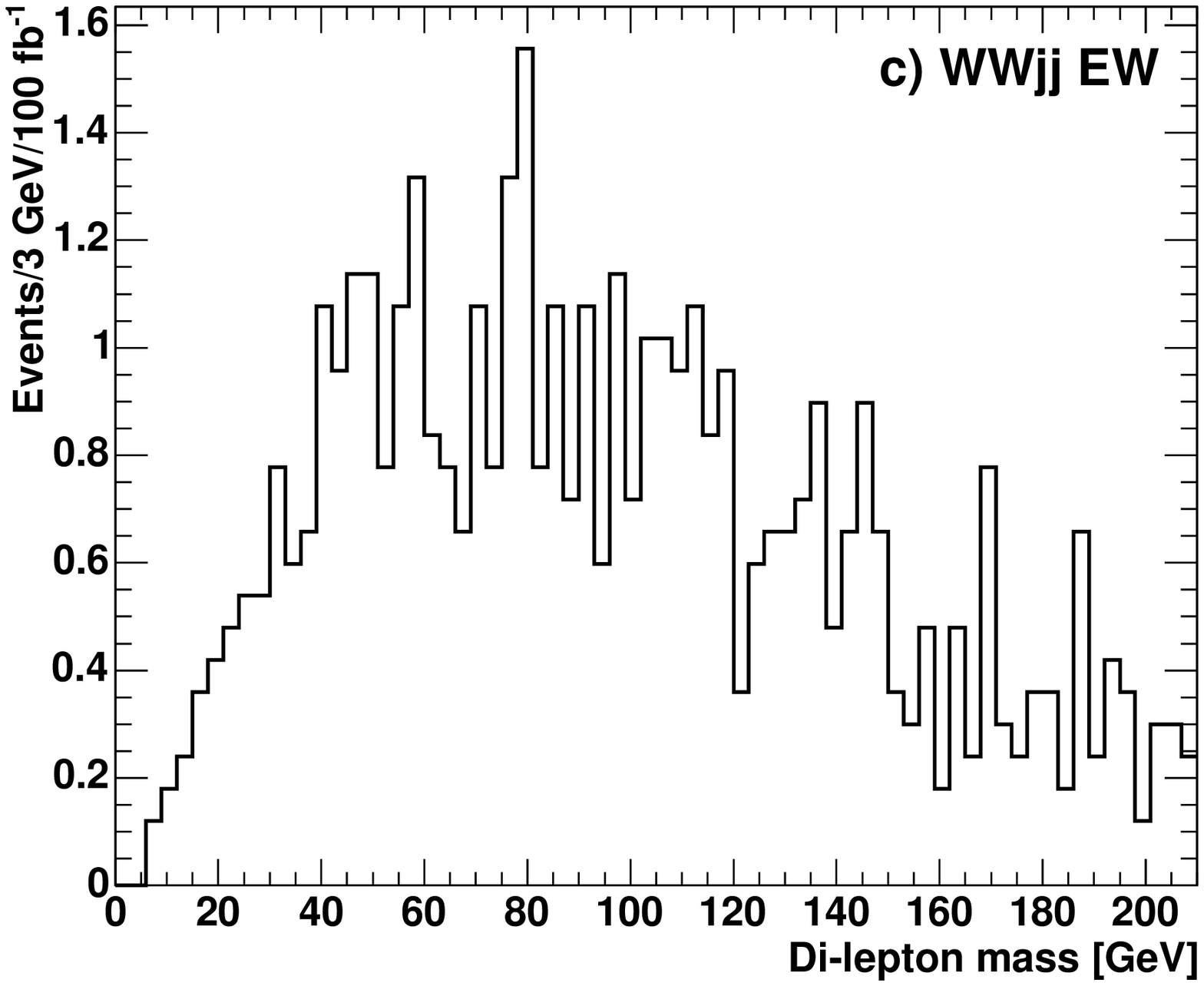}}
    \resizebox{9cm}{!}{\includegraphics{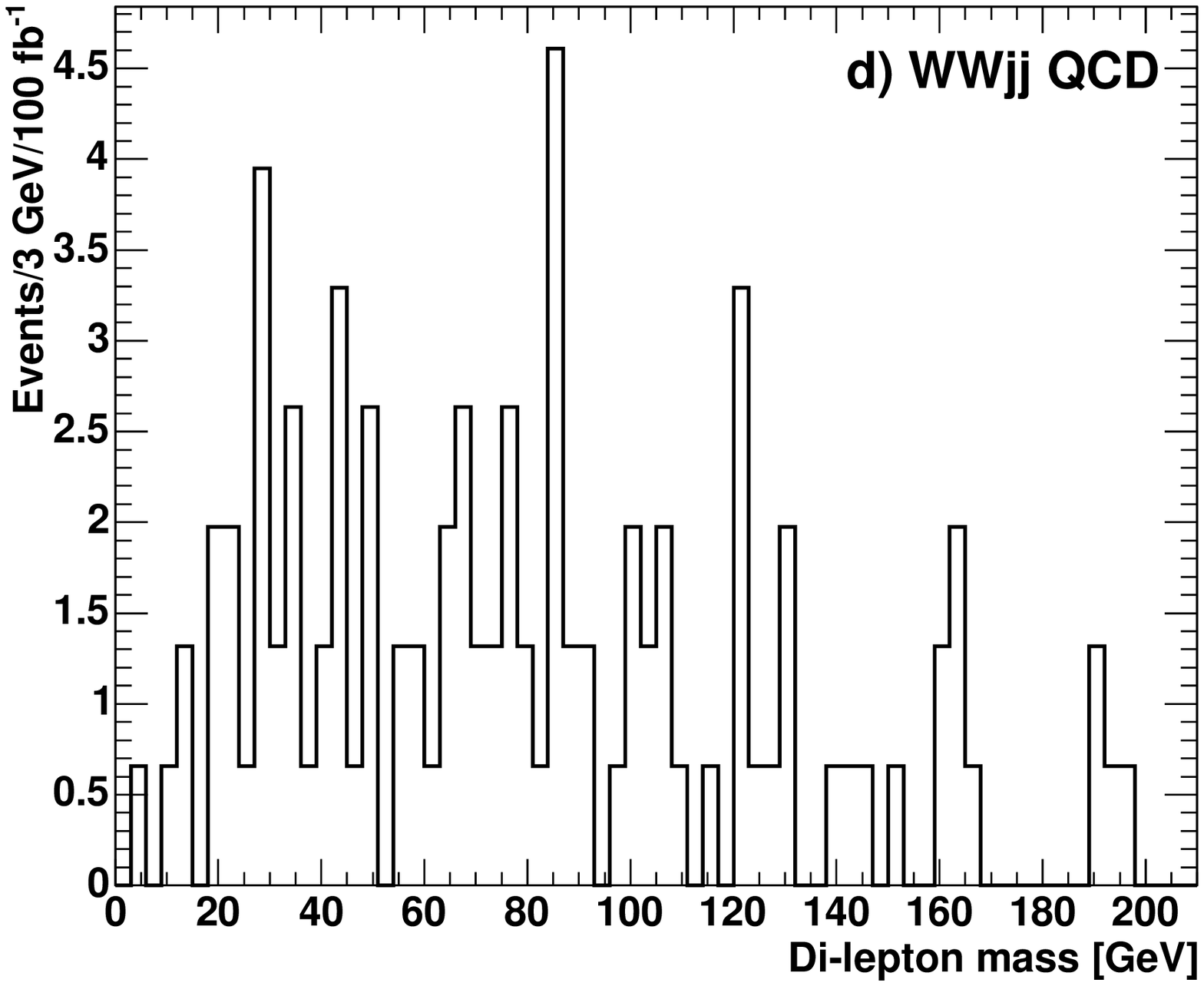}}
     \caption{\label{fig:mll}\em
     Di-lepton invariant mass distribution after jet and lepton cuts, for a) qqH, $m_{H}=120$ GeV 
and backgrounds b$t\bar{t}j$, c) EW WWjj and d) QCD WWjj.}
\end{figure*}
\begin{figure*}
    \resizebox{9cm}{!}{\includegraphics{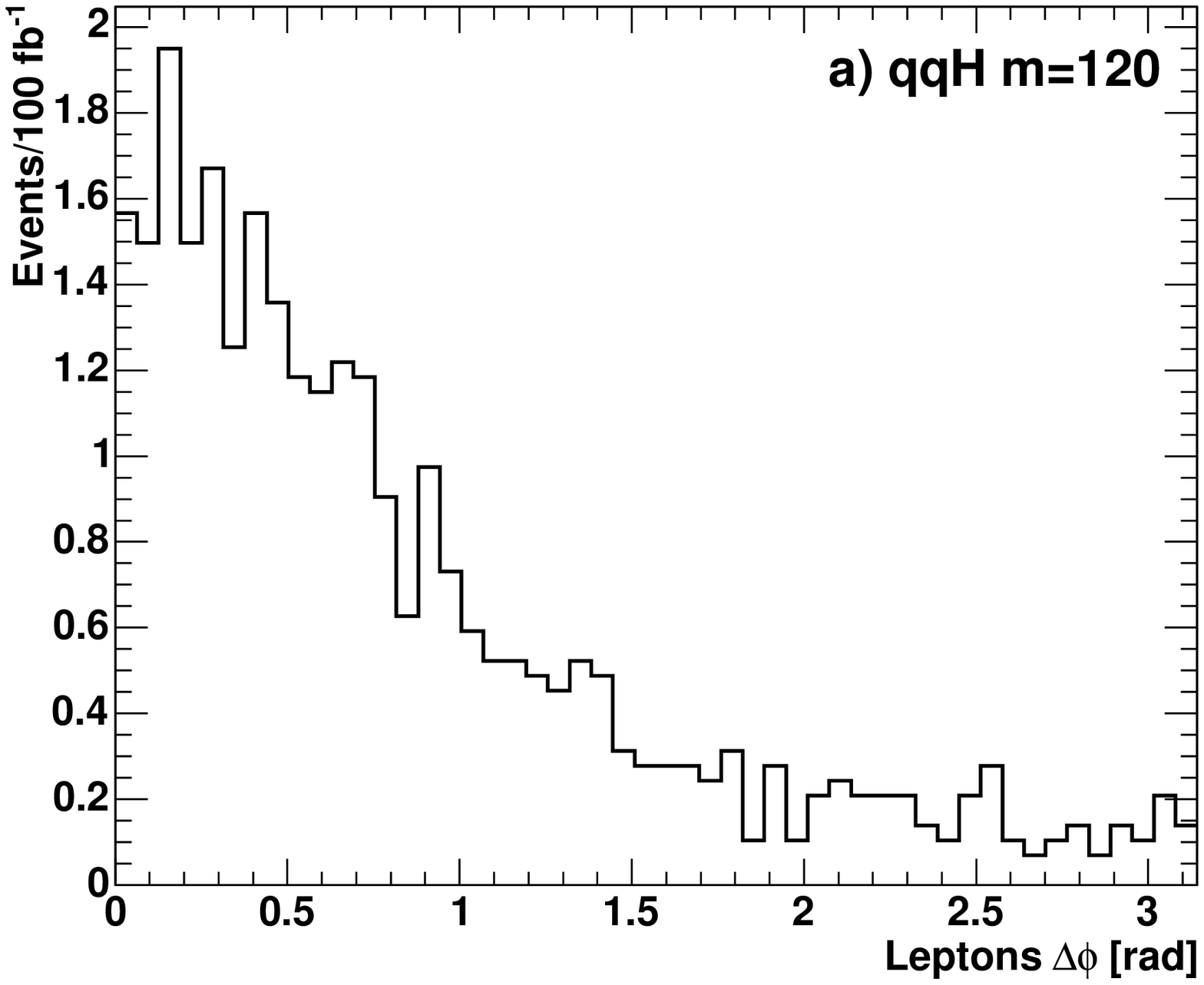}}     
    \resizebox{9cm}{!}{\includegraphics{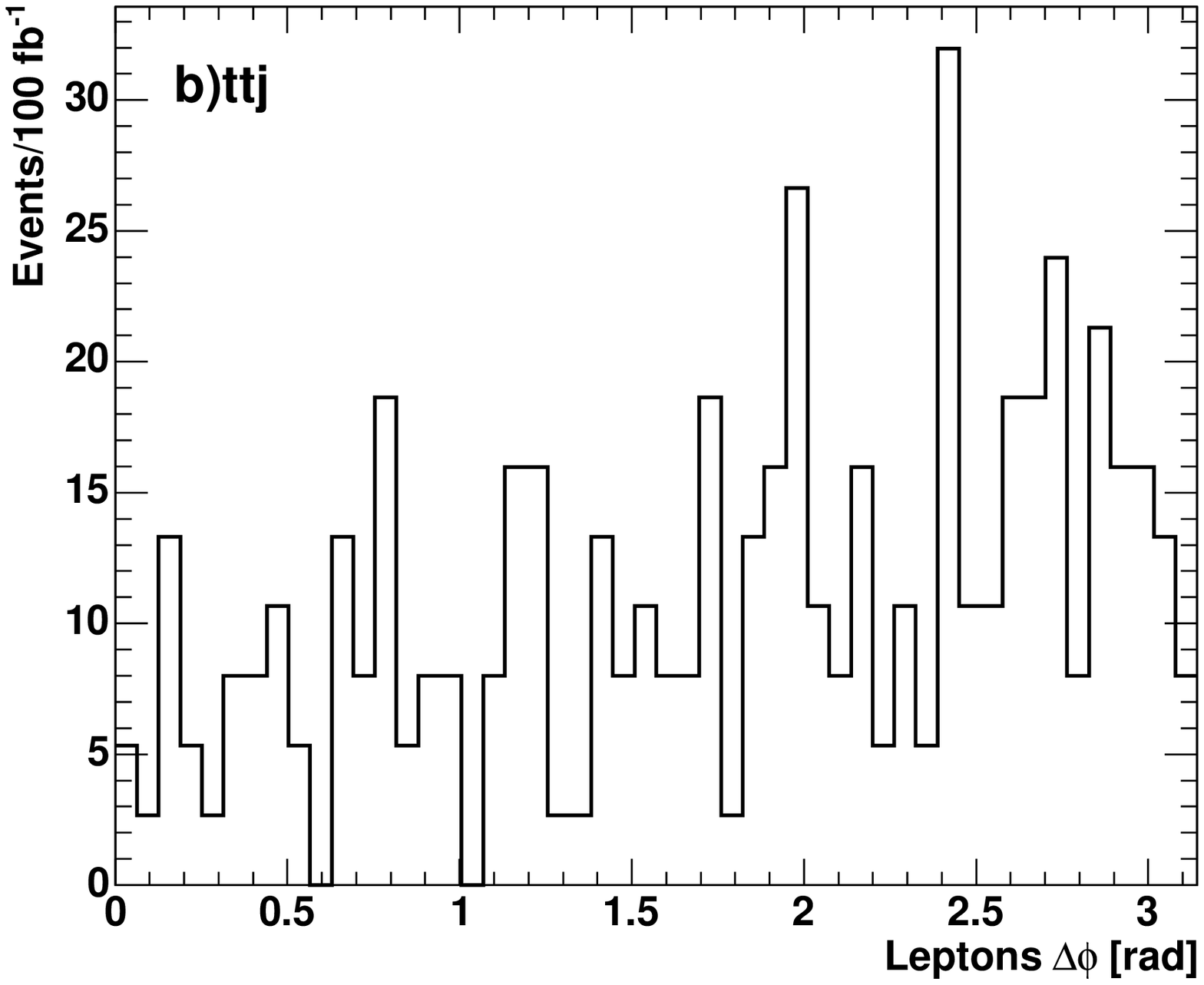}}
    \resizebox{9cm}{!}{\includegraphics{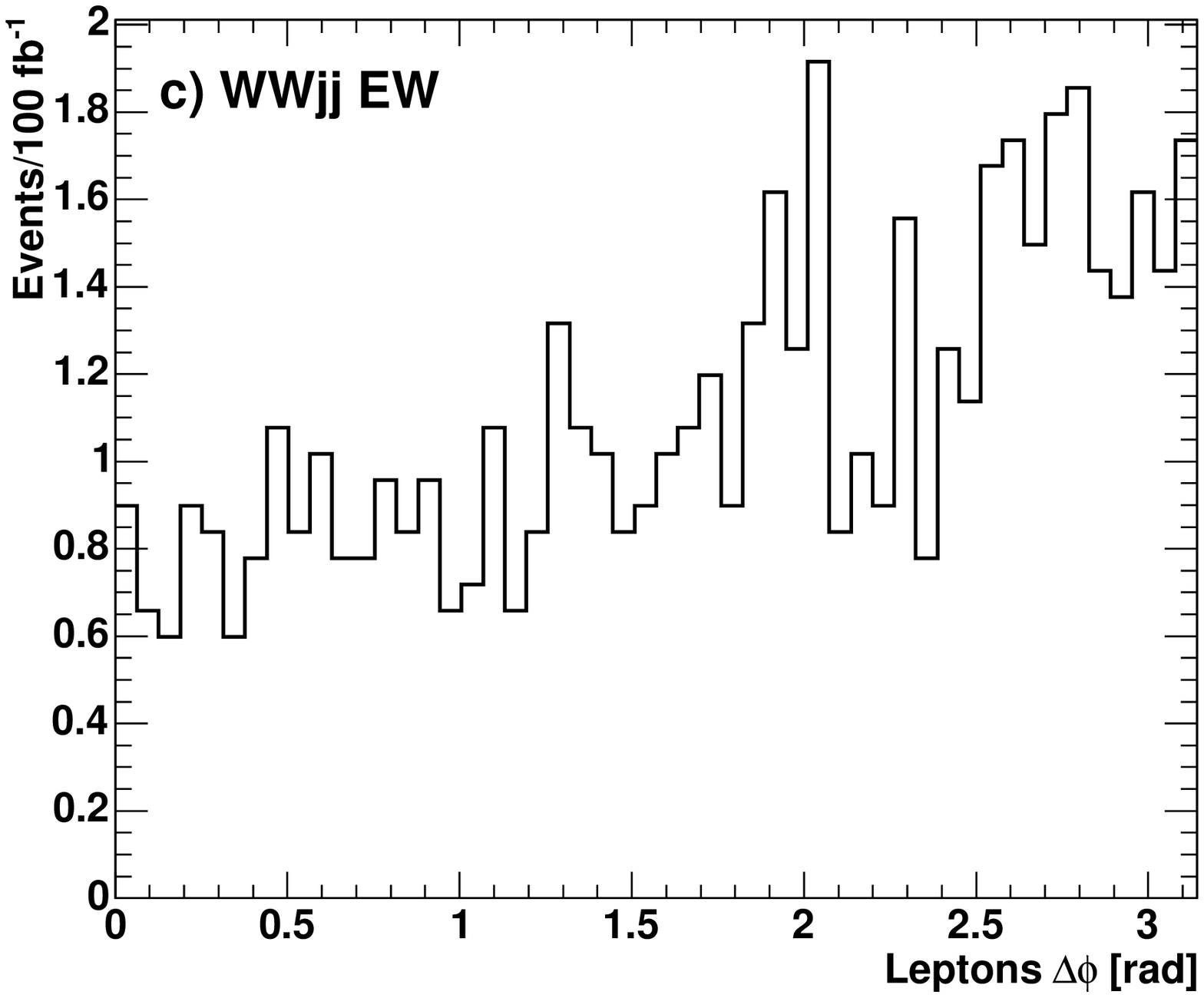}}
    \resizebox{9cm}{!}{\includegraphics{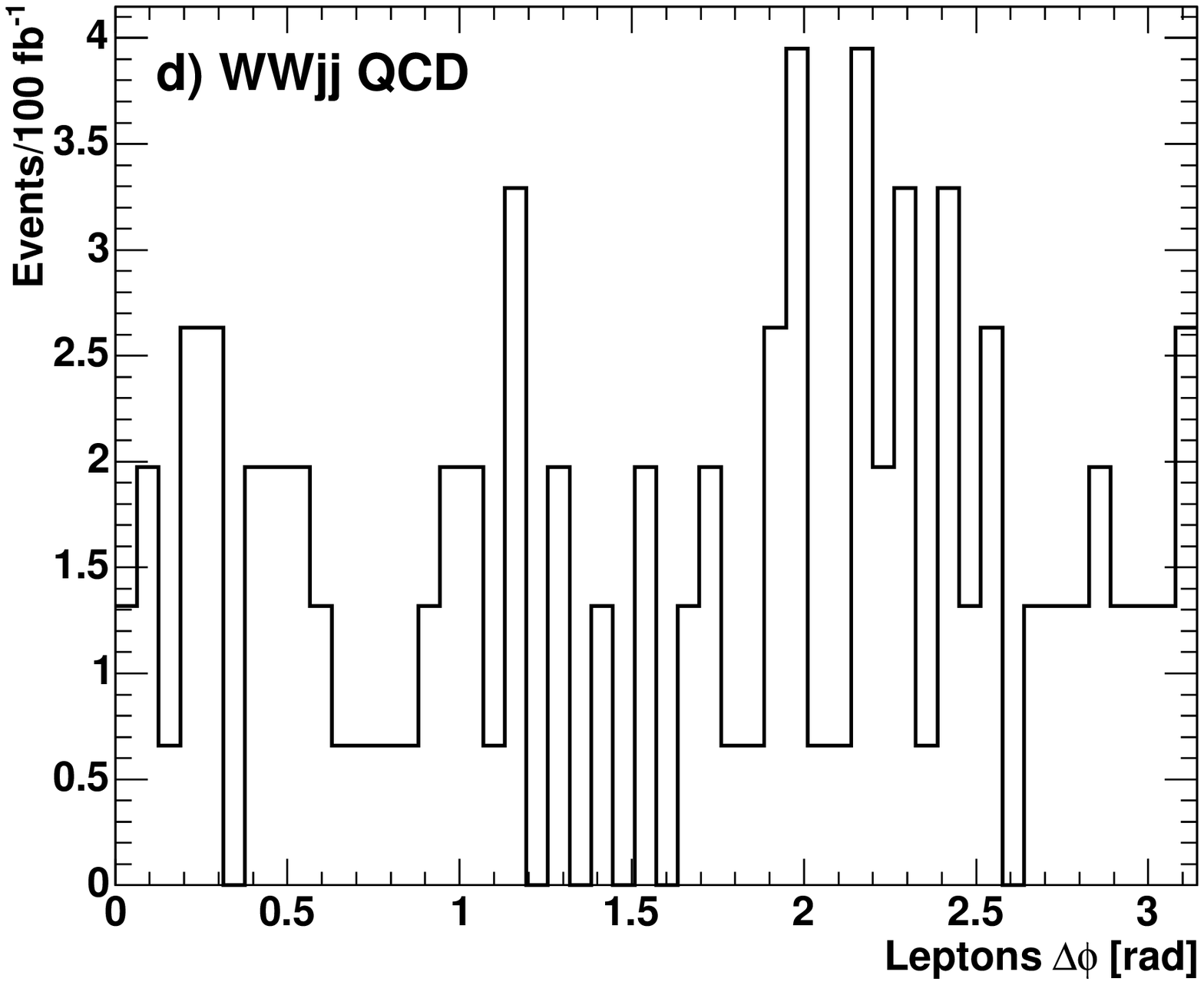}}
     \caption{\label{fig:dphi}\em
     The distribution of the difference in azimuthal angle between the two leptons, $\Delta\phi$ 
after jet and lepton cuts, for a) signal events, qqH, $m_{H}=120$ GeV and backgrounds
b$t\bar{t}j$, c) EW WWjj and d) QCD WWjj.}
\end{figure*}
\begin{figure}
    \resizebox{9cm}{!}{\includegraphics{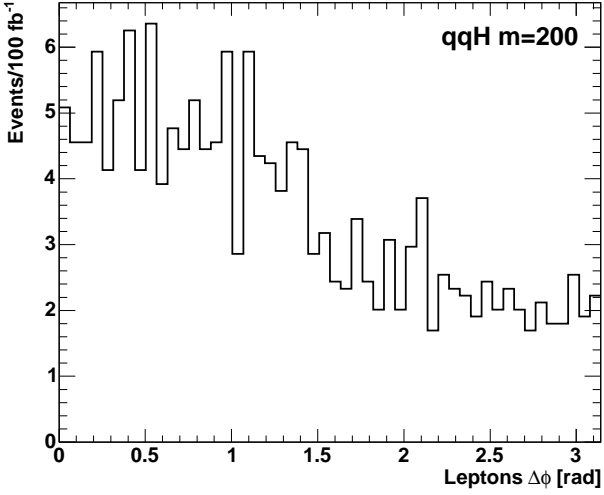}}
     \caption{\label{fig:dphi200}\em
     The $\Delta\phi$ distribution between the two leptons after jet and lepton
cuts for qqH, $m_H$ = 200 GeV}
\end{figure}
\begin{figure*}
    \resizebox{9cm}{!}{\includegraphics{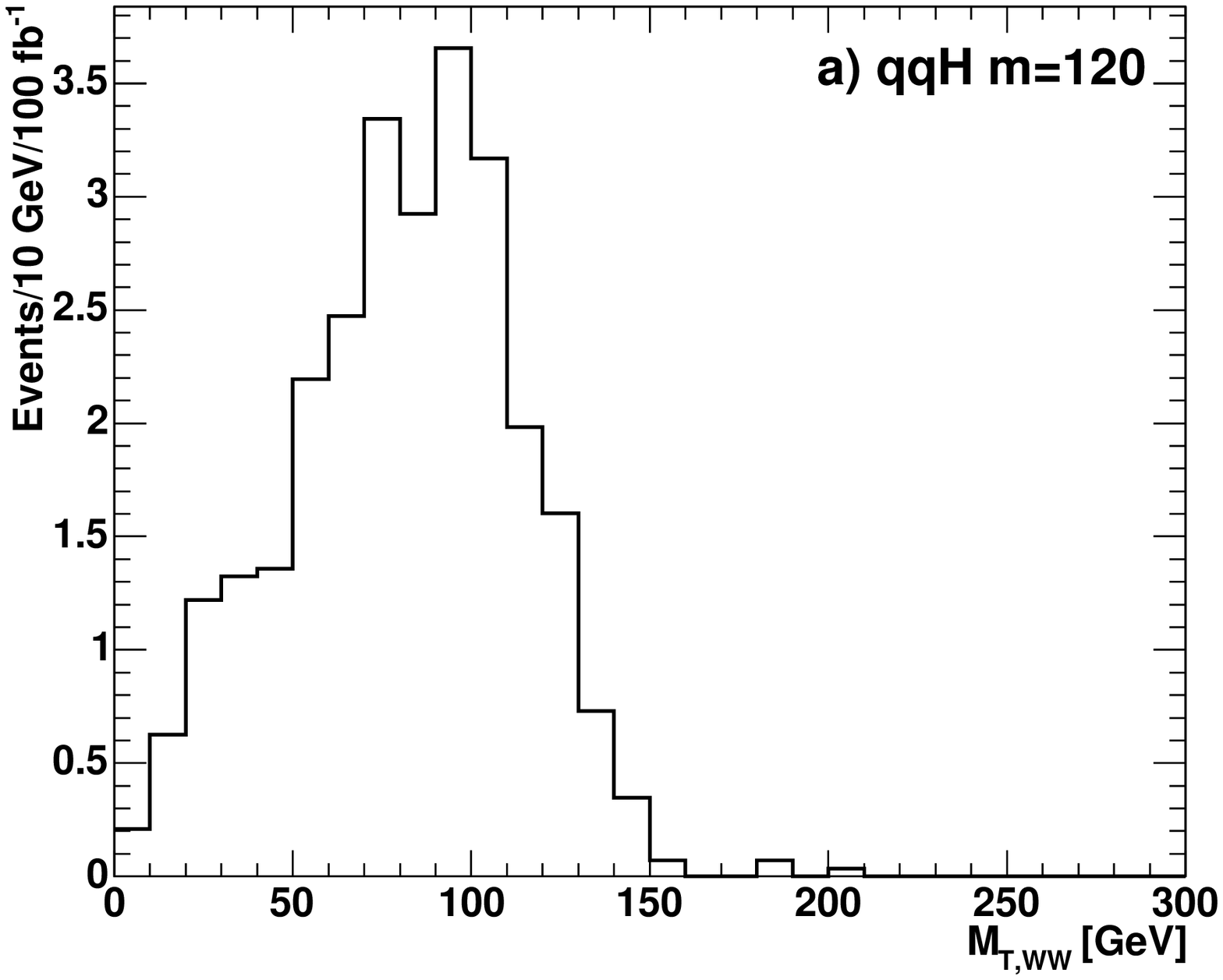}}
    \resizebox{9cm}{!}{\includegraphics{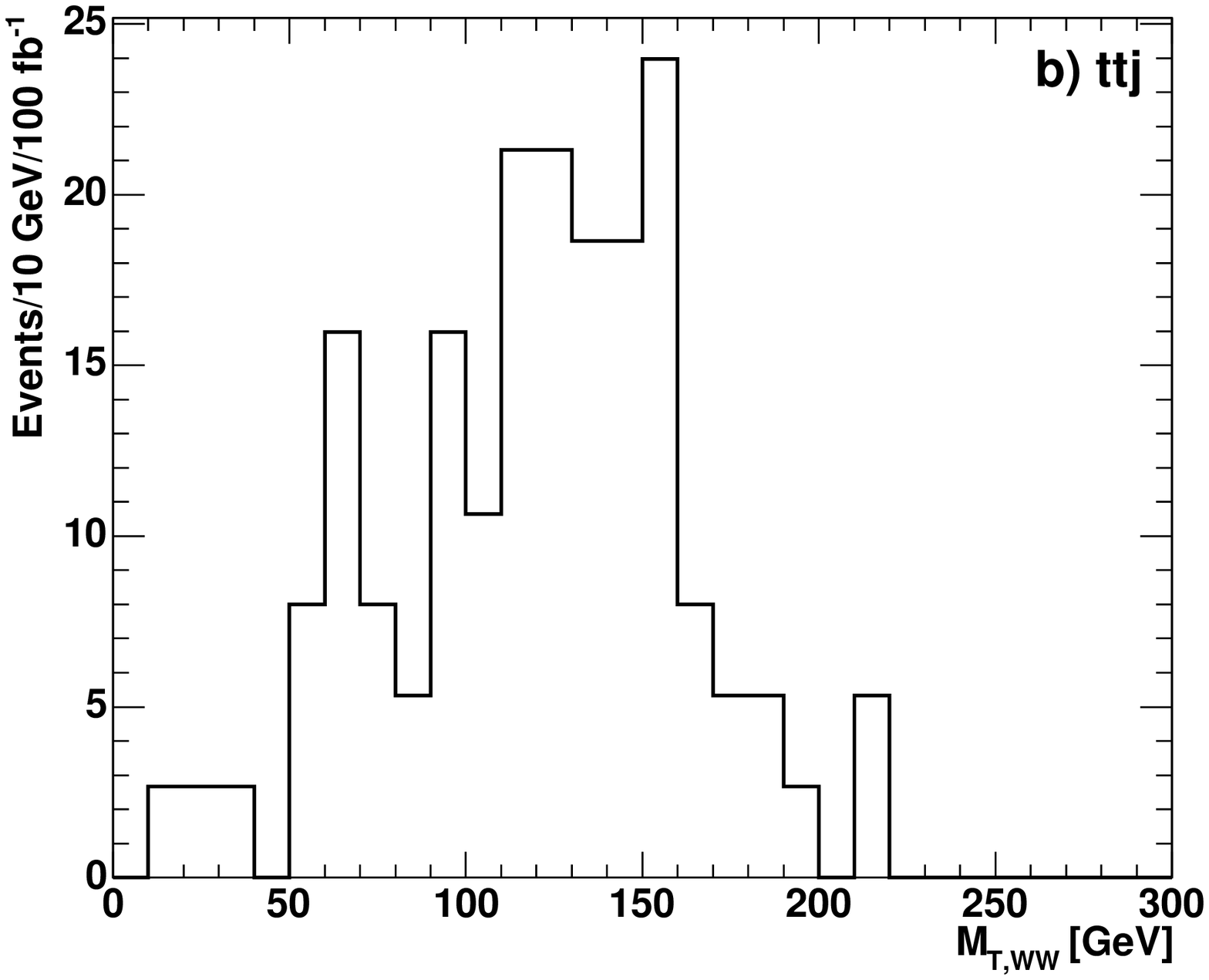}}
    \resizebox{9cm}{!}{\includegraphics{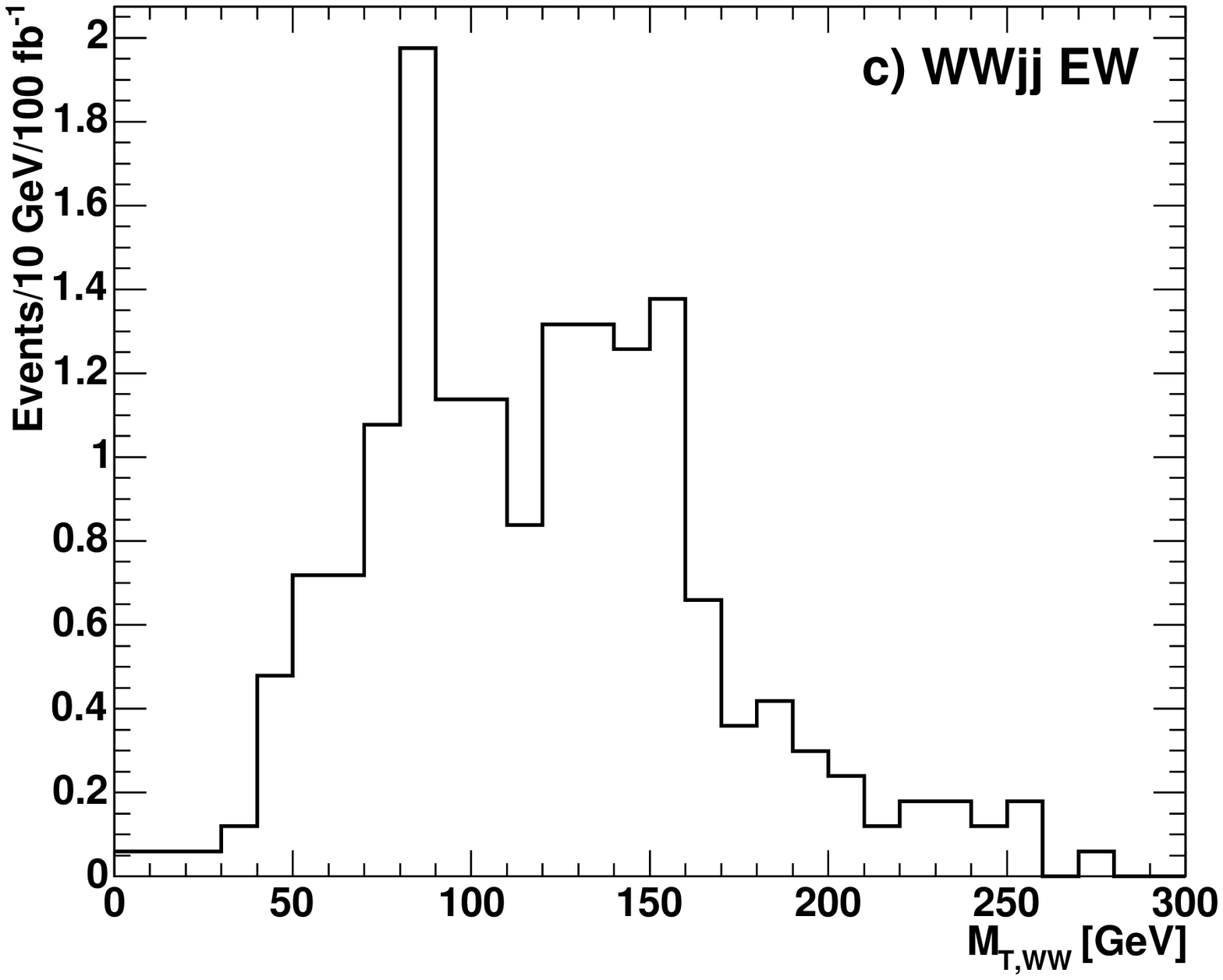}}
    \resizebox{9cm}{!}{\includegraphics{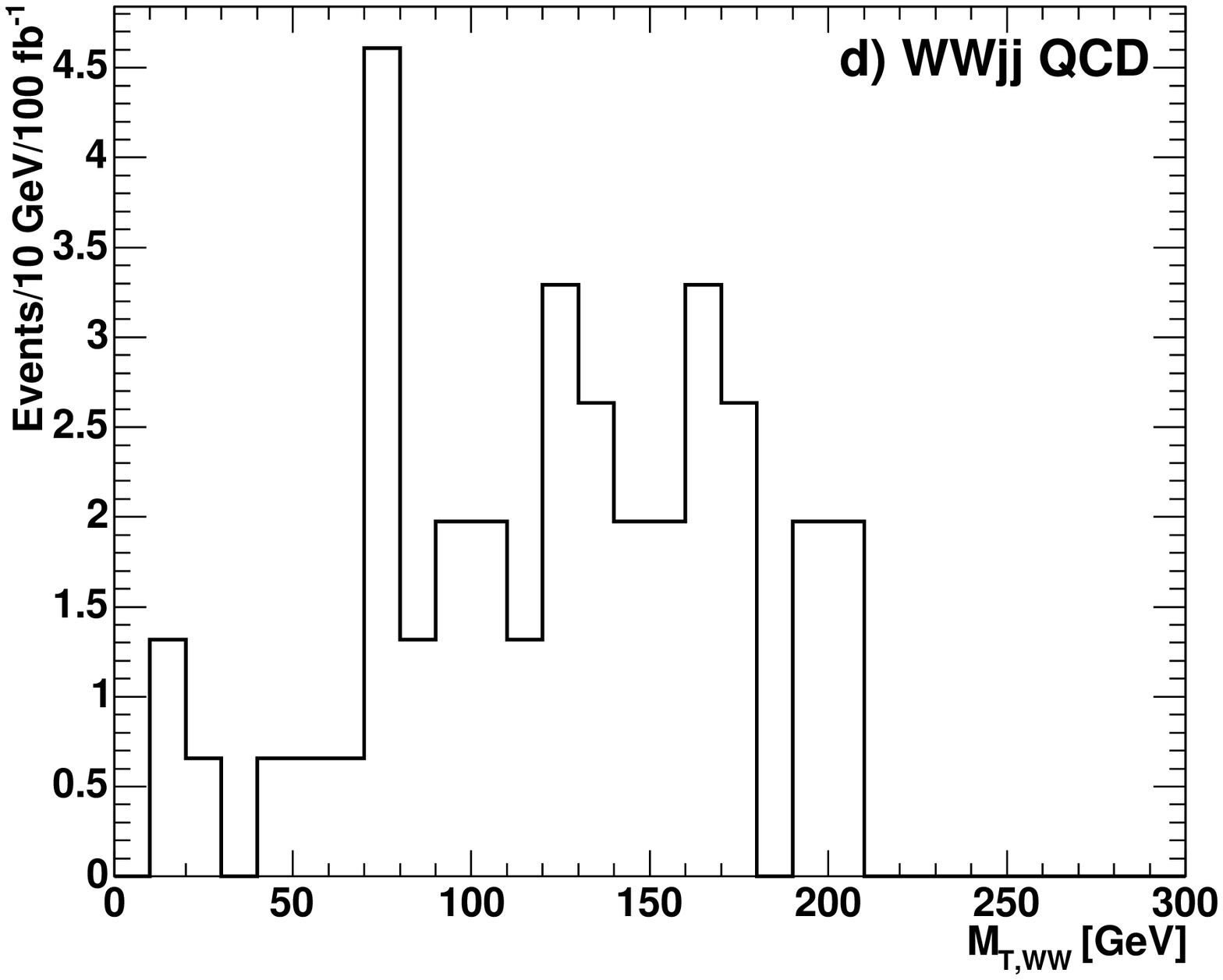}}
     \caption{\label{fig:MTWW}\em
    The transverse mass of the two $W$ bosons, $M_{T,WW}$, for a) signal events, qqH, $m_{H}=120$ GeV 
and backgrounds b) $t\bar{t}j$, c) EW WWjj and d) QCD WWjj.}
\end{figure*}
\begin{figure*}
    \resizebox{9cm}{!}{\includegraphics{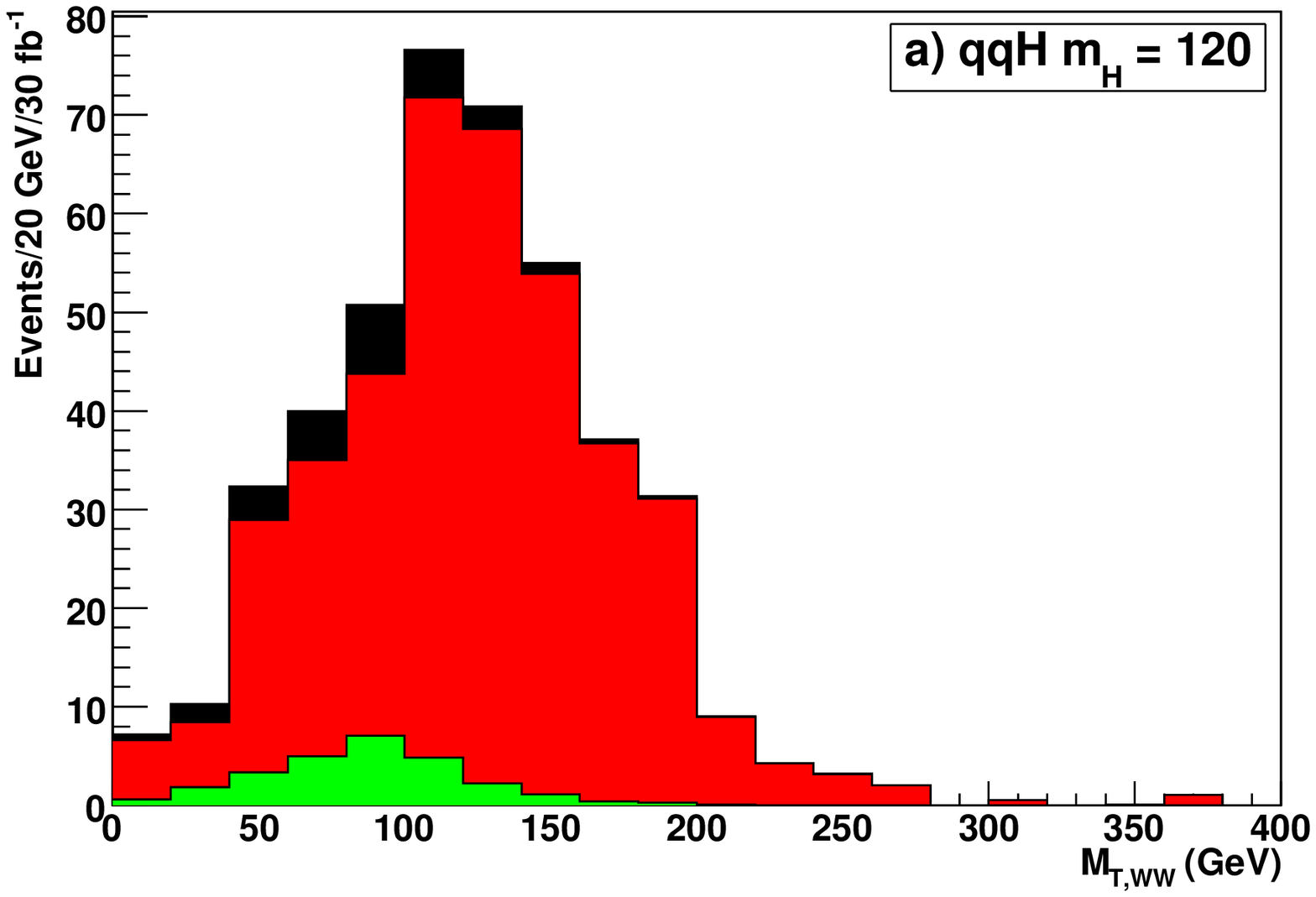}}     
    \resizebox{9cm}{!}{\includegraphics{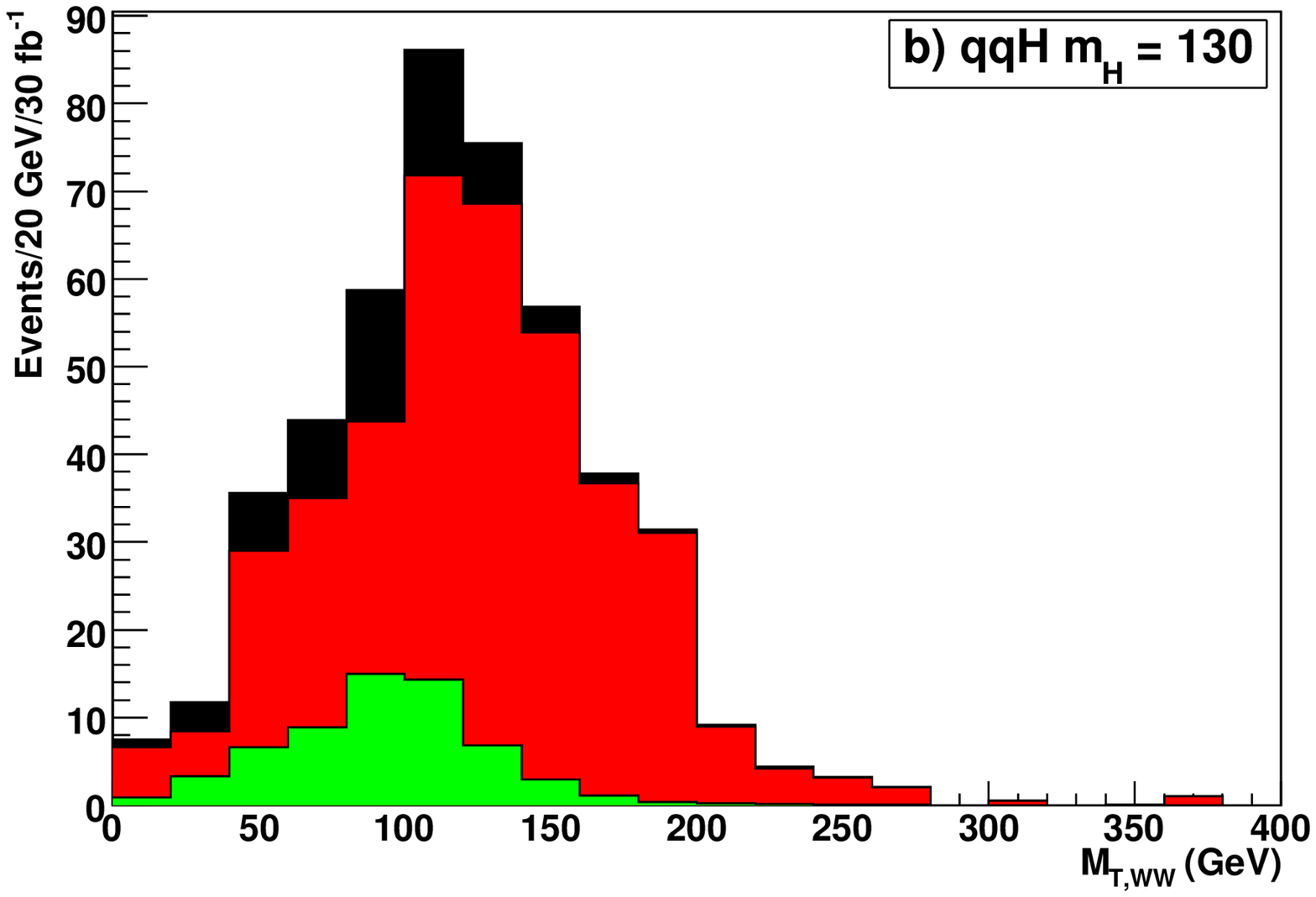}}
    \resizebox{9cm}{!}{\includegraphics{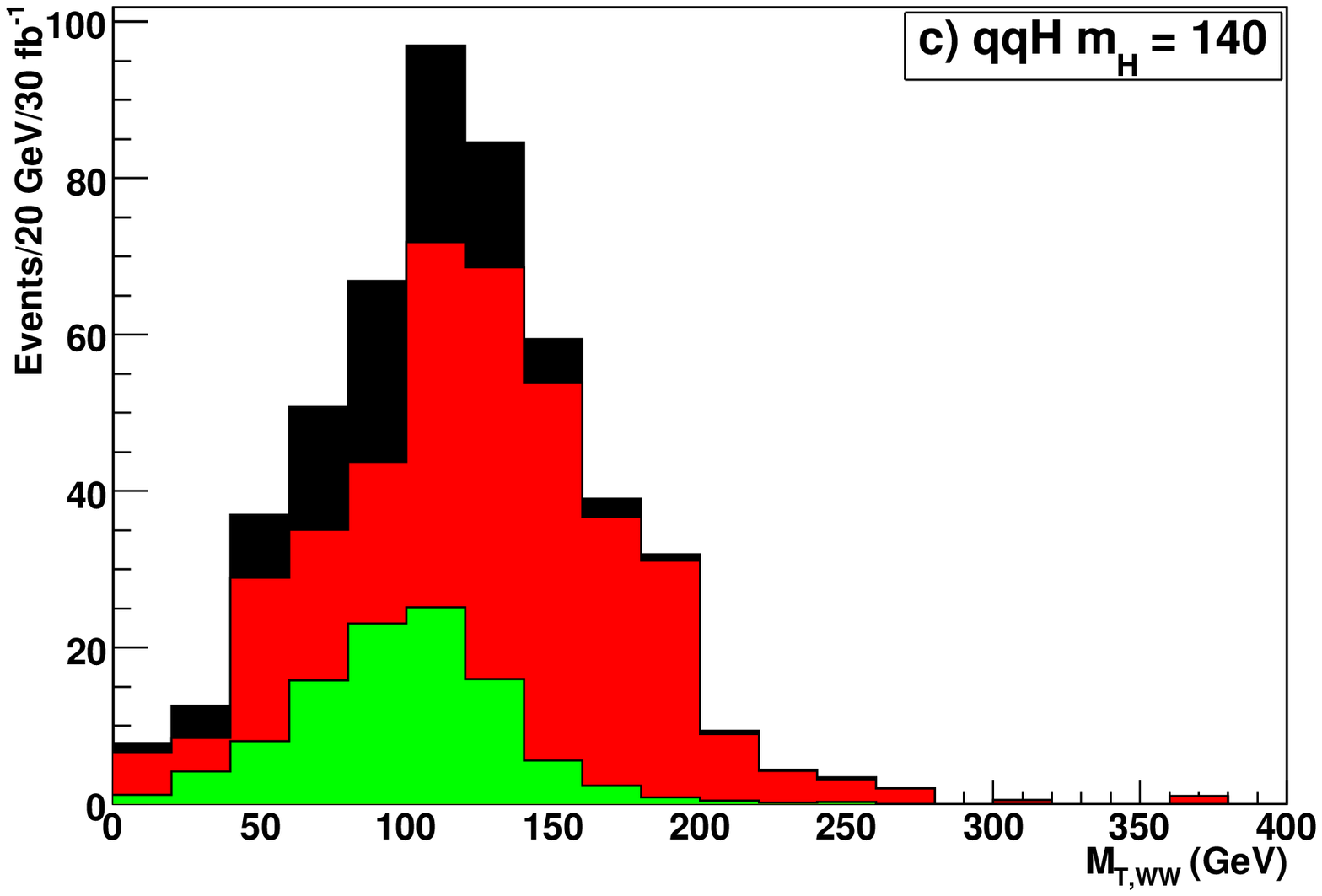}}
    \resizebox{9cm}{!}{\includegraphics{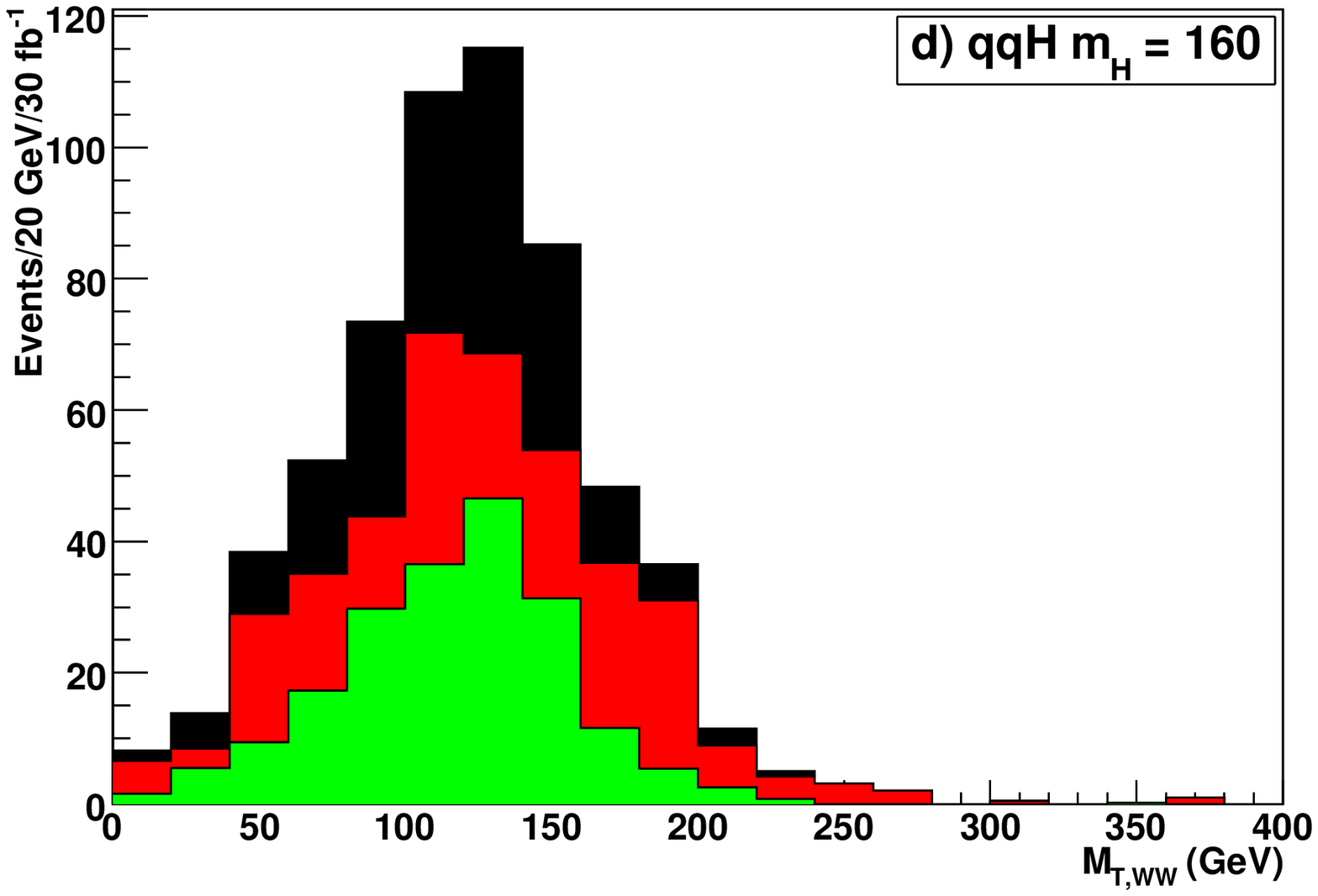}}
    \resizebox{9cm}{!}{\includegraphics{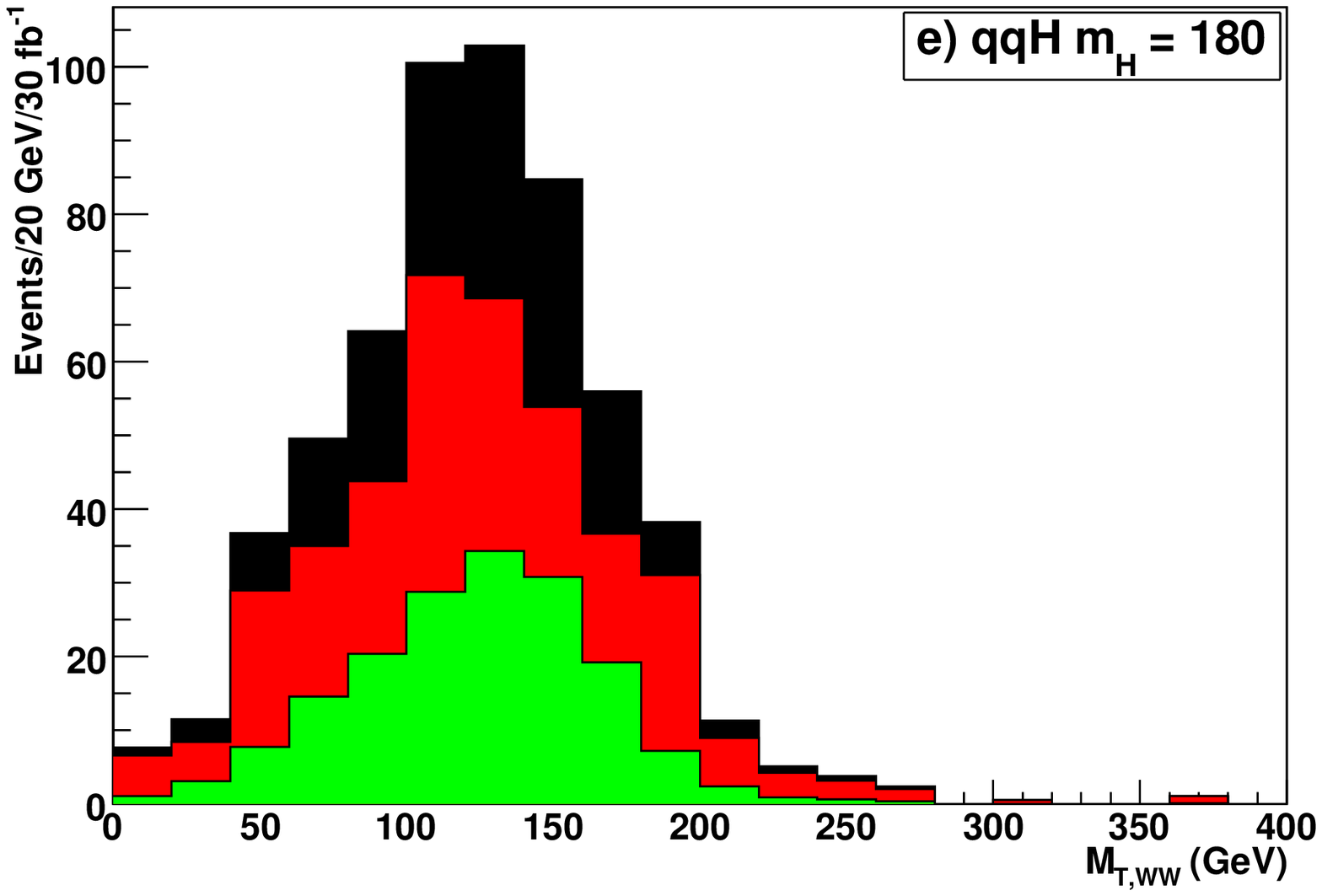}}
    \resizebox{9cm}{!}{\includegraphics{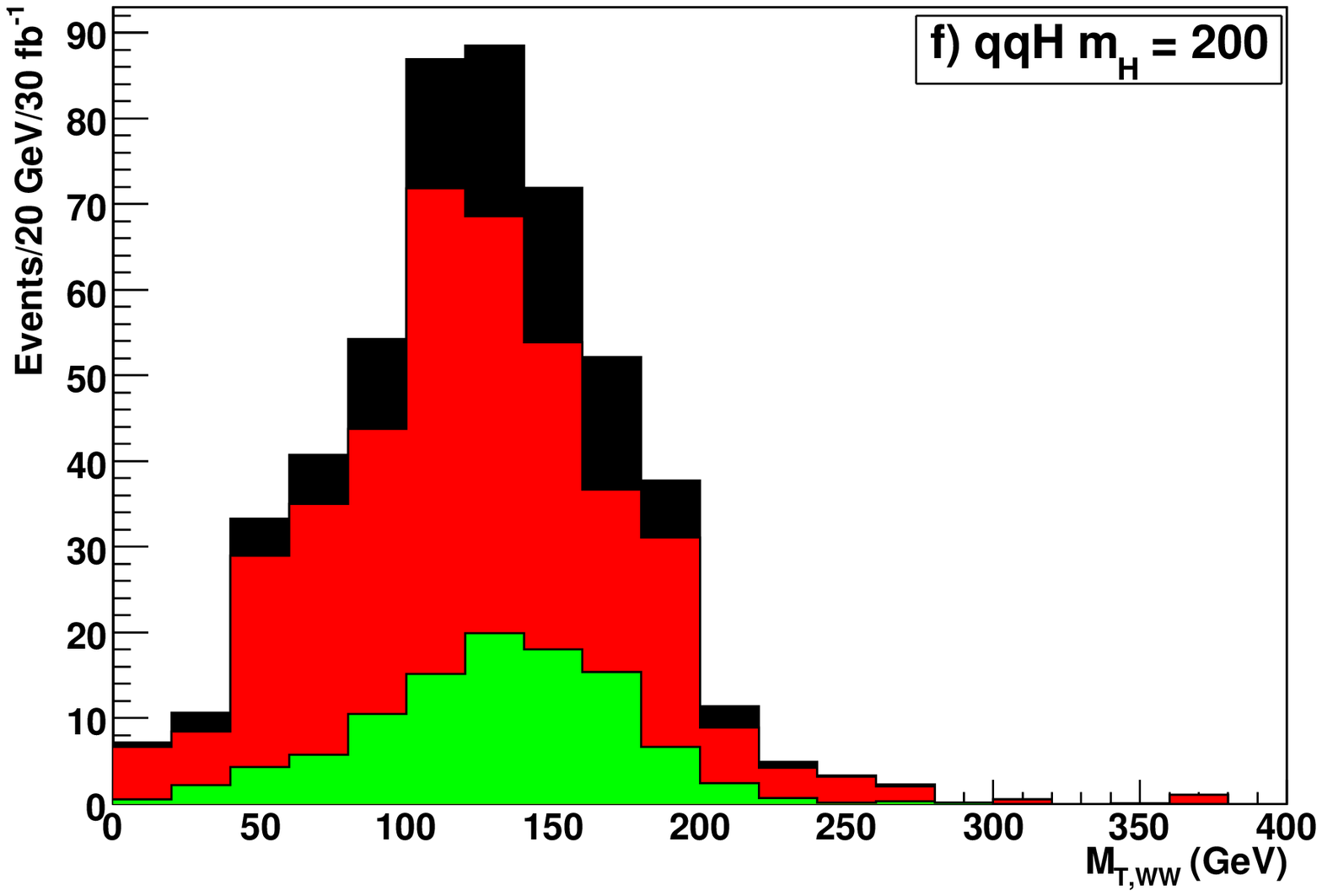}}
     \caption{\label{fig:mttots}\em
The transverse mass, $M_{T,WW}$, distributions for signal and background, with Higgs mass = 120, 130, 
140, 160, 180 and 200 GeV respectively shown in a),b),c),d),e),f).
The Lower plot (light grey) is the signal, the middle plot(dark grey) is the background, and the black
histogram is the sum.}
\end{figure*}

When it comes to the leptons, we require a di-lepton mass 
$M_{\ell\ell} < 80$~GeV (see Fig.~\ref{fig:mll}). This value
is lower than the Z-mass, so that leptonic Z-decays do not affect
the current analysis. A useful distinction arises in the relative azimuthal angle of
the two leptons due to the spin-0 nature of the Higgs boson
(see Fig.~\ref{fig:dphi},~\ref{fig:dphi200}). We take advantage of this 
discriminant and require $\Delta\phi < 2.4$~radians.
Finally, we require that the ``$WW$~transverse-mass'' be not too high
when looking for Higgs bosons with mass below 150~GeV.
The cut is that $M_{T,WW} < 125$~GeV, where $M_{T,WW} \equiv \sqrt{ 
(\not\!\!\!E_T+p_{T,\ell\ell})^2 - (\not\!\!\!{\vec{E}}_T+\vec{P}_{T,\ell\ell})^2 }$.
See Fig.~\ref{fig:MTWW} and \ref{fig:mttots} for distributions of this 
quantity.

\subsection{Additional Cuts}
\begin{figure}
    \resizebox{9cm}{!}{\includegraphics{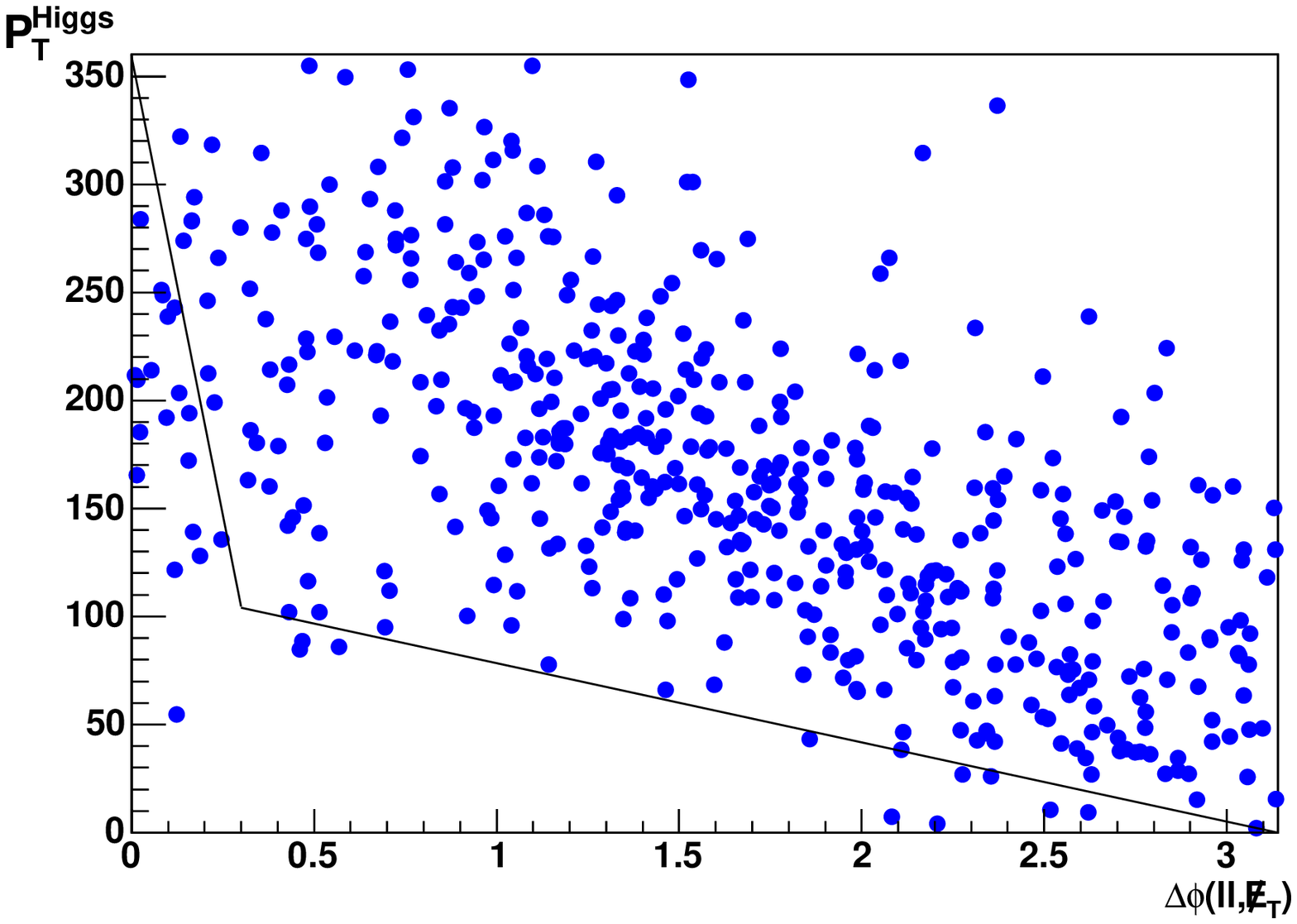}}
     \caption{\label{fig:dphillmet}\em
     The azimuthal angle difference in radians between the dilepton momentum vs. the 
missing $E_T$ vs $p_{T}^{Higgs}$ for qqH with $m_H$ = 120 GeV. 
The lines correspond to the cuts $57.29\Delta\phi(ll,\not\!\!\!E_T)+1.5p_T^{Higgs}>180$,
$12\times57.29\Delta\phi(ll,\not\!\!\!E_T)+p_T^{Higgs}>360$. 
}
\end{figure}

Additional cuts may be required in order that $bbjj$ and $\tau\tau jj$ 
backgrounds not pose a problem.   
The additional cuts $57.3\Delta\phi(\ell\ell,\not\!\!\!E_T) + 1.5p_T^{\mathrm{Higgs}} > 180$ and
$12\times57.29\Delta\phi(\ell\ell,\not\!\!\!E_T) + p_T^{\mathrm{Higgs}} > 360$ (where $\Delta\phi(\ell\ell,\not\!\!\!E_T)$ is in radians and $p_T^{\mathrm{Higgs}}$ is in GeV units), 
and also $\not\!\!\!E_T > 30~GeV$ if $p_T^{\mathrm{Higgs}} < 50~GeV$, are imported from Ref.~\cite{krplrwzep01}. 
Here, $p_T^{\mathrm{Higgs}}$ is the vector sum of the transverse energy of tag jets. 
The distribution of signal events in the $\Delta\phi(\ell\ell,\not\!\!\!E_T)$-$p_T^{\mathrm{Higgs}}$ plane
is displayed in Fig.~\ref{fig:dphillmet}. 
\par
The Drell-Yan production of di-lepton pairs, $\gamma^*\rightarrow\ell^+\ell^-$,
has a large cross-section.  In order to reduce this background sufficiently,
we impose a di-lepton mass cut $M_{\ell\ell} > 10~GeV$
and we require $\not\!\!\!E_T > 30~GeV$ when the leptons have the same
flavor 
(see Ref.~\cite{krplrwzep01}). 
\par
\begin{figure*}
    \resizebox{9cm}{!}{\includegraphics{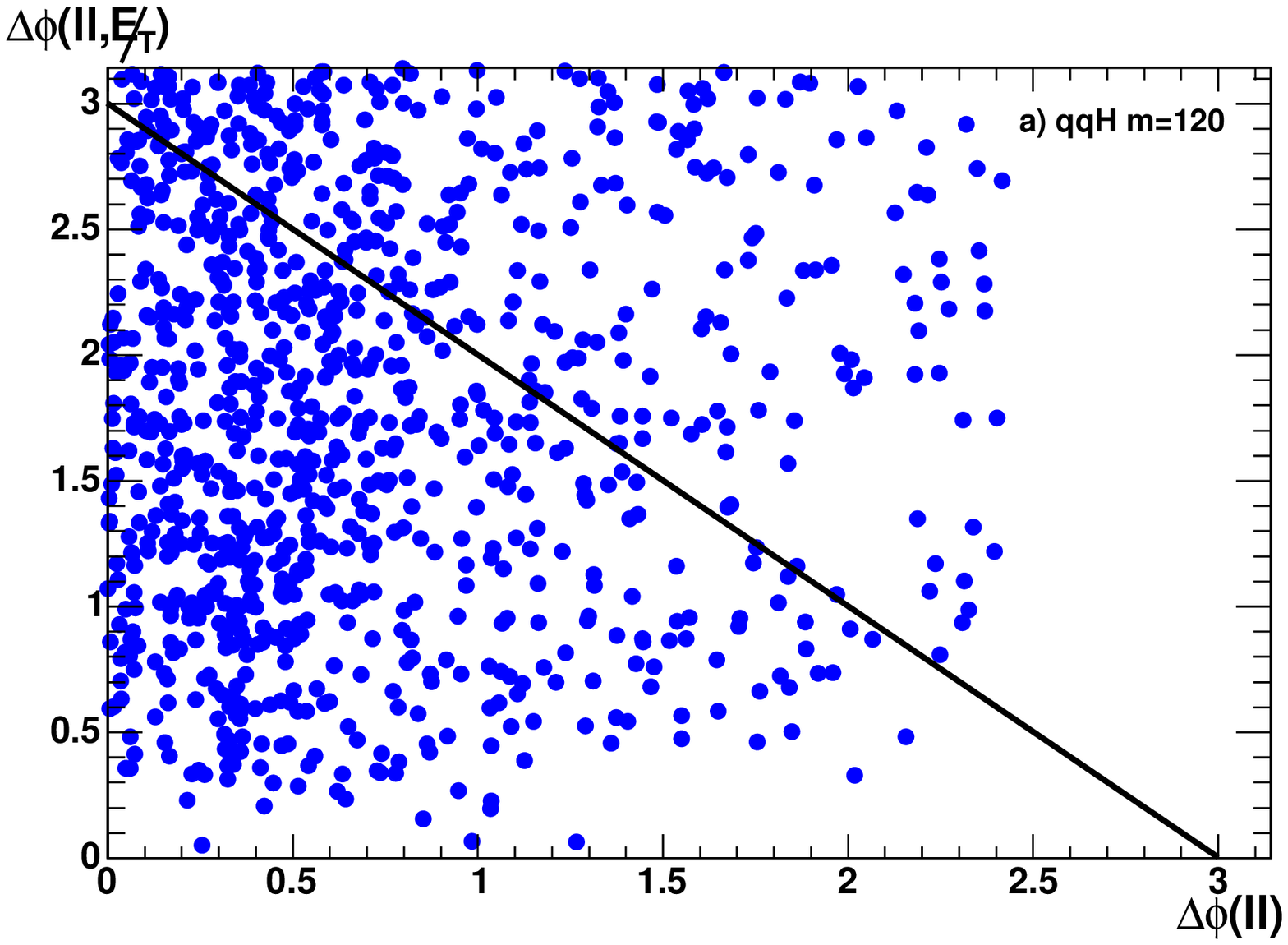}}     
    \resizebox{9cm}{!}{\includegraphics{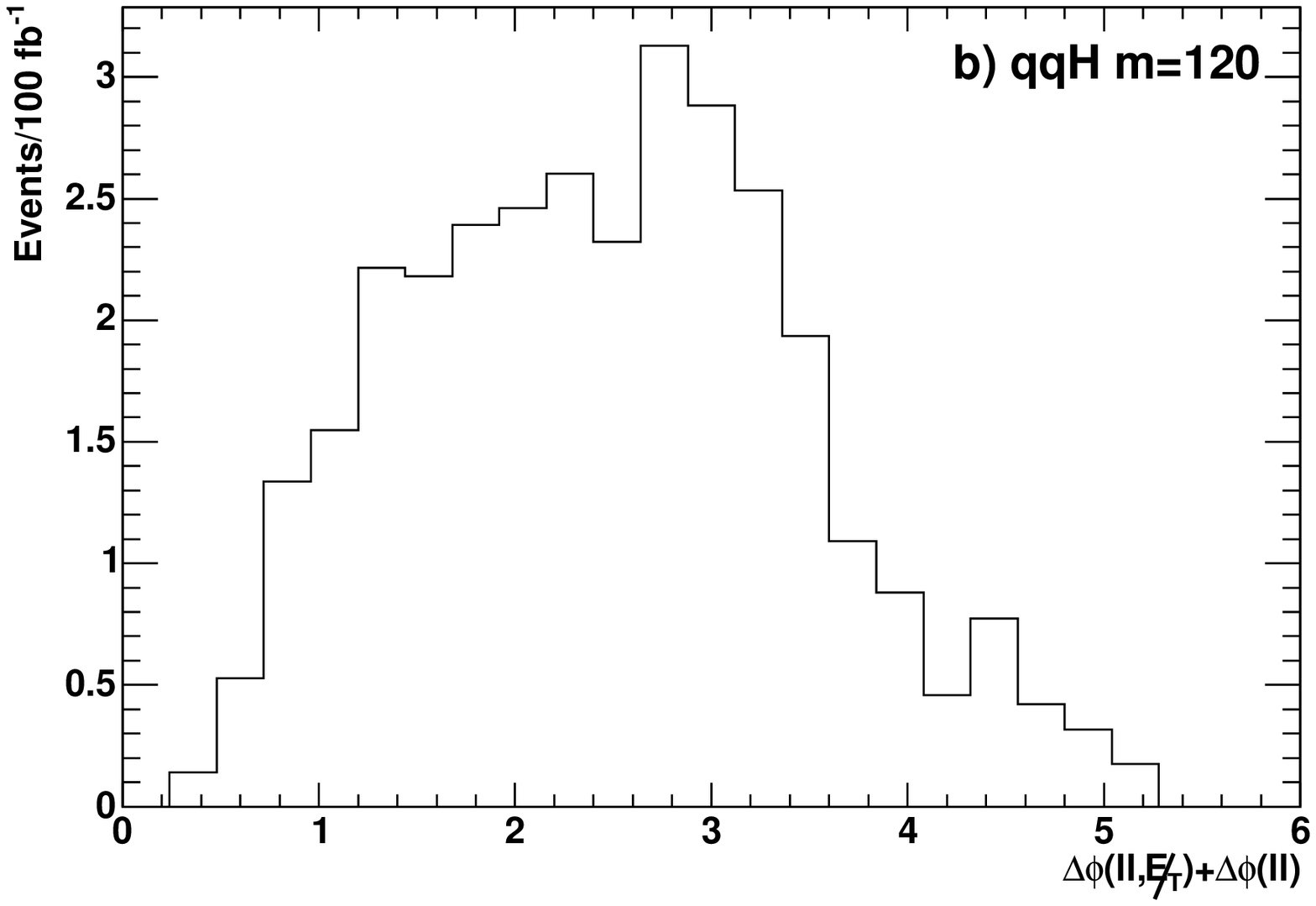}}
    \resizebox{9cm}{!}{\includegraphics{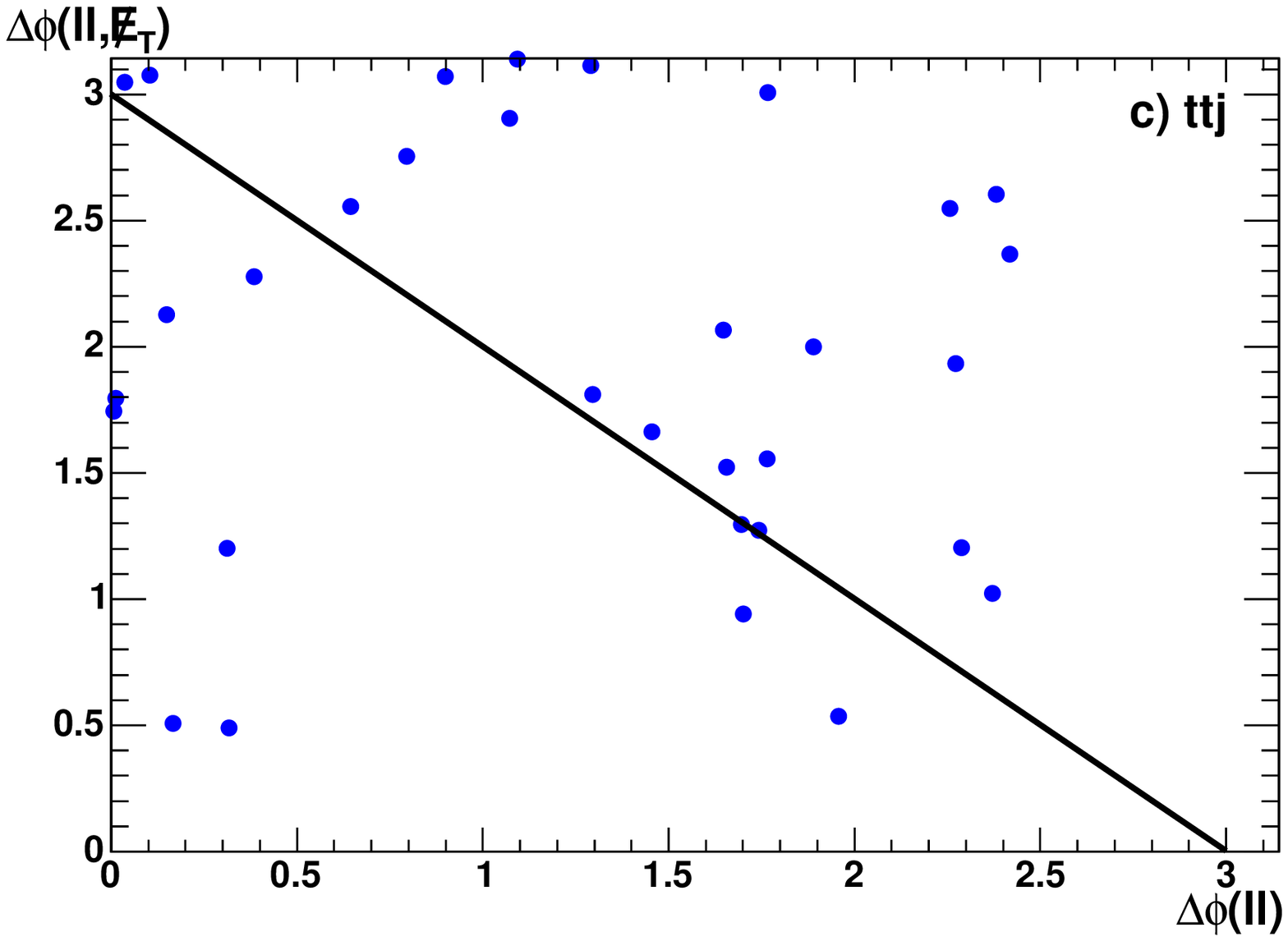}}
    \resizebox{9cm}{!}{\includegraphics{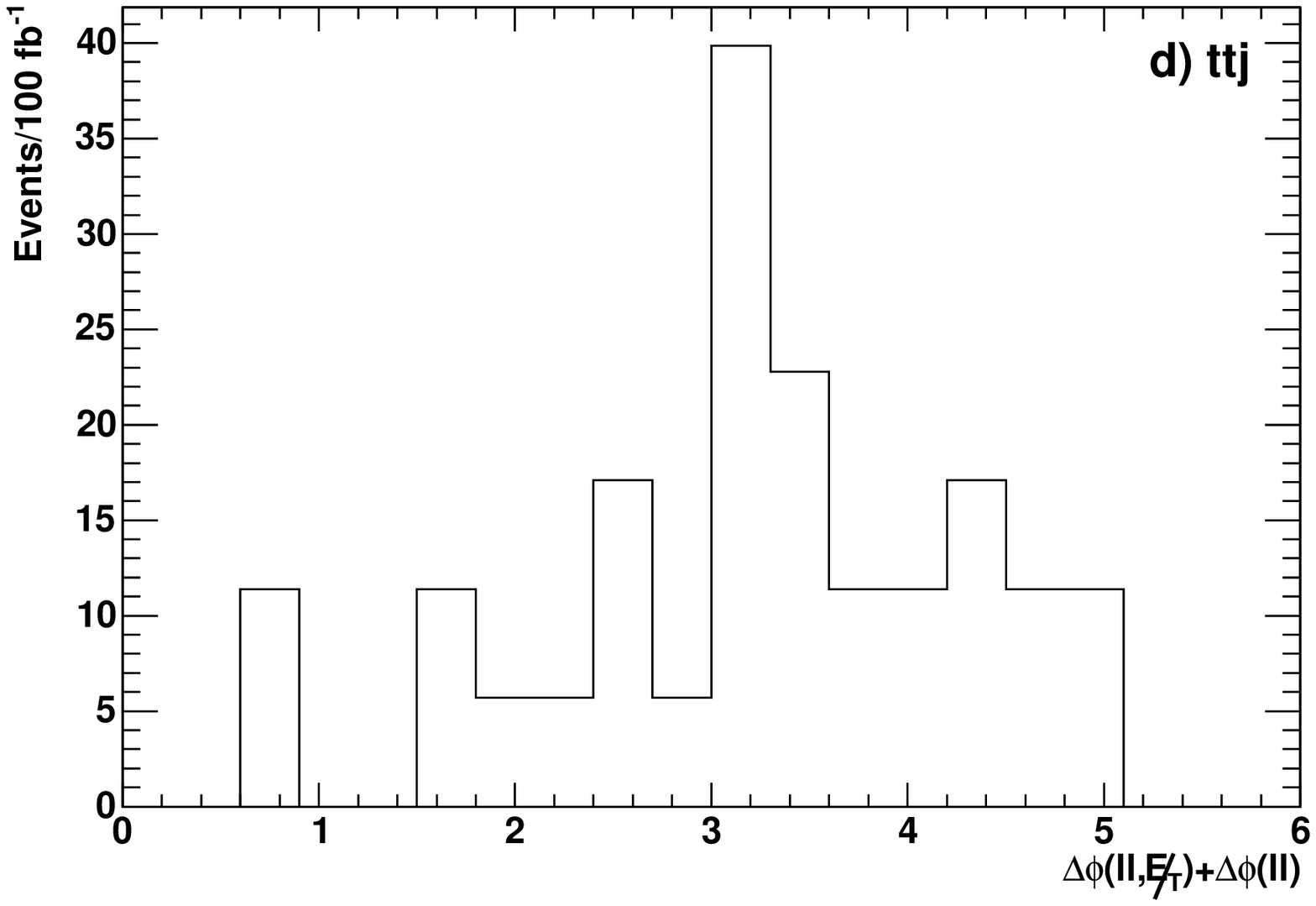}}
     \caption{\label{fig:dhillmetdphill}\em
  a) $\Delta\phi(ll,\not\!\!\!E_T)$ vs. $\Delta\phi_{ll}$ and 
b) the som of $\Delta\phi_{ll}$ and $\Delta\phi(ll,\not\!\!\!E_T)$ for qqH $m_H$ = 120 
and for ttj bacground c) and d). The lines correspond to 
$\Delta\phi_{ll}+\Delta\phi(ll,\not\!\!\!E_T) = $ 3 radians. }
\end{figure*}

 Finally, we impose the cut $\Delta\phi(\ell\ell,\not\!\!\!E_T) + \Delta\phi(\ell\ell) < 3$ radians, which
increases the signal-to-background ratio.
Fig.~\ref{fig:dhillmetdphill} shows distributions of this quantity.
The resolution of the quantity $\Delta\phi(\ell\ell,\not\!\!\!E_T)$ is improved by the 
$\not\!\!\!E_T$ correction. The additional cuts 
imposed after the transverse mass cut were determined for generator level 
analysis. Therefore, we did not include these cuts in the significance, 
background or mass estimation and their effect is seperately shown in 
Tables 4-6. Work is in progress to confirm their effect after full detector 
simulations.  
 
\section{Results}
\par
The total accepted signal cross-sections range from about $0.8$~fb
up to $7.2$~fb, depending on the Higgs mass.  They are listed in
Table~\ref{table:acccs}.   The contributions from
the $e^+e^-$ and $\mu^+\mu^-$ channel are very similar, and the
$e^\pm\mu^\mp$ channels are twice as large due to branching ratios.
The total efficiency is $3$--$6\%$, depending on ~$m_H$.
The background cross-sections are somewhat larger, and there
are two background values corresponding to the ``low-mass'' and the
``high-mass'' cuts -- see Table~\ref{table:acccs}.
\par
We computed the significance $S_{cP}$ of an excess of events over
the $t\bar{t}j$ and $W^+W^-jj$ backgrounds, assuming an integrated 
luminosity of   ${\cal{L}} = 10,~30$ and $100~\mathrm{fb}^{-1}$.
$S_{cP}$ is the probability calculated assuming a Poisson distribution with 
$N_B$ background events to observe equal or greater than a total number 
of signal and background events ($N_S+N_B$), converted to an equivalent 
number of sigmas for a Gaussian distribution~\cite{scp}.
The code to calculate $S_{cP}$ is taken from Ref.~\cite{scpcode}.
\par
The background uncertainty is included in the calculation. 
This uncertainty comes from the statistical error in the background 
estimation and amounts to about $12\%$ at $10~\mathrm{fb}^{-1}$,
$7\%$ at $30~\mathrm{fb}^{-1}$ and $4\%$ at $100~\mathrm{fb}^{-1}$.
See Section~\ref{sec:bgest} for a discussion of the background estimation. 
\par
\begin{figure}
\resizebox{9cm}{!}{\includegraphics{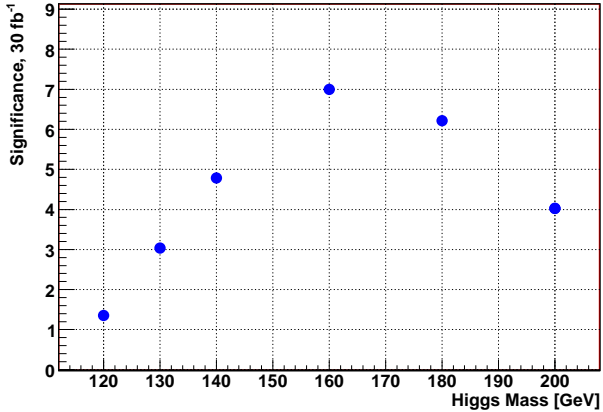}}
\caption[.]{\label{fig:signif}\em
       Significance of the Higgs signal as a function of Higgs mass for a 30 $fb^{-1}$ integrated 
luminosity.}
\end{figure}
\begin{figure}
\resizebox{9cm}{!}{\includegraphics{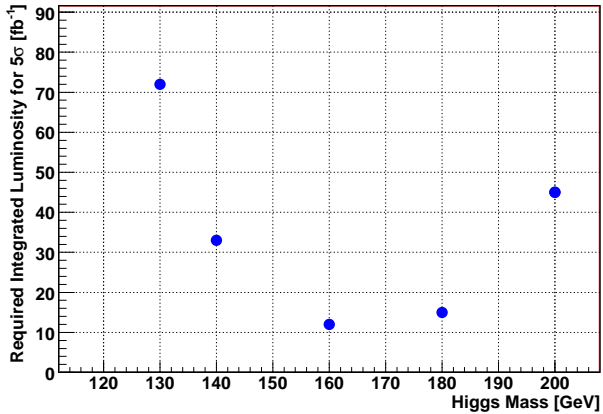}}
\caption[.]{\label{fig:minlum} Minimum integrated luminosity ($\mathrm{fb}^{-1}$) needed to obtain a $5\sigma$ excess over the $t\bar{t}j+W^+W^-jj$ background as a function of the Higgs mass.}
\end{figure}

The results are summarized in Table~\ref{table:signif}.  Even for a Higgs
mass as low as 130~GeV, a $5\sigma$ signal can be obtained with
a reasonable integrated luminosity.  For higher Higgs masses,
a very strong signal would be expected, and prospects for a
measurement of the cross section for $pp\rightarrow qqH$ become more promising.
Fig.~\ref{fig:signif} shows the significance  for an integrated 
luminosity of $30~\mathrm{fb}^{-1}$  as a function of~$m_H$,
and Fig.~\ref{fig:minlum} shows the minimum integrated luminosity
needed for a $5\sigma$ signal also as a function of $m_H$. The individual cut efficiencies
with respect to the starting cross-section 
for~$120$ and~$160~GeV$ Higgs bosons and the backgrounds are 
shown in Tables~\ref{table:resultee},\ref{table:resultem},\ref{table:resultmm}
for each channel.

\begin{table}
\caption[.]{\label{table:acccs}
Summary of accepted cross sections, in fb.
A series of assumed Higgs boson masses is shown, as well
as the backgrounds for the ``low-mass'' and ``high-mass'' cuts.}
\begin{center}
\begin{tabular}{lcccc}
 & \multicolumn{4}{c}{accepted cross-sections (fb)} \cr
channel & $e^+e^-$ & $e^\pm\mu^\mp$ & $\mu^+\mu^-$ & sum  \cr
\hline\hline
 \multicolumn{5}{c}{``low'' mass} \cr
$qqH$, $m_H = 120$~GeV & 0.183    & 0.400    & 0.253    & 0.836   \cr
$qqH$, $m_H = 130$~GeV & 0.387    & 0.854    & 0.601    & 1.842   \cr
$qqH$, $m_H = 140$~GeV & 0.617    & 1.341    & 0.955    & 2.913   \cr
$t\bar{t}j$                 & 1.139    & 2.621    & 1.065    & 4.825 \cr
$W^+W^-jj$ (EWK)          & 0.081    & 0.144    & 0.092    & 0.317  \cr
$W^+W^-jj$ (QCD)          & 0.093    & 0.207    & 0.119    & 0.419 \cr
all backgrounds        &          &          &          & 5.561  \cr
\hline\hline
 \multicolumn{5}{c}{``high'' mass} \cr
$qqH$, $m_H = 160$~GeV & 1.587    & 3.497    & 2.102   & 7.186   \cr
$qqH$, $m_H = 180$~GeV & 1.362    & 3.089    & 1.837   & 6.288   \cr
$qqH$, $m_H = 200$~GeV & 0.815    & 1.703    & 1.087   & 3.605   \cr
$t\bar{t}j$                 & 2.088    & 4.216    & 2.024   & 8.328 \cr
$W^+W^-jj$ (EWK)          & 0.127    & 0.245    & 0.165   & 0.537 \cr
$W^+W^-jj$ (QCD)          & 0.192    & 0.394    & 0.252   & 0.838 \cr
all backgrounds        &          &          &         & 9.703 \cr
\end{tabular}
\end{center}
\end{table}

\begin{table}
\caption[.]{\label{table:signif}
Significance of an excess as a function of Higgs mass,
for three assumed integrated luminosities.  The last column shows the
minimum luminosity required for a $5\sigma$ excess.}
\begin{center}
\begin{tabular}{ccccc}
Higgs mass   &  \multicolumn{3}{c}{significance} & ${\cal{L}}_{\mathrm{min}}^{5\sigma}$ \cr
   (GeV)     &   $10~\mathrm{fb}^{-1}$ & $30~\mathrm{fb}^{-1}$ & $100~\mathrm{fb}^{-1}$ & ($\mathrm{fb}^{-1}$) \cr
\hline\hline


 120  & {0.72}  & {1.35} & {2.60} & {340}  \cr
 130  & {1.77}  & {3.04} & {5.85} & {72}  \cr
 140  & {2.68}  & {4.79} & {8.33} & {33} \cr
 160  & {4.54}  & {7.00} & {13.0} & {12}  \cr
 180  & {3.95}  & {6.22} & {11.6} & {15}  \cr
 200  & {2.31}  & {4.03} & {6.99} & {45} \cr

\end{tabular}
\end{center}
\end{table}

\newpage
\begin{table*}
\caption[.]{Accepted signal (for m$_{H}$=120,160 GeV) and major background cross-sections in fb for the
$ee\nu\nu$ final state.}
\begin{center}
\begin{tabular}{llllll} \hline
           Cut                 		&  qqH    &  qqH  & ttj      & WWjj  & WWjj \cr
					&    120  &  160  &          &  EW   & QCD \cr
\hline\hline
                          		&  {5.261} & {26.97} & {8617.}    & {10.74}    &  {514.3}         \cr
$E_{T1}>50$,$E_{T2}>30$ GeV   		&  {3.742} & {18.70} & {6743.}    & {8.838}    &  {296.4}         \cr
$\Delta\eta >$ 4.2        		&  {1.217} & {6.067} & {184.2}    & {2.195}    &  {12.22}         \cr
$\eta_1\times\eta_2 < 0$  		&  {1.215} & {6.054} & {183.1}    & {2.193}    &  {12.18}         \cr
$M_{jj} > 600$       GeV     		&  {1.073} & {5.367} & {147.2}    & {2.071}    &  {9.052}         \cr
P$_T-balance$ cut         		&  {0.653} & {3.353} & {54.89}    & {1.021}    &  {3.298}         \cr
Central Jet Veto          		&  {0.401} & {2.309} & {15.04}    & {0.631}    &  {1.490}         \cr
$\geq$ 2 good leptons w opp. charge	&  {0.269} & {1.915} & {10.98}    & {0.483}    &  {0.695}         \cr
$p_T>20,10$ or $p_T>26,12$ GeV		&  {0.250} & {1.838} & {10.59}    & {0.475}    &  {0.675}         \cr
$|\Delta R(j,l)| > 0.7$   		&  {0.250} & {1.830} & {10.33}    & {0.471}    &  {0.662}         \cr
Req. leptons between jets 		&  {0.235} & {1.712} & {4.990}    & {0.417}    &  {0.430}         \cr
$M_{ll} < 80$ GeV             		&  {0.235} & {1.683} & {2.386}    & {0.144}    &  {0.205}         \cr
$\Delta\phi_{ll} < 2.4$   	        &  {0.220} & {1.587} & {2.088}    & {0.127}    &  {0.192}         \cr
$M_{T,WW} < 125$ GeV          	        &  {0.183} &         & {1.139}    & {0.081}    &  {0.093}         \cr
\hline
$57.29\Delta\phi(ll,\not\!\!\!E_T)+1.5p_T(H)>180$  \& &          &         &            &            &                \cr
$12\times57.29\Delta\phi(ll,\not\!\!\!E_T)+p_T(H)>360$&  {0.161} &         & {0.936}    & {0.069}    &  {0.073}      \cr
$M_{ll}>10 \& \not\!\!\!E_T>30 (ee,\mu\mu)$      &  {0.115} &         & {0.800}    & {0.053}    &  {0.060}         \cr
$\Delta\phi(ll,\not\!\!\!E_T)+\Delta\phi_{ll}<3$   &  {0.090} &     & {0.420}    & 
{0.031}    &  {0.033}              \cr
\hline
High Mass Cuts                          &          &         &            &            &                \cr
No $M_{T,WW}$ Cut	           	&          &  {1.587}& {2.088}    & {0.127}    &  {0.192}         \cr
\hline
$57.29\Delta\phi(ll,\not\!\!\!E_T)+1.5p_T(H)>180$ \&  &          &         &            &            &                \cr
$12\times57.29\Delta\phi(ll,\not\!\!\!E_T)+p_T(H)>360$    &          &  {1.501}& {1.885}    & {0.114}    &  {0.172} \cr
$M_{ll}>10~GeV \& \not\!\!\!E_T>30 (ee,\mu\mu)$ GeV&          &  {1.303}& {1.736}    & {0.098}    &  {0.152}         \cr
$\Delta\phi(ll,\not\!\!\!E_T)+\Delta\phi_{ll}<3$   &      &  {0.862}& {0.651}    & {0.052}    &  {0.046}              \cr
\hline
      \end{tabular}
    \label{table:resultee}
\end{center}
\end{table*}


\begin{table*}
    \caption[.]{Accepted signal (for m$_{H}$=120,160 GeV) and major background cross-sections in fb for the
$e\mu\nu\nu$ final state.}
\begin{center}
\begin{tabular}{llllll} \hline
           Cut                 		&  qqH  & qqH    & ttj        & WWjj   & WWjj \cr
					& 120 & 160 &  &  EW & QCD \cr
\hline\hline
                          		& {10.57}  & {53.24}   &  {17230.}  & {21.48}    & {1029.}         \cr
$E_{T1}>50$,$E_{T2}>30$ GeV   		& {7.290}  & {35.54}   &  {13320.}  & {17.22}    & {537.1}         \cr
$\Delta\eta >$ 4.2        		& {2.458}  & {12.56}   &  {358.5}   & {4.533}    & {24.39}          \cr
$\eta_1\times\eta_2 < 0$  		& {2.454}  & {12.55}   &  {355.5}   & {4.526}    & {24.25}          \cr
$M_{jj} > 600$    GeV        		& {2.149}  & {11.08}   &  {282.0}   & {4.299}    & {18.28}          \cr
P$_T-balance$ cut         		& {1.398}  & {7.390}   &  {117.4}   & {2.405}    & {8.287}          \cr
Central Jet Veto          		& {0.879}  & {5.128}   &  {32.70}   & {1.502}    & {4.123}          \cr
$\geq$ 2 good leptons w opp. charge   & {0.670}  & {4.388}   &  {25.07}   & {1.186}    & {2.102}          \cr
$p_T>20,10$ GeV				& {0.544}  & {4.079}   &  {23.47}   & {1.131}    & {1.975}          \cr
$|\Delta R(j,l)| > 0.7$   		& {0.539}  & {4.052}   &  {21.71}   & {1.100}    & {1.881}          \cr
Req. leptons between jets 		& {0.506}  & {3.748}   &  {10.60}   & {0.920}    & {1.068}          \cr
$M_{ll} < 80$  GeV           		& {0.505}  & {3.685}   &  {5.014}   & {0.301}    & {0.447}          \cr
$\Delta\phi_{ll} < 2.4$   	        & {0.480}  & {3.497}   &  {4.216}   & {0.245}    & {0.394}          \cr
$M_{T,WW} < 125$  GeV         	        & {0.400}  &           &  {2.621}   & {0.144}    & {0.207}          \cr
\hline
$57.29\Delta\phi(ll,\not\!\!\!E_T)+1.5p_T(H)>180$ \&  &          &           &            &            &           \cr
$12\times57.29\Delta\phi(ll,\not\!\!\!E_T)+P_T(H)>360$& {0.329}  &           &  {1.880}   & {0.109}    & {0.153}    \cr
$\not\!\!\!E_T>30$ GeV if $p_T(H)<50$ GeV        & {0.323}  &           &  {1.823}   & {0.105}    & {0.153}          \cr
$\Delta\phi(ll,\not\!\!\!E_T)+\Delta\phi_{ll}<3$ & {0.239}  &       &  {0.798}   & {0.066}    & {0.08}       \cr  
\hline
High Mass Cuts                          &          &           &            &            &                \cr
No $M_{T,WW}$ Cut	           	&          &  {3.497}  &  {4.216}   & {0.245}    & {0.394}          \cr
\hline
$57.29\Delta\phi(ll,\not\!\!\!E_T)+1.5p_T(H)>180$ \&  &          &           &            &            &         \cr
$12\times57.29\Delta\phi(ll,\not\!\!\!E_T)+P_T(H)>360$&          &  {3.105}  &  {3.418}   & {0.202}    & {0.334}    \cr
$\not\!\!\!E_T>30$ GeV if $p_T(H)<50$ GeV        &          &  {3.084}  &  {3.361}   & {0.199}    & {0.334}          \cr
$\Delta\phi(ll,\not\!\!\!E_T)+\Delta\phi_{ll}<3$ &          &  {2.003}  &  {1.709}   & {0.107}    & {0.173}               \cr
\hline
      \end{tabular}
    \label{table:resultem}
\end{center}
\end{table*}


\begin{table*}
    \caption[.]{\em Accepted signal (for m$_{H}$=120,160 GeV) and major background cross-sections in fb for the
$\mu\mu\nu\nu$ final state.}
\begin{center}
\begin{tabular}{llllll} \hline
           Cut                 		&  qqH&  qqH & ttj & WWjj & WWjj \cr
					& 120 & 160 &  &  EW & QCD \cr
\hline\hline
                          		& {5.133}  & {29.44}  & {8617.}    & {10.77}    &   {512.7}        \cr
$E_{T1}>50$,$E_{T2}>30$ GeV   		& {3.357}  & {18.31}  & {6621.}    & {8.332}    &   {232.5}        \cr
$\Delta\eta >$ 4.2        		& {1.271}  & {7.391}  & {178.0}    & {2.365}    &   {12.11}        \cr
$\eta_1\times\eta_2 < 0$  		& {1.268}  & {7.375}  & {176.7}    & {2.360}    &   {12.06}        \cr
$M_{jj} > 600$ GeV           		& {1.109}  & {6.522}  & {139.7}    & {2.251}    &   {8.988}        \cr
P$_T-balance$ cut         		& {0.854}  & {4.947}  & {55.75}    & {1.585}    &   {5.768}        \cr
Central Jet Veto          		& {0.562}  & {3.523}  & {19.55}    & {1.007}    &   {3.139}        \cr
$\geq$ 2 good leptons w opp. charge   & {0.430}  & {2.891}  & {16.11}    & {0.772}    &   {1.472}        \cr
$p_T>20,10$ GeV				& {0.327}  & {2.605}  & {14.30}    & {0.716}    &   {1.324}        \cr
$|\Delta R(j,l)| > 0.7$   		& {0.319}  & {2.537}  & {11.59}    & {0.680}    &   {1.186}        \cr
Req. leptons between jets 		& {0.290}  & {2.298}  & {5.461}    & {0.556}    &   {0.548}        \cr
$M_{ll} < 80$    GeV         		& {0.290}  & {2.226}  & {2.371}    & {0.190}    &   {0.271}        \cr
$\Delta\phi_{ll} < 2.4$   	        & {0.273}  & {2.102}  & {2.024}    & {0.165}    &   {0.252}        \cr
$M_{T,WW} < 125$    GeV       	        & {0.253}  &          & {1.065}    & {0.092}    &   {0.119}        \cr
\hline
$57.29\Delta\phi(ll,\not\!\!\!E_T)+1.5p_T(H)>180$  \&   &   	   &          &            &    	&          \cr
$12\times57.29\Delta\phi(ll,\not\!\!\!E_T)+p_T(H)>360$& {0.200}  &          & {0.826}    & {0.075}    &   {0.095}  \cr
$M_{ll}>10~GeV \& \not\!\!\!E_T>30 (ee,\mu\mu)$ GeV& {0.159}  &          & {0.746}    & {0.060}    &   {0.076}        \cr
$\Delta\phi(ll,\not\!\!\!E_T)+\Delta\phi_{ll}<3$   & {0.134}  &      & {0.426}    & {0.051}    &   {0.062}         \cr
\hline
High Mass Cuts                          &          &          &            &            &                \cr
No $M_{T,WW}$ Cut	           	&          &  {2.102} & {2.024}    & {0.165}    &   {0.252}        \cr
\hline
$57.29\Delta\phi(ll,\not\!\!\!E_T)+1.5p_T(H)>180$  \& &          &          &            &            &             \cr
$12\times57.29\Delta\phi(ll,\not\!\!\!E_T)+p_T(H)>360$&          &  {1.908} & {1.785}    & {0.147}    &   {0.229}   \cr
$\not\!\!\!E_T>30$ GeV if $p_T(H)<50$ GeV        &          &  {1.681} & {1.678}    & {0.132}    &   {0.205}        \cr
$\Delta\phi(ll,\not\!\!\!E_T)+\Delta\phi_{ll}<3$ &      &  {1.229} & {0.746}    & {0.092}    &   {0.119}         \cr
\hline

      \end{tabular}
    \label{table:resultmm}
\end{center}
\end{table*}


\par
Concerning systematics, we have first considered the impact of the jet energy scale.
The expected jet energy scale uncertainty in CMS is about $3\%$. 
For the $t\bar{t}j$ background the scale uncertainty after correction is about $5\%$
for $p_T>30$ GeV. In this analysis, the two tag jets are required to have
$Ep_{T1}>$50 GeV and $E_{T2}>$30 GeV and we reject additional jets in the central 
region if their $E_T>20$ GeV. For the jets with $E_T\sim20$ GeV,
the cross-section uncertainty after jet correction is about $10\%$.
We re-computed all yields after scaling the raw jet energies up and down by~$10\%$.  
In general, signal and background yields correlate, so the impact
on the significance with a $10\%$ jet energy scale uncertainty
 is less than $\sim8-10\%$ at 30 $fb^{-1}$.

We also tested our results for the significances to errors in the $\not\!\!\!E_T$ scale.
Increasing the $\not\!\!\!E_T$ scale by 10$\%$ decreases the significance 
by $9$ -- $11\%$. Decreasing the $\not\!\!\!E_T$ scale by 10$\%$ increases the 
significance by $0.3$ -- $3.4\%$ depending on~$m_H$. This is a systematic uncertainty
on the signal cross section.
\par
We also used the Pythia event generator for our signal as an alternative
to MadGraph.  For $m_H = 120~GeV$, the significance obtained with
Pythia is higher by~$30\%$ for a luminosity of~$100~\mathrm{fb}^{-1}$, while
for $m_H = 160~GeV$, it is higher by $10~\%$.
\par
 We found that the production cross-section depends on the choice of scale
(renormalization scale$\times$factorization scale) for the $t\bar{t}j$ background. 
The $t\bar{t}j$ cross-section is 736.5 pb as reported in Table 1,
with the definition of the scale $\Sigma m_T^2$, where $m_T^2=m_{top}^2+p_T^2$
and the sum is over {\it final state light partons}. 
However, if we change the definition of the
above sum to include {\it all the final state partons including the heavy quarks},
then the cross-section decreases to 530 pb. These two definitions of scale
are the defaults in AlpGen 1.3.3 and 2.0.x respectively.
We found that the choice of scale does not affect the kinematics of $t\bar{t}j$ at all.
Moreover, the cross-section and kinematics of the $qqH$ process are not affected by
the choice of scale.
The significance with the new scale choice is $\sim18\%$ higher.
Therefore, the uncertainties in the computed $t\bar{t}j$ background make it very important 
to measure the background directly in the experiment. 

It should be pointed out that the statistical significance
of our analysis is generally a factor of~$\sim2.6$--$3.2$
lower than the significance reported in the study for the
ATLAS detector~\cite{atlas}.  There are several reasons
for this difference. First of all, the $t\bar{t}j$ cross-section
used in Ref~\cite{atlas} is smaller than the cross-section
we use by about a factor of~$0.7$. Furthermore, the ATLAS
study includes the gluon-gluon fusion channel for Higgs production which
increases the signal by about $10\%$.

Another important difference between the two analyses concerns the central jet veto.  
Our signal simulation generates a larger number of central jets compared to
the ATLAS study, which used the PYTHIA Monte Carlo
event generator.  When we compare the signal efficiency
after all cuts using PYTHIA instead of MadGraph,
we find a difference of $\sim5-50\%$.
Finally, the very definition of significance ($S_{cP}$)
differs between the two studies.  The ATLAS study
used a definition which gives a value which is
$\sim9$--$14\%$ higher for the same number of signal
and background events.  If the number of background
events is reduced, the apparent improvement in the
significance increases more dramatically than for
our measure of significance. Thus the uncertainty of
$\sim9$--$14\%$ should be taken as a lower limit
for this particular factor.
Considering all of the above, the differences between
our results and those reported in Ref.~\cite{atlas}
can be understood.  Nonetheless, these considerations
show that there still are uncertainties in the modelling
of this channel which should be investigated by both experiments.

\subsection{Background Estimation from the Data}
\label{sec:bgest}
\begin{figure}
    \resizebox{9cm}{!}{\includegraphics{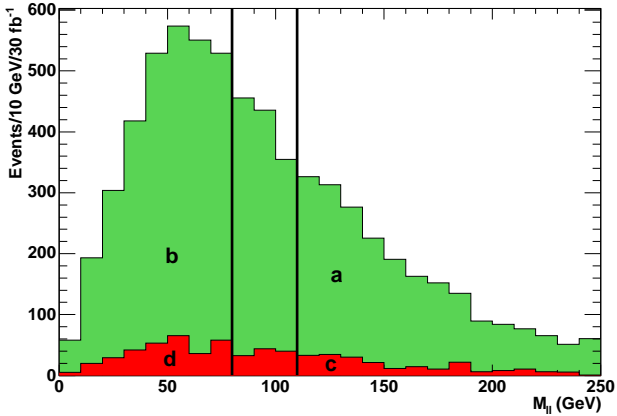}}     
     \caption{\label{fig:mllcontrol}\em
The $M_{\ell\ell}$ distribution computed with looser cuts and full analysis cuts. 
}
\end{figure}

\begin{figure}
    \resizebox{9cm}{!}{\includegraphics{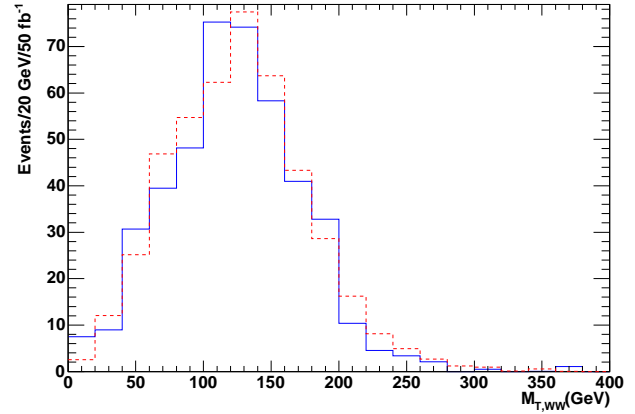}}
     \caption{\label{fig:mttt}
The transverse mass, $M_{T,WW}$, distribution for estimated(dashed) and real(solid) background. 
}
\end{figure}

For the Higgs masses considered here, there is practically no signal 
with $M_{\ell\ell} > 110~GeV$ -- see Fig.~\ref{fig:mll}.  For the
present discussion we define this as the {\sl signal-free region}.
Fig.~\ref{fig:mllcontrol} shows the $M_{\ell\ell}$ distribution computed
with looser cuts (no central jet veto, no $p_{T}$-balancing cut,
$|\Delta\eta|>3.5$, 
$\eta_{\mathrm{lo}}+0.3 < \eta_{\ell} < \eta_{\mathrm{hi}}-0.3$)
and the full analysis cuts. 
The number of events with $M_{\ell\ell} > 110~GeV$ is designated by ``$a$''
for the distribution with looser cuts and by ``$c$'' for the full analysis
cuts. The number of events for $M_{\ell\ell} < 80~GeV$ is designated by
``$b$'' for the distribution with looser cuts and by ``$d$'' for the full 
analysis cuts. The region $80 < M_{\ell\ell} < 110~GeV$ is excluded from the calculation
in order to avoid any background coming from $Z \rightarrow \ell^+\ell^-$.
Since $M_{\ell\ell} > 110~GeV$ represents the signal-free region, we can use
the numbers $a$, $c$ and $b$ to estimate the number of background events 
in the region where we expect the signal (i.e.,~$d$). 
Using the simulations, we find that $c/a = 0.097$ and $d/b = 0.098$.
The error on this estimation is dominated by the statistical uncertainty which is 
$\sqrt{c}/c \approx 7\%$.  In order to obtain the background
distribution in $M_{T,WW}$, we take the distribution obtained with
the looser cuts and scale it by the factor of~$0.098$.  A comparison
of the real and rescaled background distributions is given in
Fig.~\ref{fig:mttt} which indicates that this "data driven" works quite well. 

\subsection{Sensitivity to the Higgs Mass}
\par
\begin{figure*}
    \resizebox{9cm}{!}{\includegraphics{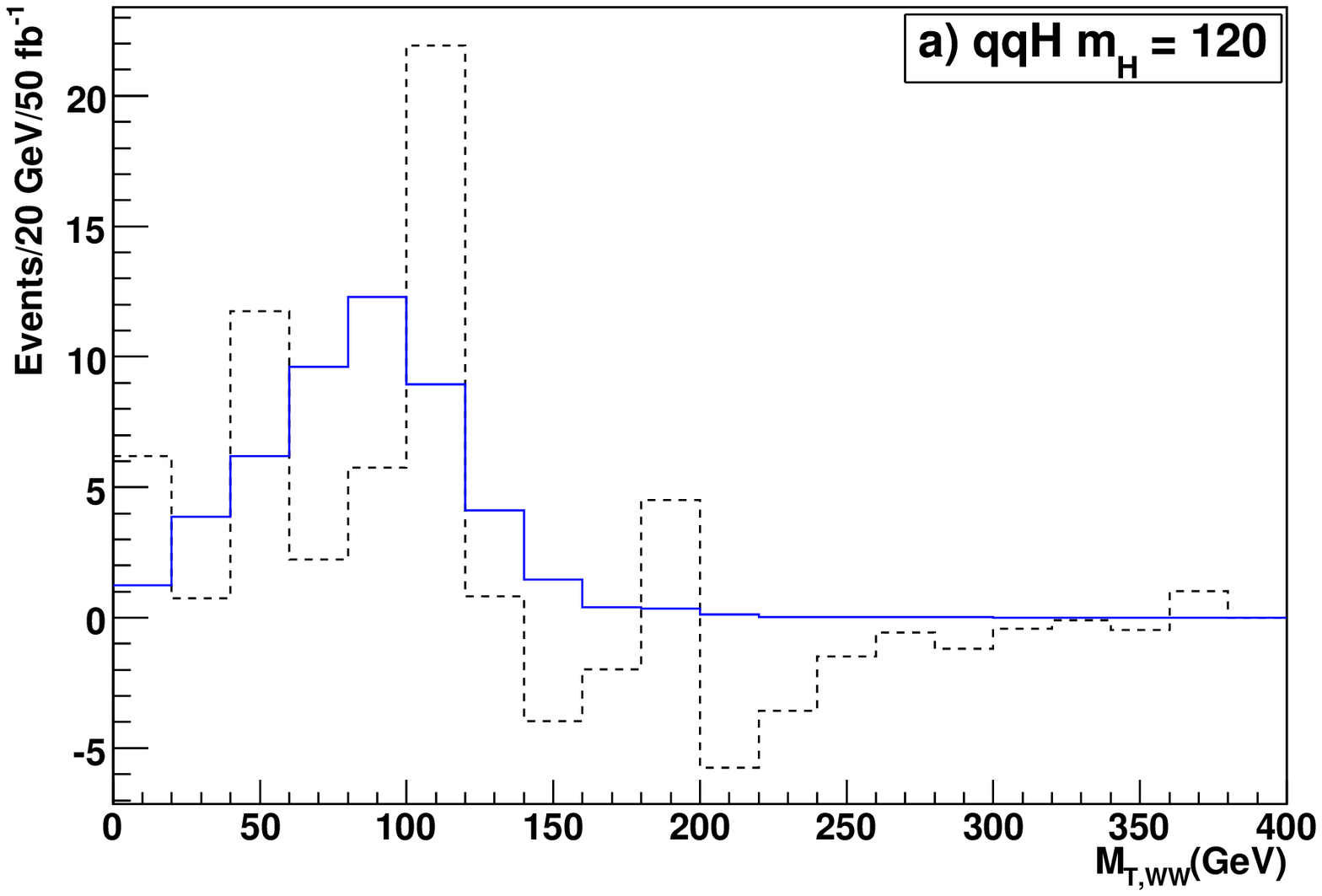}} 
    \resizebox{9cm}{!}{\includegraphics{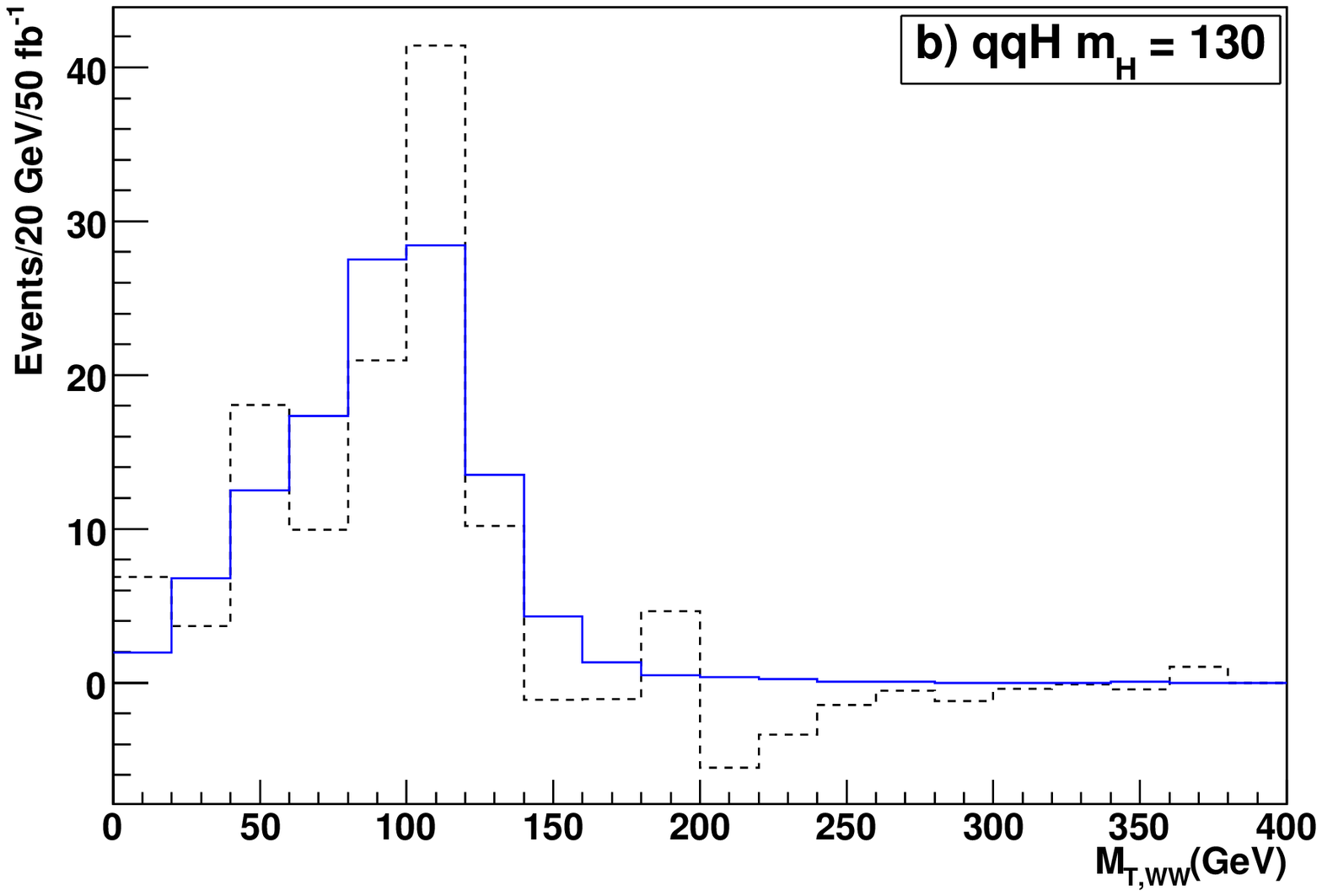}}
    \resizebox{9cm}{!}{\includegraphics{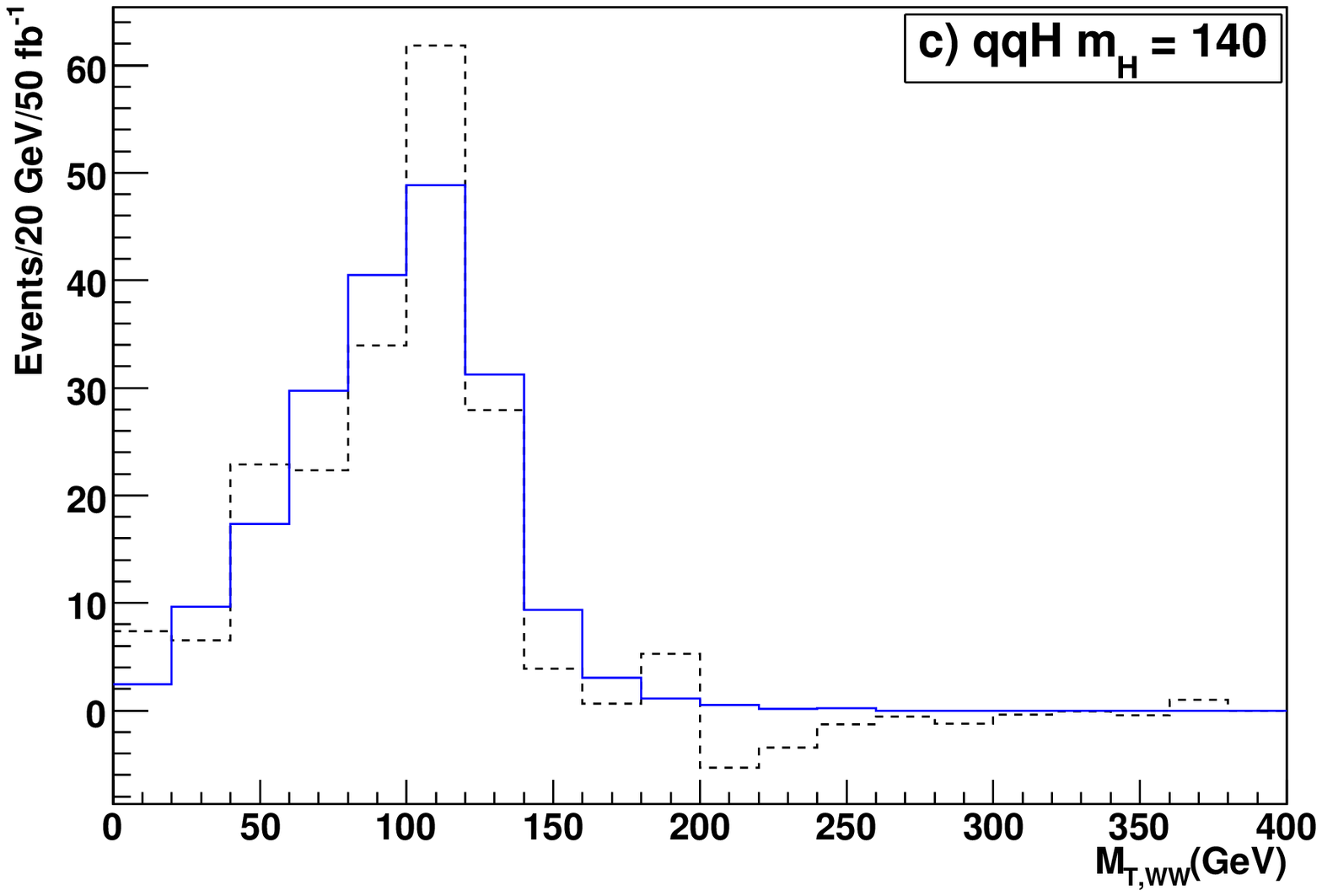}}
    \resizebox{9cm}{!}{\includegraphics{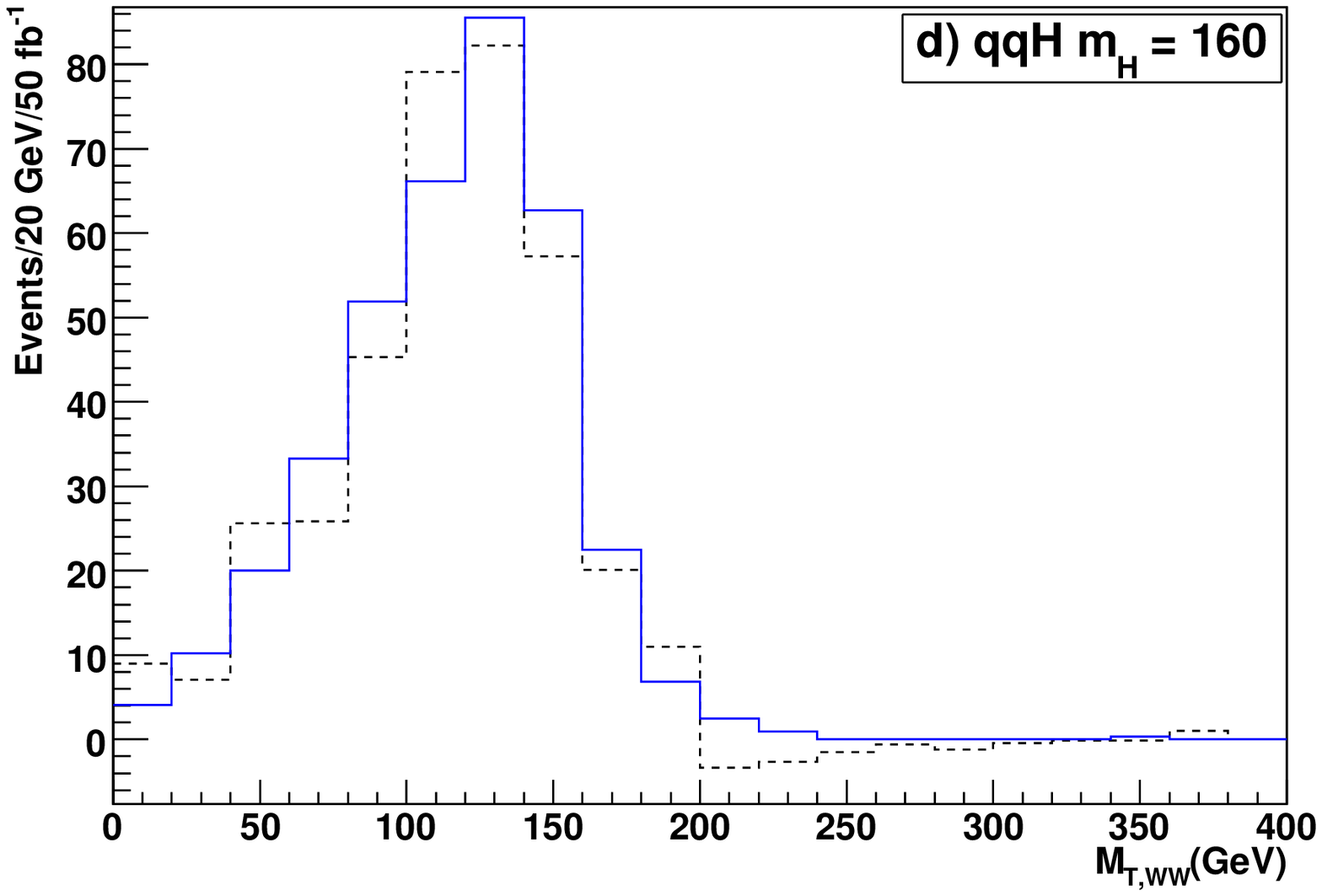}}
    \resizebox{9cm}{!}{\includegraphics{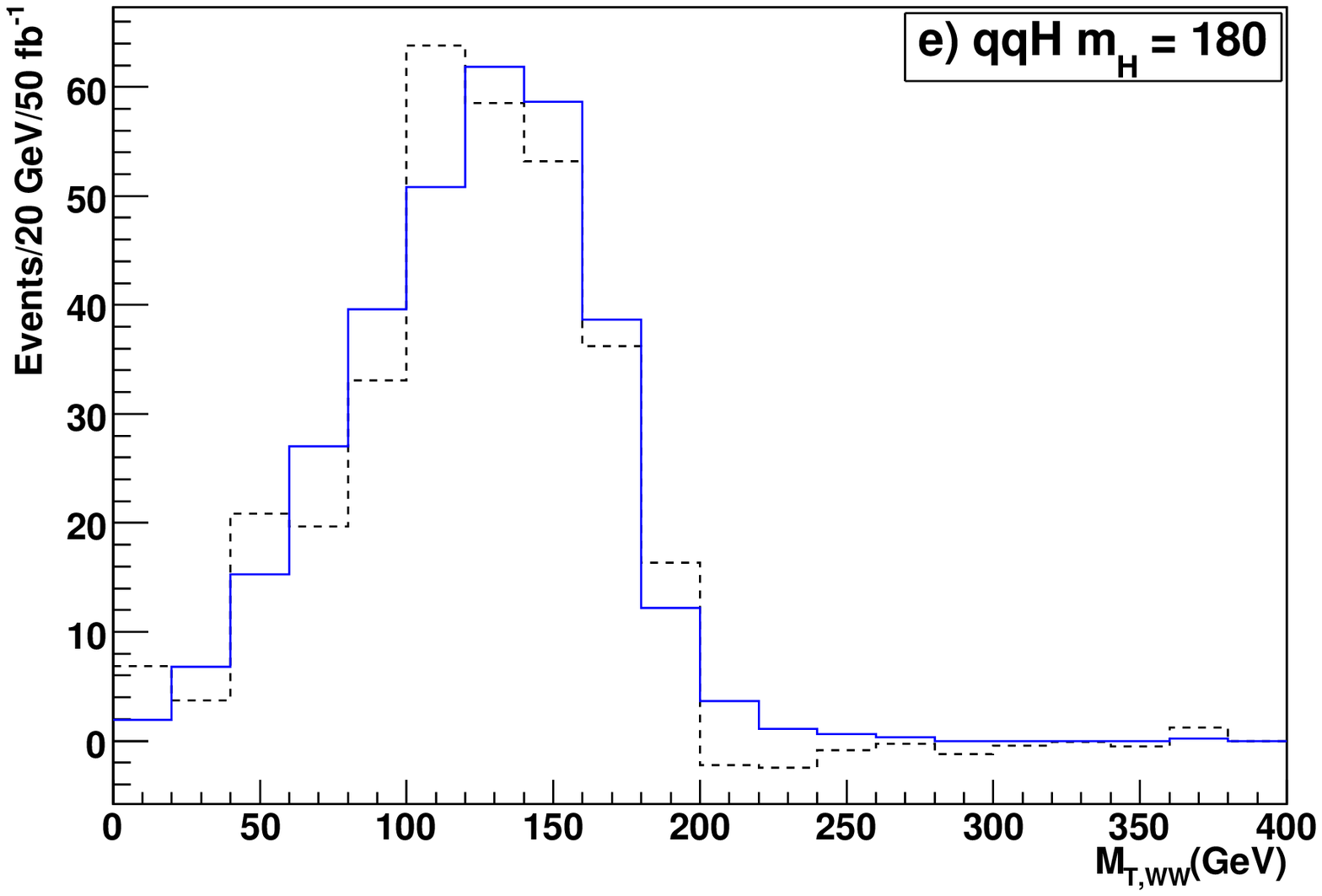}}
    \resizebox{9cm}{!}{\includegraphics{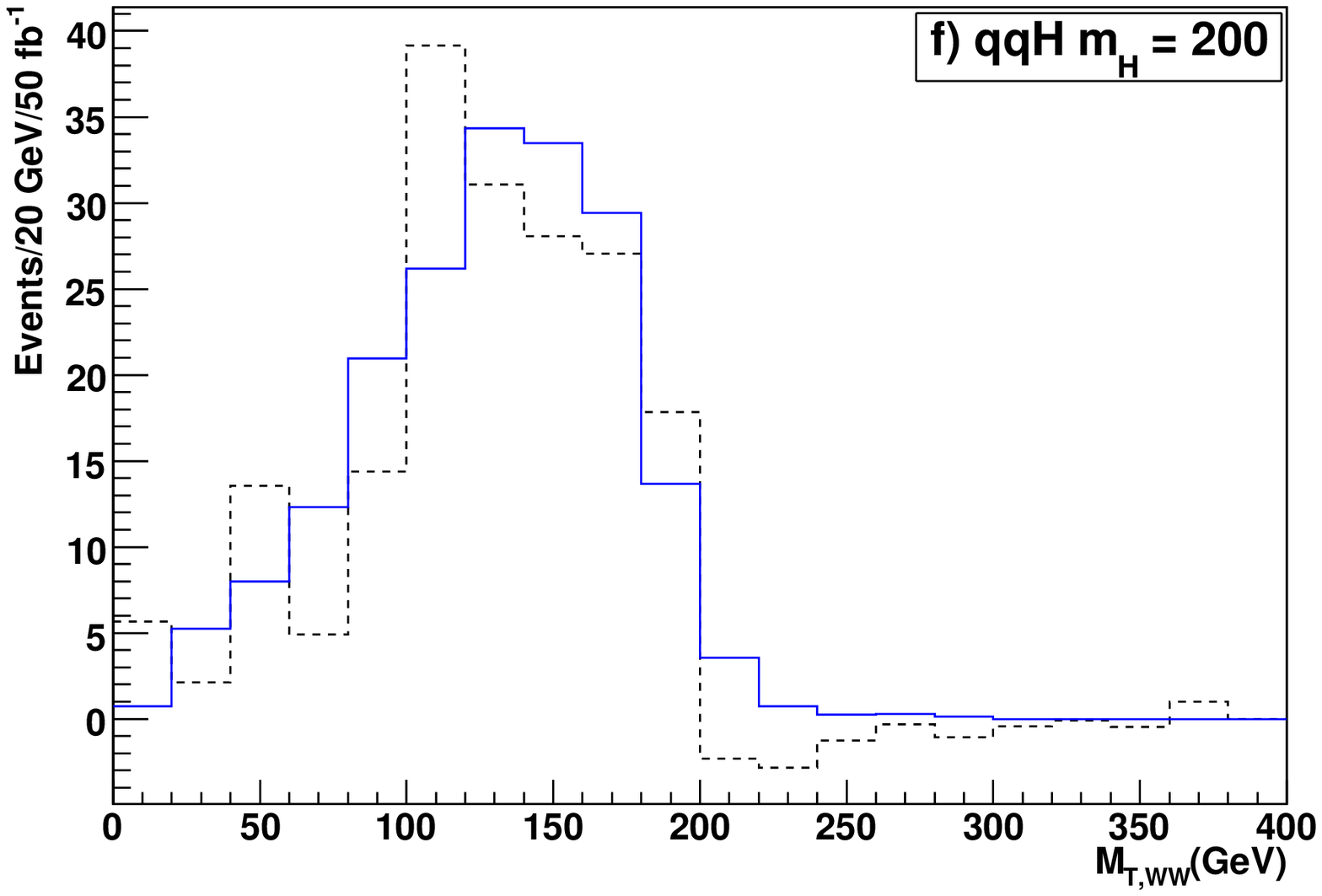}}
     \caption{\label{fig:mtestimated}\em
Estimated(dashed) and real(solid) $M_{T,WW}$ distributions for signal events, with Higgs mass of
120,130,140,160,180 and 200 GeV shown in a),b),c),d),e) and f) respectively.
}
\end{figure*}

The above significance estimates are for a pure``counting experiment''.
We can, in addition, use the information contained in the distribution of
$M_{T,WW}$ with regard to the Higgs mass.
We infer the mass of the Higgs boson from the observed
distribution in $M_{T,WW}$ by subtracting the data-driven estimate
of the background $M_{T,WW}$ distribution from the distribution obtained
with the full set of analysis cuts.  The estimated and real $M_{T,WW}$ 
distributions for signal events are shown in Fig.~\ref{fig:mtestimated}
for several different Higgs boson masses.  The inferred and the real
mean values and shapes approximately agree. 
\par

\begin{figure*}
    \resizebox{9cm}{!}{\includegraphics{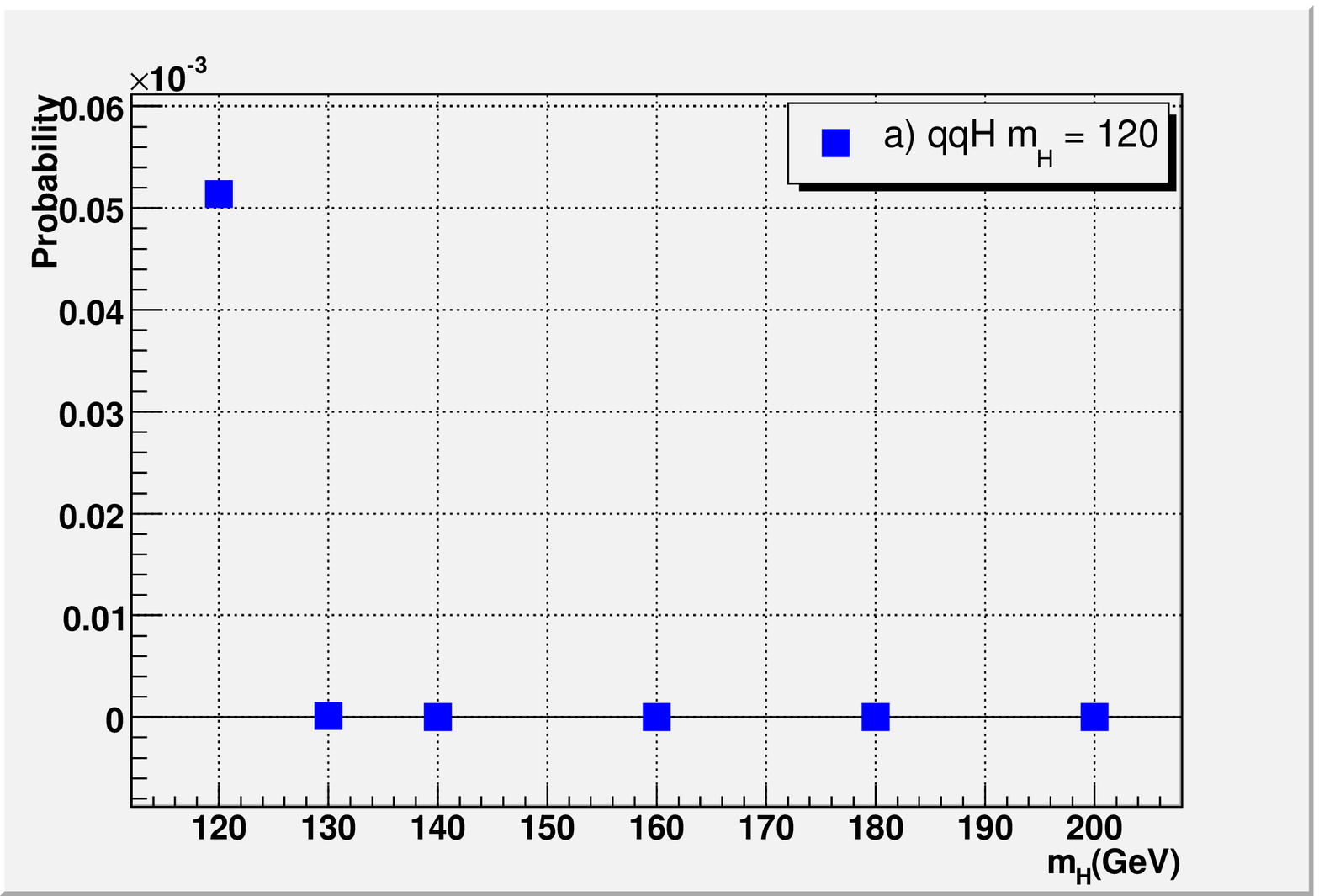}} 
    \resizebox{9cm}{!}{\includegraphics{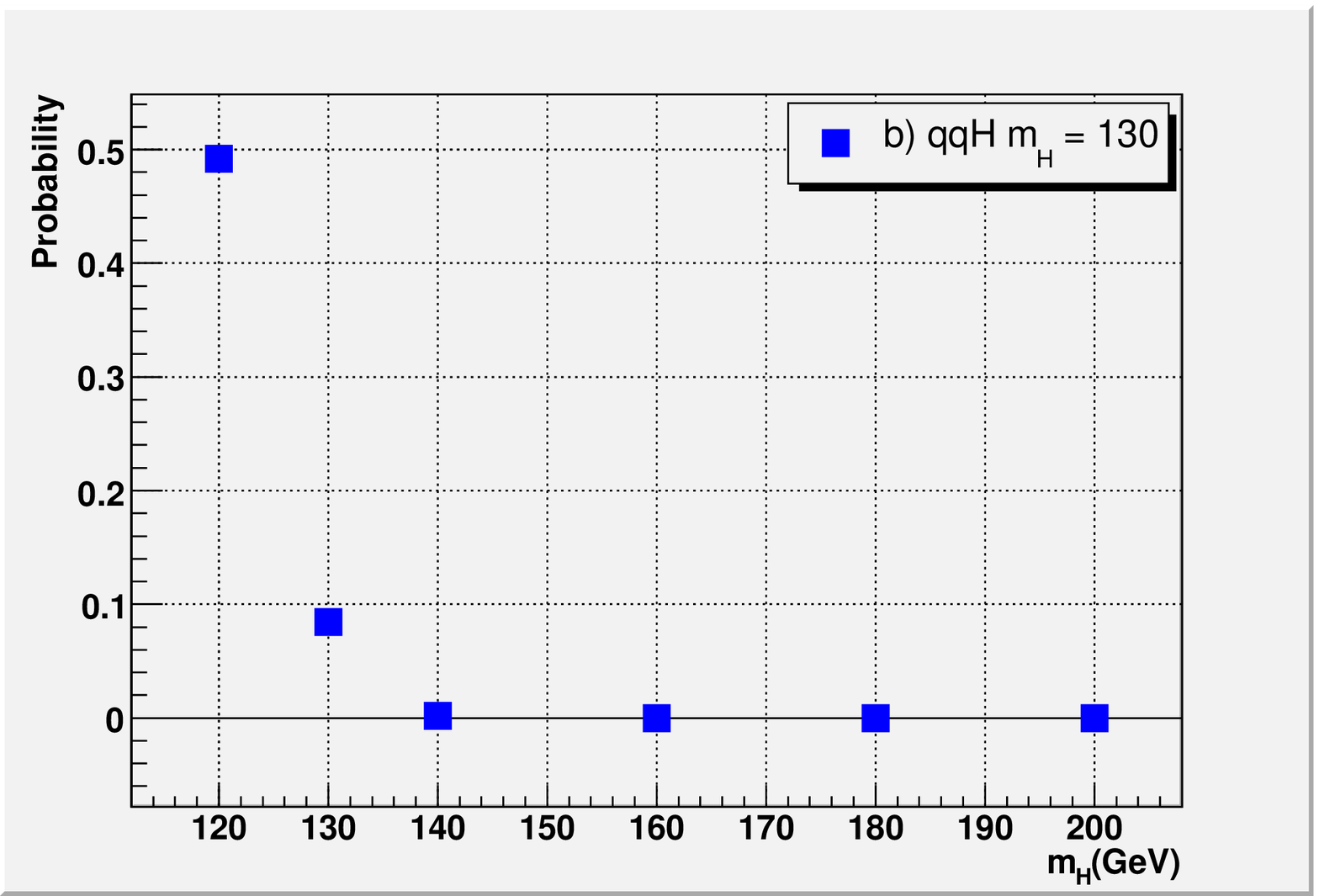}}
    \resizebox{9cm}{!}{\includegraphics{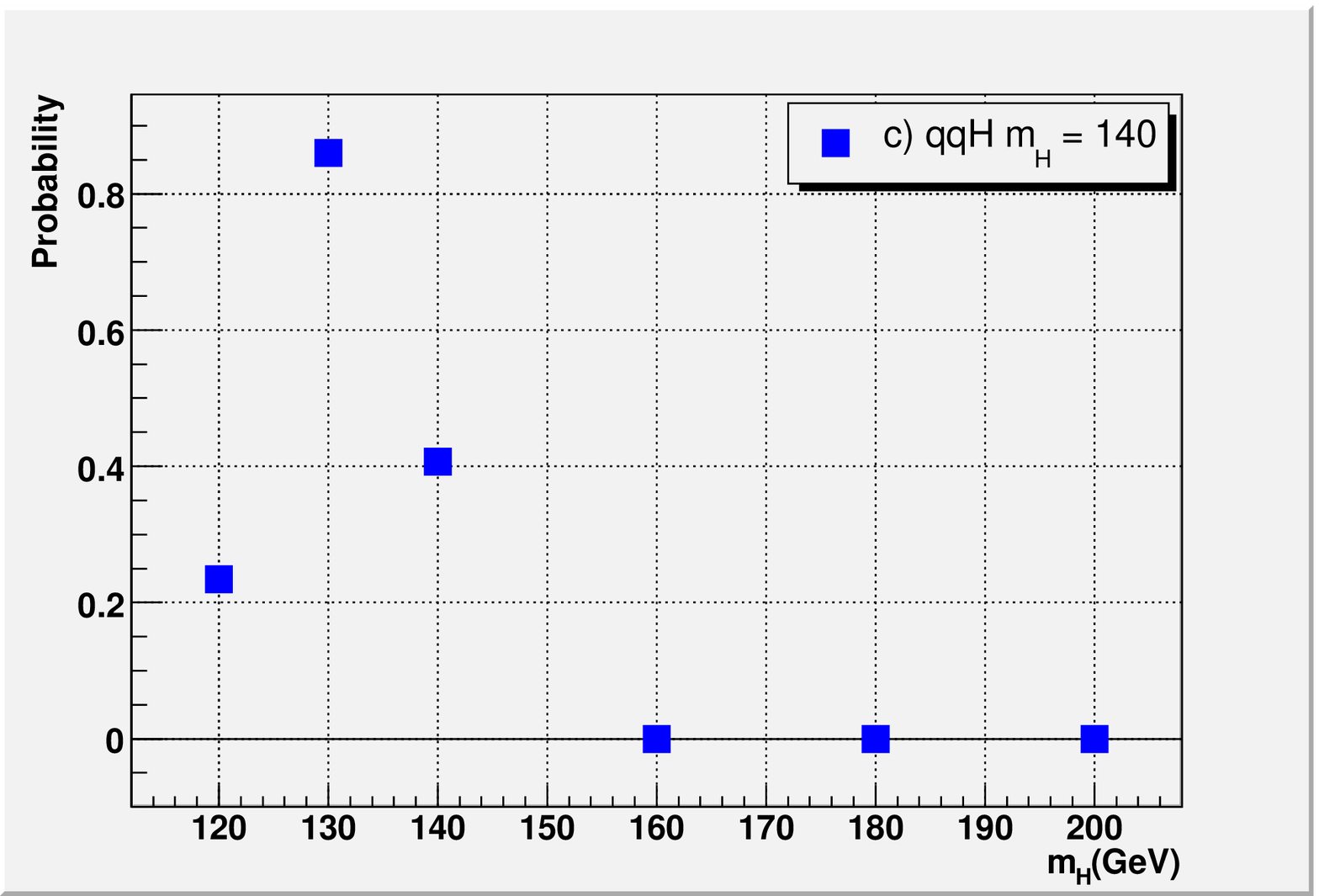}}
    \resizebox{9cm}{!}{\includegraphics{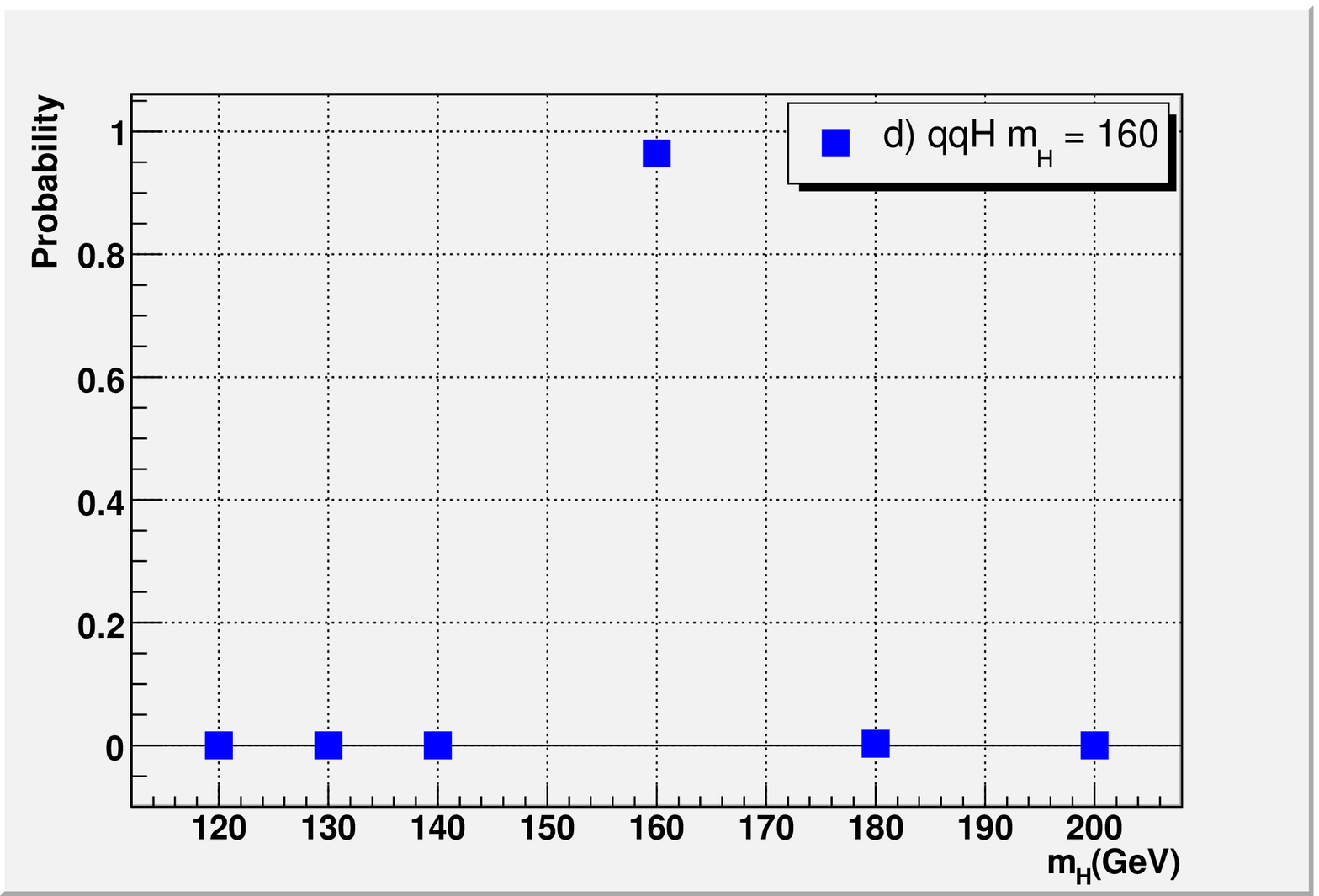}}
    \resizebox{9cm}{!}{\includegraphics{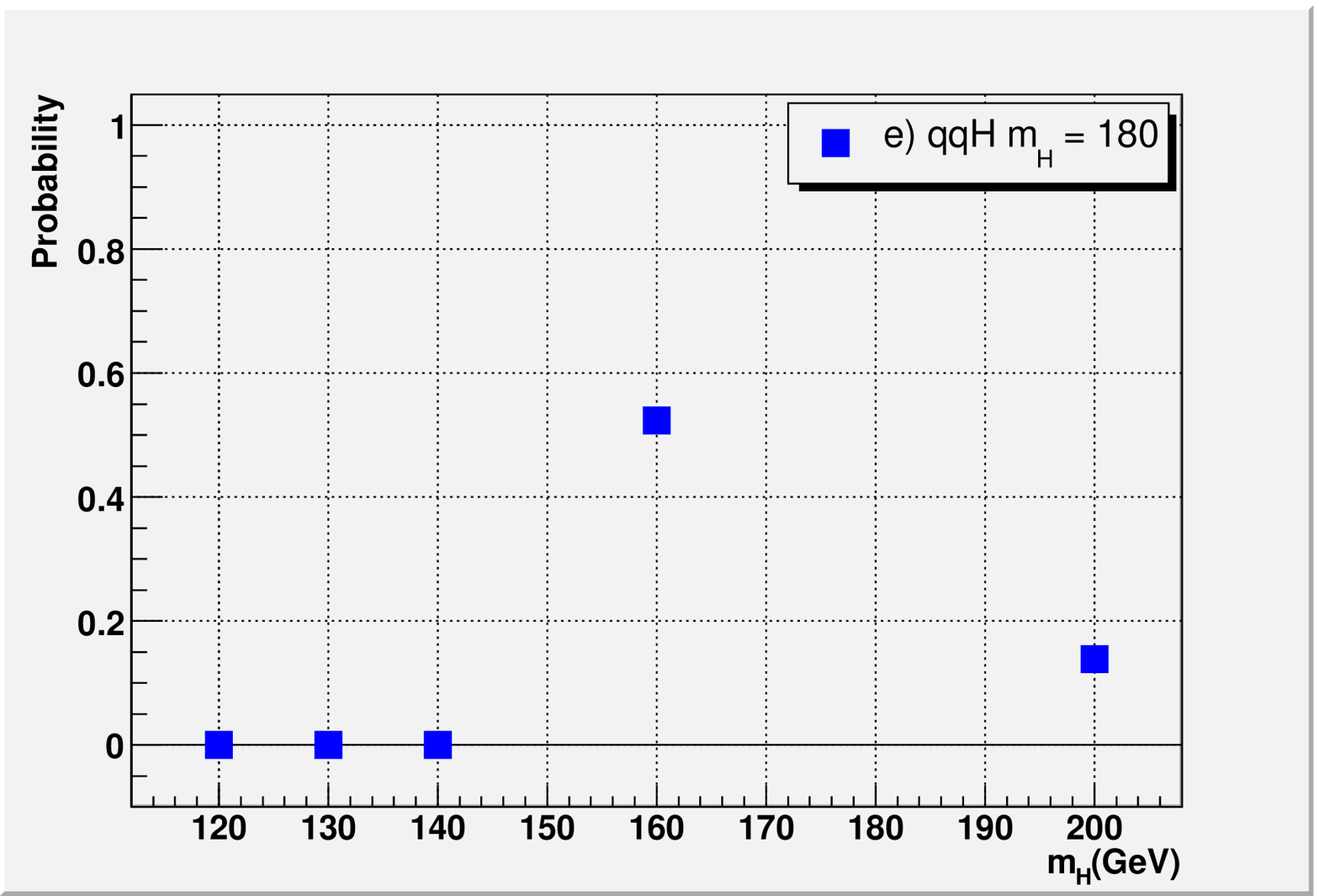}}
    \resizebox{9cm}{!}{\includegraphics{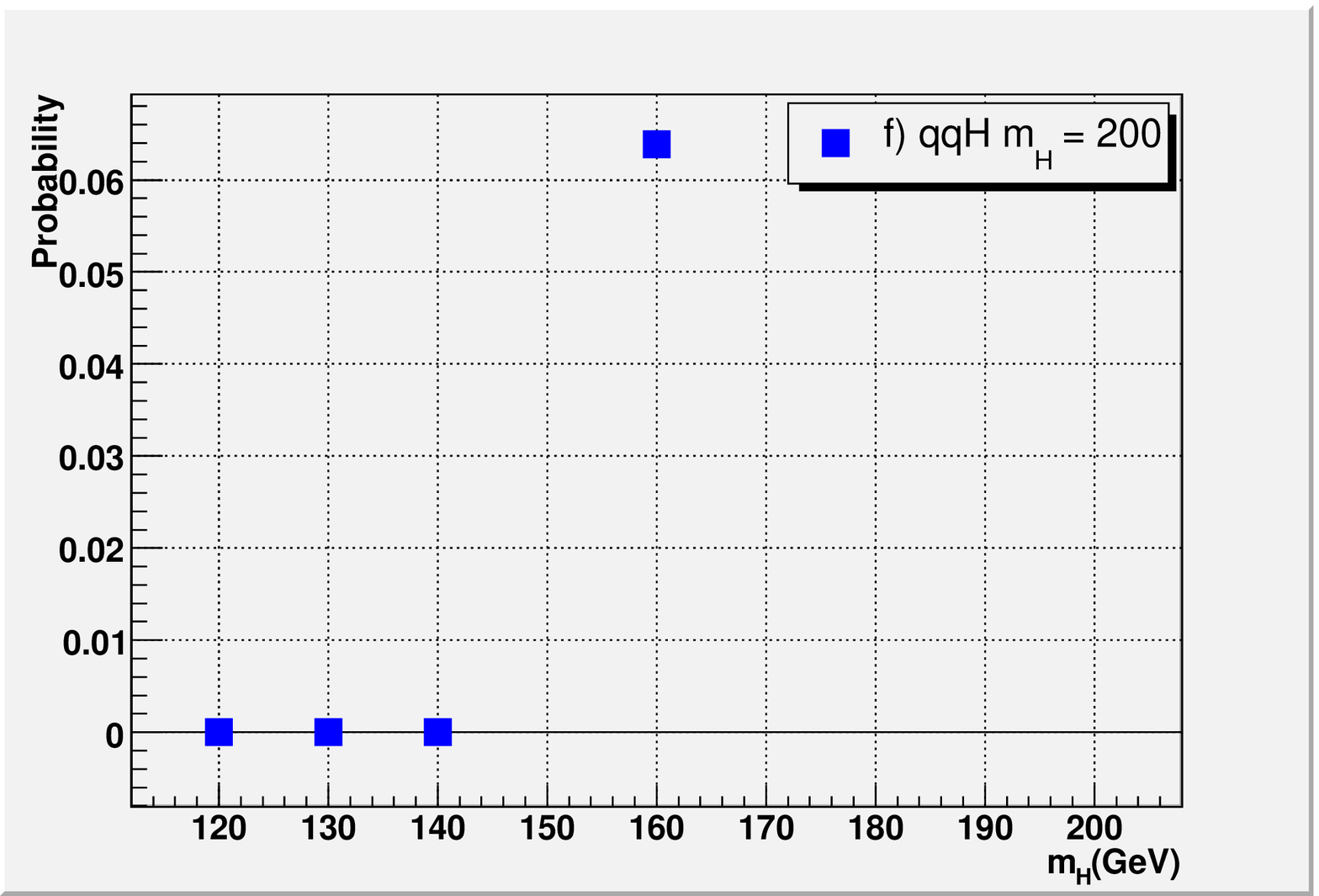}}
     \caption{\label{fig:kolmogorov}\em
Kolmogorov test function for estimating the Higgs boson mass for Higgs masses of 
120,130,140,160,180 and 200 GeV shown in a),b),c),d),e) and f) respectively.}
\end{figure*}

In an effort to obtain a quantitative measure of $m_H$, we can use
signal $M_{T,WW}$ distributions as templates to be compared to the
observed distribution.  The comparison is done using the
Kolmogorov-Smirnov test, and the results are shown in 
Fig.~\ref{fig:kolmogorov}. A value close to one indicates a good match 
between the shapes.  Comparing the means and shapes of the observed and 
template distributions, we can differentiate between Higgs boson masses for
the cases of 160, 180 and~$200~GeV$, and for low masses ($120$ -- $140~GeV$).
To differentiate between the cases of 120, 130 and $140~GeV$ Higgs mass,
we must reduce the $t\bar{t}j$ background more or we must have data corresponding 
to an integrated luminosity greater than $50~\mathrm{fb}^{-1}$.

\section{Conclusions}
We have presented an analysis meant to isolate a discovery signal for
a Standard Model Higgs boson in the vector-boson fusion channel.
We utilize the final state in which both $W$ bosons decay to electrons
or muons.  Our study is based on a full simulation of the CMS detector
and an up-to-date version of the reconstruction codes.  Furthermore,
we have generated the main backgrounds, $t\bar{t}j$ and $W^+W^-jj$, as
accurately as is presently possible.  
\par
The results of our study are encouraging, and indicate that an excess signal
with a statistical significance of over $5\sigma$ can be obtained with
an integrated luminosity of $>11~\mathrm{fb}^{-1}$ and $<72~\mathrm{fb}^{-1}$ for Higgs masses in the
range $130 < m_H < 200$~GeV. Our analysis also shows that the background
can be measured to  $7\%$ accuracy directly from the data.
This uncertainty is dominated by statistics for~$30~\mathrm{fb}^{-1}$.
Finally, we suggest a method to obtain information on the 
Higgs mass using the shape of the $M_{T,WW}$ distribution.

\section{Acknowledgments}
We are grateful to A. Nikitenko for his valuable help and useful 
comments. We would like to thank M. Zielinsky for his assistance in 
using tower  thresholds for jets and Y. Gerstein for his assistance in 
electron identification and selection. We also would like to thank 
N. Hadley, P. Bloch, R. Vidal and Albert DeRoeck for their comments, suggestions and criticisms. 
%
%
%
%
%

\clearpage

\end{document}